\documentclass[aps,prb,twocolumn]{revtex4-1}
\usepackage{graphics,graphicx,epsfig,color,hyperref,latexsym,float,amsmath}
\usepackage{times}
\input pdfcolor.tex

\begin{document}

\author{Sauri Bhattacharyya$^{1}$, 
Saurabh Pradhan$^{1,2}$ and Pinaki Majumdar$^{1}$}

\affiliation{
$^1$~Harish-Chandra Research Institute, HBNI,
Chhatnag Road, Jhunsi, Allahabad 211 019, India\\
$^2$~Department of Physics and Astronomy, Uppsala University,
751 05 Uppsala, Sweden
}

\title{Thermal dynamics of lattice modes near a polaronic crossover: \\
from the dilute polaron limit to a charge ordered state
}

\date{\today}

\begin{abstract}
We provide a comprehensive solution to the lattice dynamics problem in
the two dimensional Holstein model at finite electron density and finite
temperature. We work in the physically relevant adiabatic regime and vary
the electron-phonon interaction from the weak coupling perturbative window
to the strong coupling polaronic regime. We explore
three typical electron densities, dilute - where spatial correlations
between polarons is weak, intermediate - where correlations are
significant, and half-filling - where there is long range checkerboard
order at low temperature.
We use two methods both of which exploit the ``slowness'' of the phonons
to handle the problem. These are (i)~a standard random phase approximation
(RPA), adapted
to capture small quantum fluctuations on Monte Carlo generated classical
thermal backgrounds, and (ii)~a Langevin dynamics scheme, with a
simplified ``thermal noise'', that can address large amplitude dynamical
fluctuations. The Langevin scheme, as we argue in the paper, is the
superior method in the strong coupling part of the phase diagram, where
lattice distortions are large. It reveals a non trivial multi-peak
momentum resolved
spectrum with a high energy part, on the scale of the bare phonon
frequency $\Omega$, and a low energy peak at $\omega \ll \Omega$.
Below the polaronic threshold, the high energy dispersion changes
only modestly with temperature $T$, while the broadening, arising
from mode coupling, increases linearly with $T$ at low temperature.
The low energy peak shows up at strong coupling
and finite temperature and arises from the slow tunneling of polarons.
The tunneling events become spatially correlated as electron
density increases towards half-filling, and the weight becomes strongly
momentum and temperature dependent.  We suggest the analytic 
basis of these results.
\end{abstract}

\maketitle

\section{Introduction}

Strong electron-phonon interaction is a crucial ingredient 
in the physics of many functional materials\cite{pol1}. 
Among these are phonon mediated superconductors like alkali doped 
fullerides\cite{fullerides}, BiO based compounds\cite{BiO},  
the colossal magnetoresistance manganites\cite{CMR,man},
and possibly even the cuprates \cite{cpr1,cpr2}.
The presence of strong electron-phonon (EP) coupling 
is suggested in some cases by 
mid-infrared absorption \cite{fullerides,BiO} and in
some others by neutron scattering \cite{cpr1,cpr2,
weber1,weber2,weber3,weber4,kajimoto,reznik,dean,braden}. 

Strong EP interaction leaves an imprint on the phonon spectrum. 
In some metallic manganites, near a short-range charge ordered
(CO) insulator \cite{weber1,weber2,weber3,weber4}, 
one observes anomalous broadening and softening 
of bond-stretching phonons. Similar  
softening and band-splitting have 
been reported in doped nickelates\cite{kajimoto} 
near a checkerboard ordered phase. In cuprates
phonons are strongly affected \cite{reznik}
for momenta
close to the wavevector of stripe order, and recent 
experiments reveal  a non-trivial temperature 
dependence of the softening wavevector \cite{dean}. 
Finally, 
giant phonon anomalies have been observed in doped 
bismuthates \cite{braden}.
The effects above trace back to polaronic correlations
in these materials.

Polarons form as a bound state between an electron and a 
phonon cloud when the EP coupling exceeds a threshold. 
Theoretical problems in polaron physics are broadly 
to (a)~understand the `single polaron' problem,
(b)~address collective effects, {\it e.g}, 
ordering tendencies and spatial correlations among the polarons, 
(c)~probe the spectral
and conducting properties of the polaron fluid, and (d)~clarify 
the impact of polaron formation and CO 
correlations on the phonon dynamics.  (a)-(c) are reasonably 
well studied \cite{var,diag1,diag2,inst,berciu,
zeyher,millis1,millis2,ciuchi1,ciuchi2,kumar1}, 
(d), our focus, much less so.
The focus of this paper is on phonon
dynamics in electron-phonon systems when 
there are either large thermally induced 
distortions in a
metal close to a polaronic crossover or
there are polaronic distortions in the ground
state itself. In both these cases polaron tunneling leads to
a highly anharmonic signature in the dynamics.
Before discussing issues in phonon dynamics we 
quickly recapitulate the key results on (a)-(c)
to set the background.

The single polaron problem is essentially understood, 
via a combination of variational 
techniques\cite{var} - which
yields ground state energies, polaron bandwidth,
and electron effective mass,
numerical diagonalization\cite{diag1,diag2},
and instanton methods\cite{inst}.
The recently developed `momentum averaged' 
approximation\cite{berciu}
yields accurate spectral function in this
problem.
The traditional tool at {\it finite density} is
Migdal-Eliashberg (ME) theory \cite{ME theory}
but it cannot access polaron physics. 
To overcome this other approaches like 
adiabatic expansion \cite{zeyher,millis1,millis2},
or numerical schemes like quantum Monte Carlo \cite{QMC} 
(QMC) and dynamical mean field theory \cite{DMFT} (DMFT)
have been developed. DMFT suggests crossover from 
polaronic quasparticles at low temperature 
to an incoherent  high temperature phase \cite{ciuchi1} at low 
density, and a metal-insulator transition at half-filling  
\cite{ciuchi2}. 
In the strict adiabatic limit the 
ground state in two and three dimensions 
has also been established \cite{kumar1}.
 
The phonon physics is relatively less explored.
QMC investigations indicate the opening of a Peierls
gap in the strong coupling spinless 1D model 
\cite{creff}, and have probed the effect of Peierls 
transition on the dynamical structure factor \cite{hohen}. 
DMFT suggests phonon softening \cite{bulla} at strong coupling, 
but misses the momentum dependence of phonon broadening
and the connection between phonon dynamics and short-range 
static correlations. 
There are analytical studies at low filling \cite{fehske} in 1D 
and numerics at relatively weak coupling \cite{hague} in 2D, 
yielding phonon dispersion and self-energies.
Recently \cite{scalettar} the role of `bare' 
phonon dispersion in deciding 
charge density wave and superconducting 
transitions has been probed. 

The situations where the phonon dynamics remains poorly
understood are-
(i)~there are large thermally induced distortions in a
metal close to a polaronic crossover, 
(ii)~there are
polaronic distortions in the ground
state itself and the polarons undergo tunneling, and
(iii)~there are significant short range charge ordering
correlations. 

We address phonon dynamics in the context of 
the two dimensional Holstein model. We use methods
that are formulated in real space, to capture
spatial correlations, and real frequency (or real time).
The methods are described in detail later, at the moment
we just state that these are (a)~the random phase approximation
(RPA) applied on classical thermal backgrounds 
generated via a `static path approximation' (SPA) 
Monte Carlo (MC) strategy
\cite{kumar1}, 
and (b)~a dynamical approach involving a Langevin equation
driven by thermal noise \cite{chern,sauri}.
The RPA sets a reference because of its wide use in the 
community but we discover that in the
strong coupling regime the Langevin dynamics (LD) 
results capture the physics far better. 

The parameter space of our problem includes 
(i)~the ratio $\gamma$ of the phonon frequency $\Omega$
to electron hopping (indicating the degree of `adiabaticity'), 
(ii)~the ratio $\lambda$
of `polaron binding energy' $E_p$ to half of 
the electron bandwidth $W/2$,
(iii)~the electron density $n$, and (iv)~the temperature $T$.
We focus on the  
adiabatic regime, $\Omega/t \ll 1$, and
provide detailed phonon spectra over the
$\lambda-T$ parameter space at three typical densities:
(a)~$n=0.1$, where spatial correlations 
amongst polarons are weak, (b)~$n=0.4$, where
charge correlations are significant,
and (c)~$n=0.5$, where the system has a charge ordered ground state. 

While the results in the paper are organised in terms of the three
density regimes above,  a classification in terms of the 
dynamical regimes we observe is more compact. 
The relevant parameters for this are (i)~the strength
of coupling $\lambda$ with respect to the coupling
$\lambda_c(n)$ for the polaronic transition in the
ground state, and (ii)~the proximity of the density to half-filling,
$n=0.5$ (where a CO state occurs).
We denote the characteristic scale of polaronic
distortion as $x_{pol}\sim \sqrt{8\lambda t/K}$.
In terms of these our main results are below.

1.~We observe three broad dynamical regimes - 
(a)~perturbative, (b)~polaron tunneling, and (c)~large oscillations. 
Regime (a) involves small oscillations, $\Delta x \ll x_{pol}$,
about homogeneous backgrounds, accessible through both 
RPA and LD methods.  Regimes (b) and (c) involve `large amplitude' 
dynamics, where $\Delta x \gtrsim x_{pol}$.
These are inaccessible within the RPA scheme. 

2.~In the polaron tunneling regime 
the time series consists of small oscillations, 
with amplitude $\Delta x \ll x_{pol}$,
with occasional large moves, $\Delta x \sim x_{pol}$.
In the large oscillation regime there is  a 
continuum of fluctuation going even beyond $x_{pol}$. The power 
spectrum in the tunneling regime is `two branch' 
as opposed to a broad single branch for large oscillations.

3.~(i)~For $\lambda < \lambda_c(n)$, the low temperature dynamics
is accessible within RPA and falls in regime (a) and
even intermediate $T$ phonons can be described 
perturbatively, if one is not too close to $\lambda_c$.
(ii)~For $\lambda\sim\lambda_c(n)$ increasing $T$ leads to
a crossover from regime (a) to regime (c).
(iii)~For $\lambda > \lambda_c(n)$,
large distortions are present at $T=0$ and we observe 
dynamics characterized by regime (b) at low temperature, 
gradually crossing over to (c) with increasing $T$.

4.~At half-filling $\lambda_c=0$ and the ground state
is $(\pi,\pi)$ ordered. Here, RPA is valid over a wider region
at low $T$, whose dynamics is of type (a). We see a gradual evolution 
from (a) to (b) to (c) on increasing $T$.

The rest of the paper is organised as follows: in Section~II we discuss
our model and methods. Section~III provides an
overview of the dynamical regimes that emerge on varying
density, coupling and temperature.
Section~IV presents detailed results in the dilute regime,
$n=0.1$.
Section~V shows results in the `charge correlated' regime,
$n=0.4$, while Section~VI provides some results on
the charge ordered phase at $n=0.5$.
Section~VII discusses the interpretation, validity, and
applicability of our results.  We then conclude.
Some of the detailed formulae are given 
in an Appendix.

\section{Model and method}

We study the single band, spinless, Holstein model on a 
2D square lattice:
\begin{equation}
H=\sum_{<ij>}t_{ij}c^{\dagger}_{i}c_{j} 
+\sum_{i}(\frac{p^2_{i}}{2M} + \frac{1}{2}Kx^2_{i})-g\sum_{i}n_{i}x_{i}
\end{equation}

Here, $t_{ij}$'s are the hopping amplitudes. We study a 
nearest neighbour model 
with $t=1$ for three values of density, viz. 
$n=0.1$, $n=0.4$ and $n=0.5$. 
$K$ and $M$ are the stiffness constant and mass, respectively,
of the optical phonons, 
and $g$ is the electron-phonon coupling constant. 
We set $K=1$, $M$ is tuned to adjust the bare 
phonon frequency. The $M\rightarrow \infty$ limit
for fixed $K$ is the adiabatic limit. 
In this paper, we 
report studies for $\Omega = \sqrt{K/M} = 0.1$, 
which is a reasonable value for real materials. 
The chemical potential $\mu$ is varied to maintain 
the electron density at the required value.

\subsection{Random phase approximation (RPA) at $T \neq 0$}

The model above leads to the action
\begin{eqnarray}
 S &= &\int_0^{\beta}d\tau[
\sum_{ij} \bar{\psi_i}
\{(\partial_{\tau}-\mu)\delta_{ij} - t_{ij} \} \psi_j
+ \sum_i\bar{\phi_i}(\partial_{\tau}+\Omega) \phi_i \cr
&&~~~~~~~~~~~~~~  
-~ g\sqrt{\frac{\Omega}{2K}} 
\sum_{i}(\bar {\phi_{i}}+\phi_{i}){\bar\psi_{i}}\psi_{i}]
\end{eqnarray}
Here $\psi_{i}$ and $\phi_{i}$ are the fermion and coherent state Bose 
fields respectively.  and 
$\beta$ denotes the inverse temperature and we use units where
$k_{B}=1$, $\hbar=1$. The relation
between the real and coherent state fields is
\begin{equation}
x_{i}=\sqrt{\frac{1}{2M\Omega}}(\phi_{i}+\bar{\phi}_{i})
\end{equation}

In QMC \cite{QMC}, one `integrates out' the fermion fields
to construct the effective bosonic action. The
equilibrium configurations $\phi_i(\tau)$ 
of that theory are obtained via Monte Carlo
sampling.  Physical correlators are then computed as averages with
respect to these $\phi_i(\tau)$ configurations. Within DMFT 
\cite{DMFT} one maps the original action in Eq.(2) to an 
impurity problem, with parameters determined
self-consistently.

In the adiabatic limit, {\it i.e}, a static 
approximation for the phonons, 
one treats the $x_{i}$ as classical 
fields, neglecting their imaginary 
time dependence. This corresponds to the case 
of infinite $M$ for a fixed $K$. 
The quantum character 
of lattice vibrations is built in 
perturbatively. 

The full action containing the electrons and 
displacement fields can be 
rewritten in frequency space as-
\begin{eqnarray}
S &=
&S_{ph} + S_{f}\nonumber \cr
S_{ph}&=&\frac{1}{2}\sum_{i,m}\bar x_{im}(M\Omega_{m}^{2}+K)x_{im} \cr
S_{f} &=&\sum_{i,j,\alpha,\beta}\bar{\psi}_{i\alpha}
[(-i\omega_{\alpha}-\mu)\delta_{ij}\delta_{\alpha\beta}
- t_{ij}\delta_{\alpha\beta}  \cr
&&~~~~~~~~~~~~~~~~~~-g\sqrt{\beta}x_{i,\alpha-
\beta}\delta_{ij}]\psi_{j\beta}
\end{eqnarray} 
where $\omega_{\alpha}$ and $\omega_{\beta}$ are fermionic 
Matsubara frequencies, $\Omega_{m}$ is 
a Bose frequency. To get the partition function, 
one integrates over $\psi_{i\alpha}$, $x_{i0}$ and the $x_{im}$ fields.

The next step is to separate the Bose field into zero 
and non-zero Matsubara modes. We write $S=S_0 + S_1$ where 
the first part contains fermions coupling only to the 
zero frequency (static) mode and the second, the finite 
frequency modes.
We can formally `diagonalize' the fermions in presence of the 
static mode to write $S_0$ as-
\begin{equation}
	S_{0}={\sum_{l,\alpha}\bar{\xi}_{l,\alpha}
	(-i\omega_{\alpha}+\epsilon_l)\xi_{l,\alpha}} + 
\frac{1}{2}K\sum_ix^2_{i0}
\end{equation}
where $\xi$'s correspond to 
the fermionic eigenmodes in the $\{x_{i0} \}$ background. 
$S_0$ defines the static path approximation (SPA) action. 
The eigenvalues ($\epsilon_l$'s ) depend  non-trivially on the 
$\{x_{i0} \}$ background. To obtain the effective zero mode 
distribution $P\{x_{i0}\}$, one has to integrate out the 
fermions. At the SPA level, there exists an effective 
Hamiltonian $H_{eff}$, depending on $x_{i0}$, 
for the fermions and the distribution can be formally written as-
\begin{eqnarray}
P\{x_{i0}\} &= & Tr_{c,c^{\dagger}}e^{-\beta (H_{eff} +
 {1 \over 2} K x_{i0}^2) } \cr
\cr
H_{eff} & =& \sum_{<ij>}t_{ij}c^{\dagger}_{i}c_{j} - g \sum_{i}n_i x_{i0} 
\nonumber
\end{eqnarray}

Around the SPA action, we set up a cumulant expansion of $S_1$.
Owing to the disparate timescales of the phonons and electrons, we 
attempt a sequential integrating out. First, 
the fermions are traced out and one obtains 
the following expression for $S_1$
in terms of the $x_i(i\Omega_m)$:
\begin{eqnarray}
 S_1 &=
 &S^{\prime}_{ph}-Trln[\beta
((-i\omega_{\alpha}-\mu)\delta_{ij}\delta_{\alpha\beta} \nonumber 
\\  &&~~~~~~~~~~~~~~~ -t_{ij}\delta_{\alpha\beta}-g\sqrt{\beta}x_{i,\alpha-
\beta}\delta_{ij})]\nonumber\\
S^{\prime}_{ph}&=&\frac{1}{2}\sum_{i,m\neq 0}\bar
 x_{im}(M\Omega_{m}^{2}+K)x_{im}
\end{eqnarray} 
Analytic determination of the trace is impossible beyond
weak coupling. 
Our method constitutes of 
expanding the trace in finite frequency $x_{i}$ modes upto Gaussian 
level. Physically, we assume that the quantum fluctuations are small 
in amplitude, controlled by a low $\gamma=\Omega/t$ ratio. 
Retaining only 
quadratic terms in the dynamic modes allows us to analytically 
integrate them out later. The method is obviously perturbative in 
$\gamma$ as we diagonalize the fermions in presence of static 
distortions and evaluate correlation functions on this `frozen' 
background, which enter as coefficients of the quadratic Bose term 
for finite frequencies. After 
re-exponentiating this term 
the new partition function becomes- 
\begin{equation}
 Z = \int [D\bar{x}][Dx][D\bar{\xi}][D\xi]e^{-(S_{0}+S_{1})}
\end{equation}
where 
\begin{eqnarray}
		S_{0}&=&{\sum_{l,\alpha}\bar{\xi}_{l,\alpha}
(-i\omega_{\alpha}+\epsilon_l)\xi_{l,\alpha}} +\frac{1}{2}Kx_{i0}^2\\ \nonumber
		S_{1}&=&{\sum_{i,j,m>0}\bar{x}_{im}[(M\Omega^2_{m}
+K)\delta_{ij}+{g^2}\Pi_{ij}^{m}(\{x_{i,0}\})]x_{jm}}\\ \nonumber
\end{eqnarray}
Here $\Pi_{ij}^{m}$ is a one-loop fermion polarization correcting the 
free Bose propagators. The expression for this is
\begin{equation}
\Pi_{ij}^{m}(\{x_{i0}\})=\frac{1}{\beta}\sum_{\alpha}G_{ij}^{\alpha}G_{ji}
^{\alpha-m}
\end{equation} 
$G_{ij}^{\alpha}$'s are Matsubara components of real-space fermion Green's 
functions computed in an arbitrary static background. One can write a spectral 
representation of this as follows-
\begin{equation}
G^{\alpha}_{ij}(\{x_{i0}\})= \sum_n \frac{u_{in}\bar{u}_{jn}}{i\omega_{\alpha}
-\epsilon_n}
\end{equation}
where $u_{in}$ is amplitude at site $i$ for the 
$n$th eigenstate of 
the SPA Hamiltonian and $\epsilon_n$'s are 
the corresponding eigenvalues. 

The fermion 
diagonalization for arbitrary static distorted backgrounds has to be dealt 
with numerically. To access large system sizes within a reasonable simulation 
time, a cluster algorithm is used for each Metropolis update. The method has 
been extensively benchmarked earlier.
\cite{kumar2}.
Our studies are on 32$\times$32 lattices 
using 8$\times$8 clusters.

The coefficient of the quadratic Bose term of $S_1$ in (9) defines the inverse 
propagator $(D^{-1})_{ij,m}$ for the renormalized phonons. One can view the 
$\Pi^{m}_{ij}$ as a self-energy for the phonons in real space, correcting 
the bare propagator. To 
get the renormalized Green's function, we solve a Dyson's equation 
on the lattice with inhomogeneous backgrounds.
\begin{equation}
[D]^{-1}_{ij}(\omega)=[D]^{-1}_{0,ij}(\omega)+g^{2}[\Pi]_{ij}(\omega)
\end{equation}
Here the bare phonon propagator $D^{-1}_{0,ij}$ is defined 
through the equation- 
\begin{equation}
[D]^{-1}_{0,ij}(\omega)=(M\omega^2-K)\delta_{ij}
\end{equation}
where we have analytically continued the Green's functions to real
frequency and $D^{-1}_{0,ij}$ denotes the inverse propagator for
the bare phonon.
This involves 
a calculation that grows as $O(N^4)$ with increasing lattice size $N$. For 
lattices of size 18$\times$18, this computation was implemented. 

While we used the scheme above as our benchmark, we tried out an
approximation, explored earlier in the context of DMFT, where 
the polarization is replaced by its $\omega_{m}=0$ 
component for all the Bose
modes\cite{millis2}. This approximation renders a direct solution of Dyson's 
equation for the phonon propagator unnecessary. Instead, to extract the 
renormalized spectrum, one has to solve a single Hamiltonian problem for the 
bosons in real space for each background configuration. The results
reasonably match with full RPA in the weak and strong coupling windows,
but aren't reliable near the crossover.

On structurally disordered configurations, $\Pi_{ij}(\omega)$ is not 
translationally invariant. 
We take a Fourier transform of $D_{ij}$ with respect to ($\vec{R_{i}}
-\vec{R_{j}}$), compute the phonon spectrum and finally average over 
equilibrium backgrounds. From this Bose propagator, 
we also find out the phonon density of states (DOS) 
by tracing over momenta.  The expressions for $\Pi^{m}_{ij}$ 
and $D^{m}_{ij}$ in the $\omega_{m}=0$ approximation are quoted 
in the Appendix. 

This RPA method, while being complicated in its implementation,
essentially captures small amplitude quantum fluctuations about
equilibriun `backgrounds'. Hence, it's adequate in the low temperature, 
weak coupling scenario. However, one misses out on anharmonic 
`mode coupling' physics and most importantly, thermally induced 
polaron tunneling.
The latter involves `swapping' of large distortions on neighbouring
sites and are crucial in describing \textit{critical dynamics} 
at half-filling. Even at lower densities, these moves help in 
cause a very interesting 
low-frequency weight transfer in the phonon spectrum. 
To capture them, we've used a real time 
equation of motion method, described next.

\subsection{Langevin Dynamics (LD) method}

The Holstein problem can be set up in the Keldysh language in terms of
coherent state fields corresponding to $x_{i}$ and $c_{i}$ operators, 
with their full space-time 
dependence retained. We indicate how 
a Langevin-like equation of motion \cite{sauri} 
can be obtained from the Keldysh action 
in the adiabatic limit. Physically, this corresponds to a 
``small $\dot{x}$'' approximation- namely the velocity of this field is 
much smaller than Fermi velocity.  We outline this below.

The partition function for the $x_i$ `oscillators' can be written as-
\begin{equation}
Z_{osc}=\int Dx_{i,f}Dx_{i,r}e^{ i(S_0+S_1) }
\end{equation}
Here $x_{i,f}$ and $x_{i,r}$ are lattice displacement fields along
forward and return contours respectively. The expressions for $S_{0}$ 
and $S_{1}$ are- 
\begin{eqnarray}
S_{0}&=&\frac{1}{2}\int dt[\sum_{i}(M\ddot{x}_{i,f}+Kx_{i,f})x_{i,f}-
(M\ddot{x}_{i,r}+Kx_{i,r})x_{i,r}] \cr
S_{1}&=&iTr(log([\mathcal{G}^{-1}]_{ij}(t,t^{\prime}))
\end{eqnarray}
where $\mathcal{G}$ is the matrix electron Green's function (with 
dimension $N\times N$ in real space and $2\times2$ in Keldysh space) in a 
time fluctuating $(x_{i,f},
x_{i,r})$ `background'.
To facilitate the derivation, one can transform to new 
`classical' and `quantum' variables 
$$
x_{i,cl}  = \frac{x_{i,f}+x_{i,r}}{2},~~~~~
x_{i,q}   = x_{i,f}-x_{i,r}
$$

We perturbatively expand $S_{1}$ in powers of $x_{i,q}(t)$
while retaining $x_{i,cl}(t)$ non-perturbatively in the theory. 
The parameter that controls the expansion \cite{martin}
is $\Omega/t$. 
The expansion in powers of $x_{i,q}(t)$ means we're adopting a 
semiclassical picture. Expanding up to linear order gives classical 
deterministic phonon dynamics. The quadratic term carries the effect of an 
added noise. This is done following the lines of Ref.39.

The look of the effective action for the oscillators now is-
\begin{eqnarray}
S_{eff} &=& 
S_{0} + g\sum_{i}\int dt[G^{K}_{cl}]_{ii}(t,t)x_{i,q}(t) \cr
&&~
+ g^{2}\sum_{ij}\int dt dt^{\prime}[\Pi^{K}_{cl}]_{ij}(t,t^{\prime})
x_{i,q}(t)x_{j,q}(t^{\prime})
\nonumber
\end{eqnarray}
where $[G^{K}_{cl}$] is the Keldysh component of electron Green's function
$\mathcal{G}$ computed setting $x_{i,q}=0$. The quantity 
$[\Pi^{K}_{cl}$] is the Keldysh
component of electronic polarizability for $x_{i,q}=0$, related to the 
Green's functions by the relation-
$$
\Pi^{K}_{ij}(t,t^{\prime})=G^{R}_{ij}(t,t^{\prime})G^{A}_{ji}(t^{\prime},t)
+(R \leftrightarrow A)
+G^{K}_{ij}(t,t^{\prime})G^{K}_{ji}(t^{\prime},t) 
$$
$G^{R}$ and $G^{A}$ being retarded and advanced components of $\mathcal{G}$.

The coefficients of the linear and quadratic terms in $x_{i,q}(t)$ are 
thus determined through computing electronic correlation 
functions in an `arbitrary'
$x_{i,cl}(t)$ background. This calculation can be simplified by expanding
the `trajectories' $x_{i,cl}(t)$ around a reference time $t_{0}$ in powers
of the velocity $\dot{x}_{i,cl}$. The velocity independent term is interpreted
in terms of a force exerted by an instantaneous effective Hamiltonian. The
linear in $\dot{x}_{i.cl}$ term gives rise to `damping' with a frequency 
dependent kernel. 

The next stage of approximation concerns the frequency dependence of
the Keldysh component
of electronic polarizability $\Pi^{K}_{ij}(\omega)$. At equilibrium, the
frequency dependence of this quantity can be factored according to
fluctuation-dissipation theorem\cite{neq} as-
\begin{equation}
 \Pi^{K}_{ij}(\omega)=coth(\frac{\omega}{2k_{B}T})
 (\Pi^{R}_{ij}(\omega) - \Pi^{A}_{ij}(\omega))
\end{equation}
where $\Pi^{R}_{ij}$ and $\Pi^{A}_{ij}$ are the retarded and advanced
components of the polarizability respectively.
These are defined as-
\begin{equation}
\Pi^{R/A}_{ij}(t,t^{\prime})=G^{R/A}_{ij}(t,t^{\prime})G^{K}_{ji}(t^{\prime},t)
+(R/A \leftrightarrow K)
\end{equation}

Next, we make the high temperature ($k_{B}T \gg \omega$) approximation on 
the RHS. The hyperbolic cotangent gives a factor of ($2k_{B}T/\omega$), 
and the low frequency spectral part of $\Pi$ contributes $\gamma\omega$,
where $\gamma=g^{2}\frac{Im(\Pi^{R}(\omega))}{\omega}$ and we've neglected
the spatial dependence of the polarizability.

If one carefully carries out the evaluation of the linear in velocity
($\dot{x}_{i,cl}$) term, the coefficient comes out to be the the spectral
part of the polarizability $Im(\Pi^{R}_{ij}(\omega))$. Again neglecting
spatial dependences here and taking the low-frequency limit, the term
simplifies to $\gamma\omega$ and becomes the usual non-retarded Langevin
damping coefficient.

Finally, one decouples the quadratic term in $x_{i,q}(t)$ through
a Hubbard-Stratonovich transformation introducing a `noise' field 
$\xi_{i}(t)$ and then integrates over $x_{i,q}(t)$ in the 
partition function to obtain an `equation of motion' \cite{martin} 
for $x_{i,cl}(t)$.
This leads to our dynamical equation, below.
\begin{eqnarray}
M\ddot x_{i}(t) ~~~ & = &~ -\gamma\dot{x}_i(t)  - K x_i(t) - 
{ {\partial {\langle H_{el}\{ x\} \rangle }}   \over {\partial x_i }} 
+ \xi_i(t) \cr \cr
H_{el}~~~~~~ & = &~ \sum_{ij}(t_{ij} - \mu \delta_{ij}) 
c^{\dagger}_{i}c_{j} - g\sum_in_ix_i
\cr \cr
{\partial {\langle H_{el}\{ x\} \rangle }}   \over {\partial x_i }
& = &~ -g {\bar n}_i(t) \cr
\cr
{\bar n}_i(t) ~~~~~&  = &~ \sum_{\epsilon_{n}(t)}\vert U_{in}(t)\vert^2
n_f(\epsilon_{n}(t))
\end{eqnarray}
where $U_{in}(t)$ are site amplitudes of the instantaneous eigenvectors 
of $H$ (as in Eq.1) for a given $x_{i}(t)$ configuration and $\epsilon_{n}(t)$
are the corresponding eigenvalues. $n_f(\epsilon_{n}(t))$ denotes Fermi 
factors.

The first term in the right hand side describes damping while
the second and third terms are effective forces.
Note that  {\it both spatial correlations and nonlinearities in the
$x_i$}
arise from the implicit dependence of $n_i$ on the phonon background
$\{ x_i \}$.
The last term is the noise field, specified by the conditions-
$$
\langle \xi_{i}(t)\rangle =0,~~~~ 
\langle \xi_{i}(t)\xi_{j}(t^{\prime}) \rangle = 2\gamma 
k_{B}T\delta_{ij}\delta(t-t^{\prime})
$$
The unit of time is taken to 
be the inverse of the bare oscillator frequency $\tau_{0}=2\pi/\Omega$.  
For most of our simulations, we chose $\gamma=1.0t$, which sets the
damping timescale to $2M/\gamma= 3\tau_{0}$. The imaginary part of the
retarded polarizability $Im \Pi^{R}(\bf {q},\omega)$ is gapped at low $T$
in the present model. At intermediate temperatures, it picks up a low
energy contribution proportional to $\omega$. The microscopic estimate
of $\gamma$, based on $Im \Pi^{R}$, is smaller and also $T$ dependent.
To minimise parameter variation and ensure reasonably rapid
equilibriation we have set $\gamma=1$.

We integrate the equation numerically using the well-known Euler-Maruyama 
method \cite{kloeden}.
The time discretization for most calculations was set to 
$\Delta t= 1.6 \times 10^{-4} \tau_{0}$. We typically ran the simulations for 
$\sim10^7$ steps, ensuring a time span of almost a few hundred times the 
equilibration time. This ensured enough frequency points to 
analyze the power spectrum.

\subsection{Calculating indicators}

We quantify the equal-time and dynamical properties through several indicators
within Langevin dynamics.
We first define some timescales. We set an `equilibriation time' $\tau_{eq}
= 100 \tau_0$ before saving data for the power spectrum. The outer timescale,
$\tau_{max} \sim 10 \tau_{eq}$.  The `measurement time' $\tau_{meas}
= \tau_{max} - \tau_{eq}$, and the number of sites is $N$. We calculate the 
following:
\begin{enumerate}
\item
Dynamical structure factor, $D({\bf q}, \omega) = \vert
X({\bf q}, \omega) \vert^2$,
$$
X({\bf q}, \omega) = 
\sum_{ij} \int_{\tau_{eq}}^{\tau_{max}} 
dt e^{i {\bf q}. {\bf r}_i }
e^{i \omega t}
x({\bf r}_i, t)
$$
\item
The structure factor 
\begin{eqnarray}
{\bar S}({\bf q}) ~&=& {1 \over \tau_{meas}} 
\int_{\tau_{eq}}^{{\tau}_{max}} dt e^{i \omega t} S({\bf q},t) 
\cr
S({\bf q},t) &=& {1 \over N^2} 
\sum_{ij} e^{i {\bf q}. ({\bf r}_i - {\bf r}_j) }
x({\bf r}_i, t) x({\bf r}_j, t) 
\nonumber
\end{eqnarray}
\item
The distribution of distortions:
$$
P(x)= \frac{1}{N \tau_{meas}} \sum_{i} 
\int_{\tau_{eq}}^{\tau_{max}} dt \delta(x - x_{i}(t))
$$
\item
Dispersion $\omega_{\bf q}$ and damping $\Gamma_{\bf q}$: 
\begin{eqnarray}
\omega_{\bf q} &=& \int_{0}^{\omega_{max}} d\omega 
\omega  D({\bf q}, \omega) \cr 
\Gamma_{\bf q}^{2} & =  &
\int_{0}^{\omega_{max}} d\omega 
(\omega - \omega_{{\bf q}})^2 D({\bf q}, \omega) 
\nonumber
\end{eqnarray}
While calculating moments, we've normalized by ($K/8\lambda t$),
to ensure dimensional consistency.
\end{enumerate}

Within RPA, the statics is quantified through $P(x)$ and
${\bar S}({\bf q})$. The time averaging is replaced by average over equilibrium
configurations. The dynamics is quantified through 
$$
A({\bf q}, \omega)=-\frac{1}{\pi}Im(\bar{D}({\bf q},\omega))
$$
where $\bar{D}({\bf q},\omega)$ is the spatial Fourier transform
of $D_{ij}(\omega)$ as defined in Eq.(10).

\section{Overview of thermal regimes in dynamics}

Fig.1(a) features the ground state coupling-density ($\lambda-n$)
phase diagram, obtained using MC simulations. 
There are three prominent phases- (i) Fermi liquid
(FL), where the state is homogeneous, (ii) polaron liquid (PL), where
distortions have formed on the lattice, but spatial correlations
are not very significant, (iii) charge correlated (CC), where
short-range charge order is present. Shaded areas indicate
regions of phase separation (PS).
Moreover, the actual charge-ordered
(CO) phase features exactly at half-filling for all couplings.
We choose three densities (indicated through vertical lines)
for our investigation of the thermal
dynamics. These are- (i) `dilute' ($n=0.1$), which exhibits a clean 
FL-PL crossover around $\lambda=0.8$, (ii) `correlated' ($n=0.4$), 
where FL transits to CC through a PS regime 
near $\lambda=0.4$, and (iii) `ordered' ($n=0.5$). 
The choice is motivated by the character of the respective
phases at strong coupling.

\begin{figure*}[t]
\centerline{
\includegraphics[height=4.3cm,width=4.3cm]{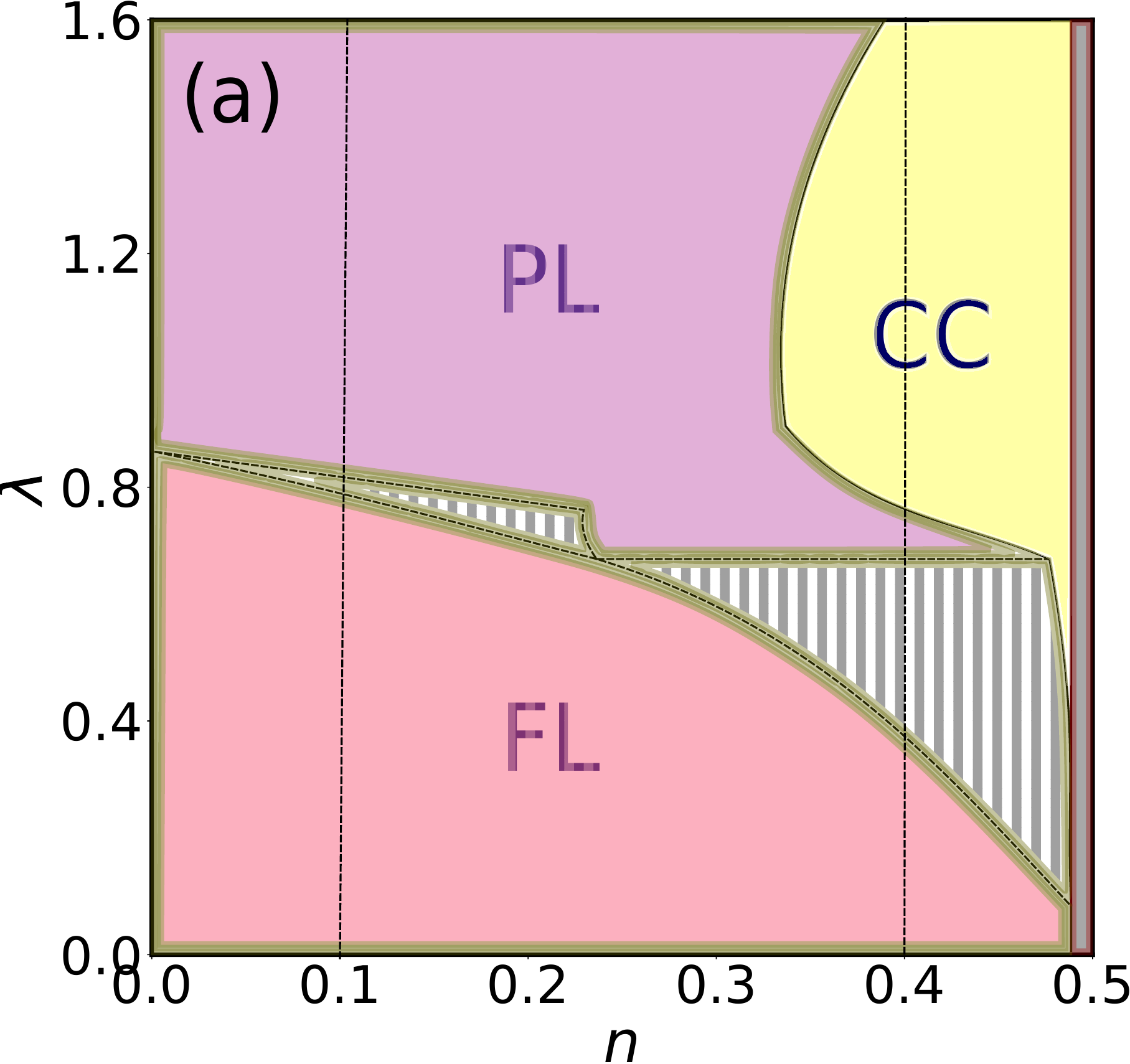}
\includegraphics[height=4.3cm,width=4.3cm]{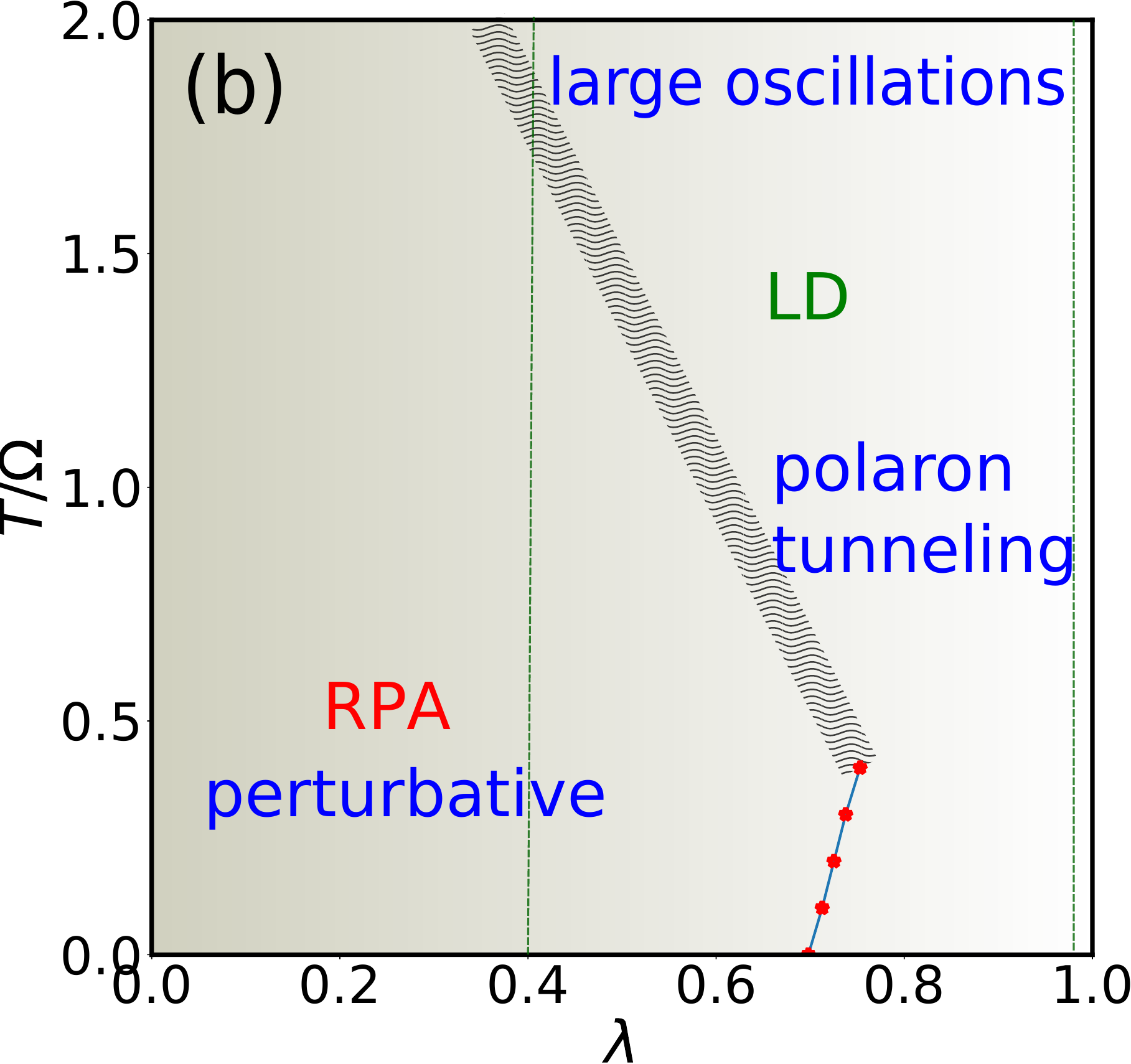}
\includegraphics[height=4.3cm,width=4.3cm]{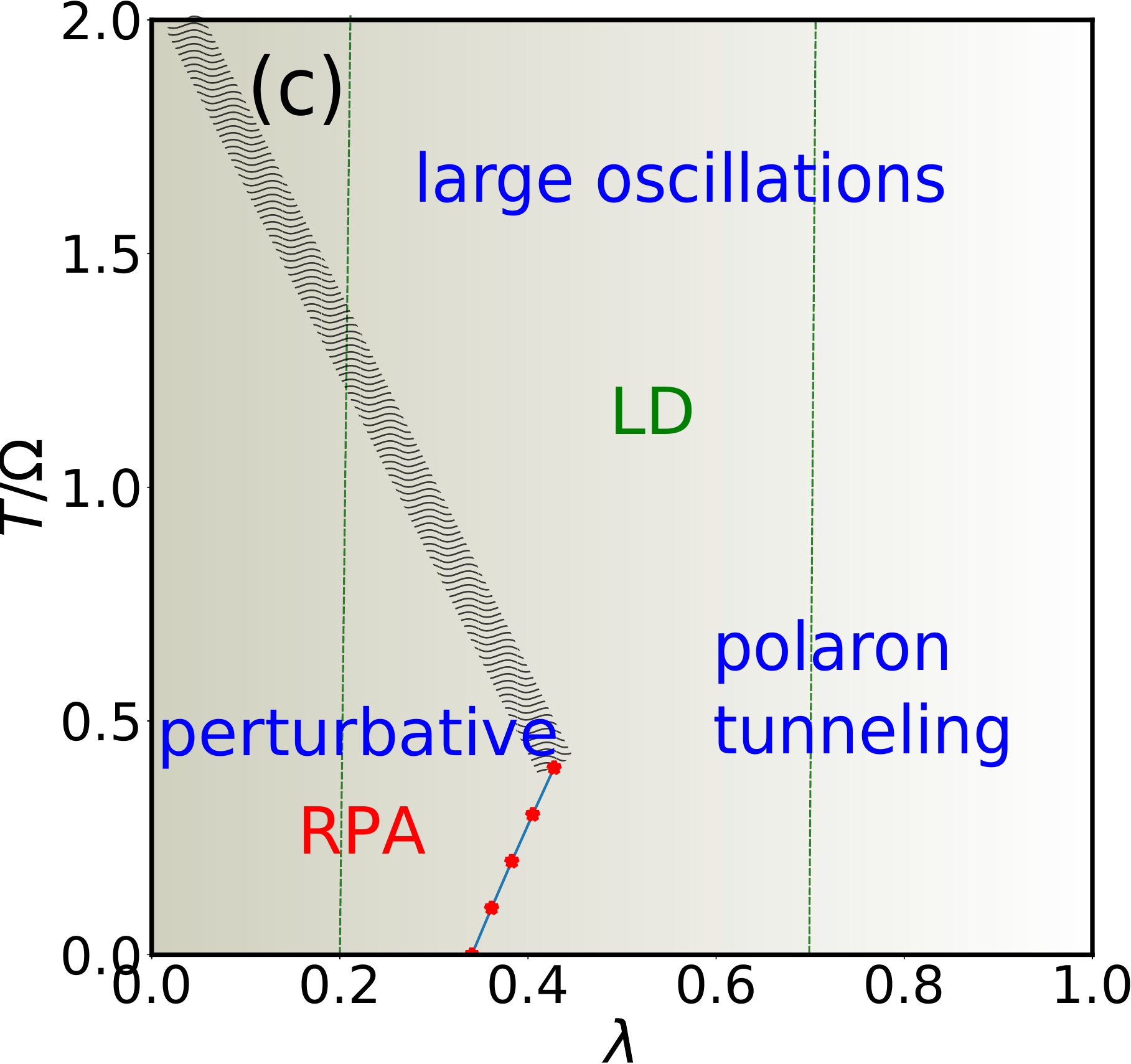}
\includegraphics[height=4.3cm,width=4.3cm]{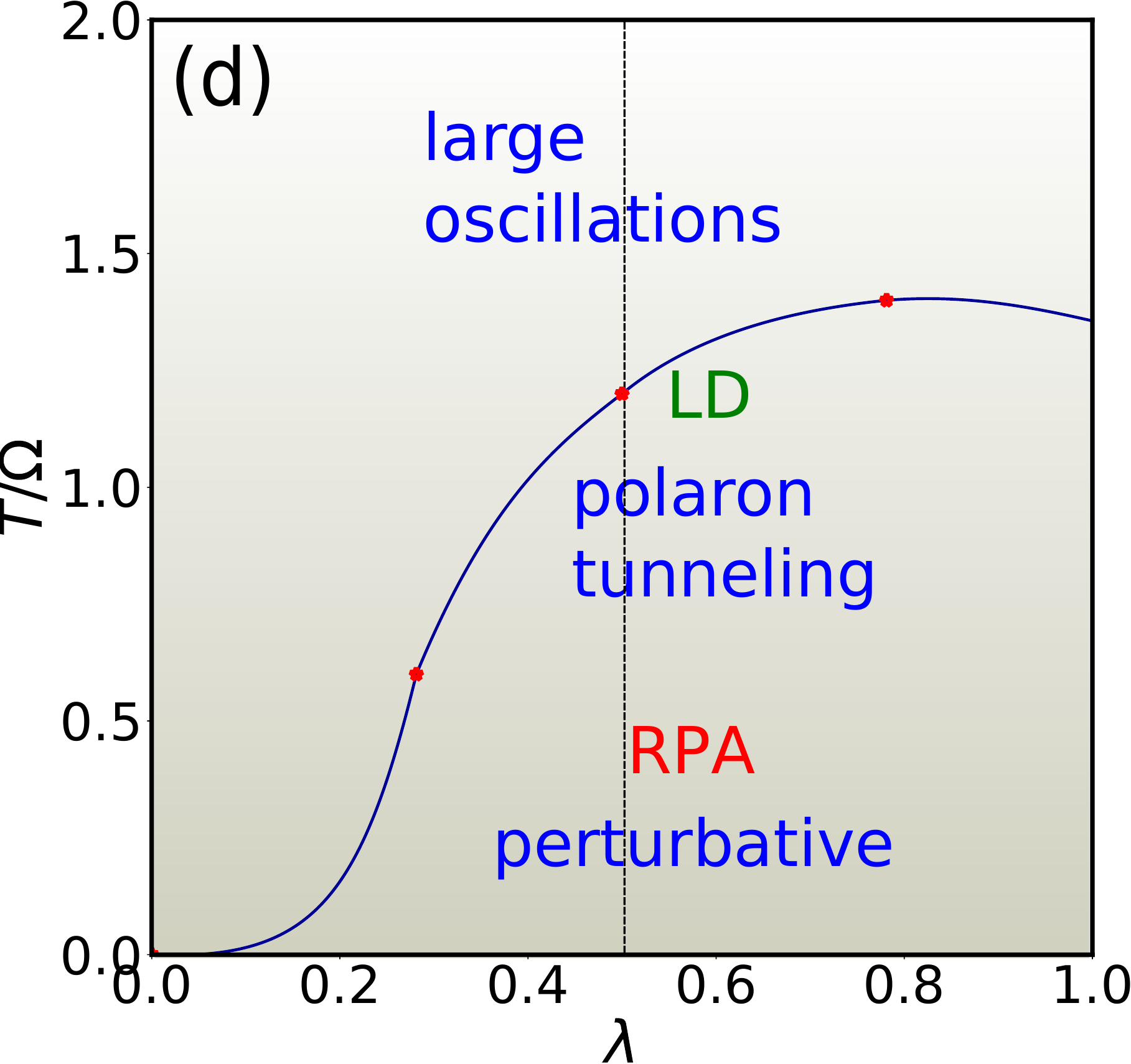}
}
\caption{2D Holstein model: adiabatic ground state and dynamical regimes.
(a): Ground state phase diagram obtained via Monte Carlo for $\Omega=0$. 
We show the undistorted (Fermi liquid - FL), distorted but `uncorrelated'
(polaron liquid - PL) and charge correlated (CC) regimes. $n=0.5$ has
long range charge order (CO). The $n=0.1,~0.4,~0.5$ cross sections where
we study finite $T$ dynamics are highlighted. 
(b)-(d): Dynamical regimes of phonon behaviour at $n=0.1,~0.4,~0.5$ 
respectively, at $\Omega/t = 0.1$. The dimensionless EP coupling is
$\lambda = g^2/(8Kt)$. The critical coupling $\lambda_c$ 
for transition from the FL to a PL or CO state collapses as $n 
\rightarrow 0.5$. We observe three broad dynamical regimes - 
(a)~perturbative, (b)~polaron tunneling, and (c)~large oscillations. 
In (a), we see small oscillations, $\Delta x \ll x_{pol}$,
about homogeneous backgrounds.
Regimes (b) and (c) involve `large amplitude' 
dynamics, where $\Delta x \gtrsim x_{pol}$.
A two-branch spectrum emerges in the former.
}
\end{figure*}

Figs.1(b)-(d) represent the phonon dynamics at these
three densities. The metallic and polaronic regimes
at low $T$ are separated by the solid blue lines
in the first two panels. The first order discontinuity
ends at a temperature $T^{*}\sim0.5\Omega$. 

We have three broad dynamical regimes- 
(i) perturbative, (ii) polaron tunneling and (iii) large
oscillations in each of them. 
The regions to the bottom left 
characterize the weak coupling phase. The phonons here
belong to regime (i) and are described reasonably well by the RPA approach.
There's a harmonic-anharmonic crossover around $T\sim T^{*}$.
Close to $\lambda_c$, heating up produces thermally induced
polarons, which feature short-range correlations for $n\sim0.5$.
The wavy lines represent the resulting dynamical crossover
from (i) to (ii) on the weak coupling side. In the polaronic
phase, the low temperature dynamics is that of regime (ii), 
which crosses over smoothly to (iii) on heating up.

At half-filling, there's a thermal transition from checkerboard
order to a disordered state. The corresponding $T_{CO}$ is sketched in 
solid blue in Fig.1(d). The low $T$ dynamics is perturbative.
Beyond $T\sim0.5T_{CO}$, correlated tunneling events start
showing up appreciably. Beyond $T\sim1.5T_{CO}$, these
events merge with large oscillations.

\section{The dilute regime}

\subsection{Static properties}

The statics is quantified using the distribution of the
displacement field $P(x,T)$ and the phonon structure factor 
$S({\bf q},T)$. The former shows a unimodal to bimodal
transition across the critical coupling ($\lambda_{c}=0.8$) at low $T$.
Figs.2(a)-(b) (top panel) features the thermal behaviour in the weak 
($\lambda/\lambda_{c}=0.5$)
and strong ($\lambda/\lambda_{c}=1.2$) coupling regimes. We see
$\sim\sqrt{T}$ broadening at weak coupling and gradual 
smearing of the large distortion peak in the strong coupling case.
The weight of the second peak is proportional to electron density.
One also observes a quantitative agreement between answers obtained
using SPA-MC and LD approaches.

The structure factor $S({\bf q})$ (bottom panel) is benign for all 
momenta as the polarons don't influence each other. 
Therefore, we focus on dynamics in the following section.

\begin{figure}[b]
\centerline{
\includegraphics[height=4cm,width=4cm]{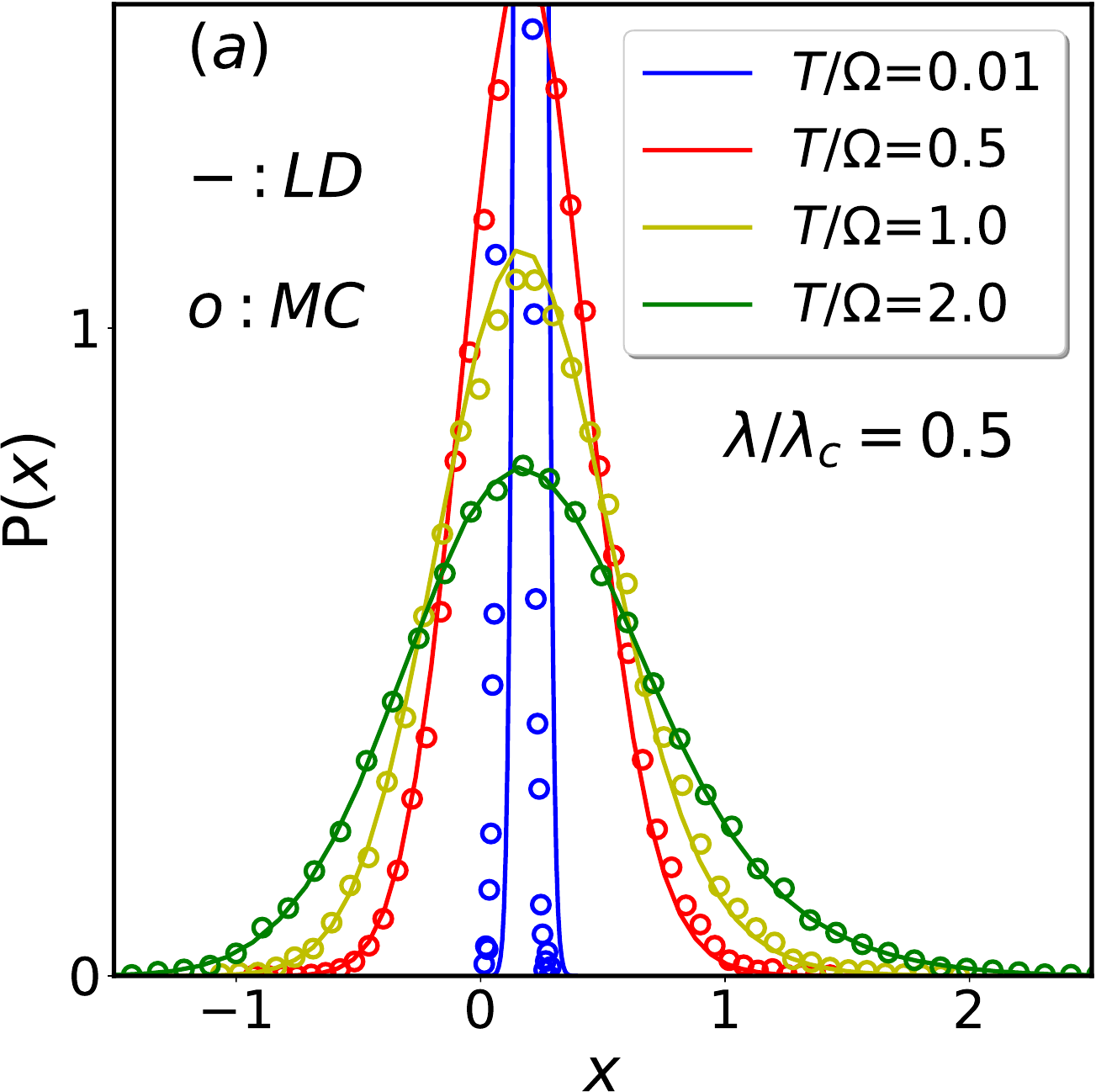}
\includegraphics[height=4cm,width=4cm]{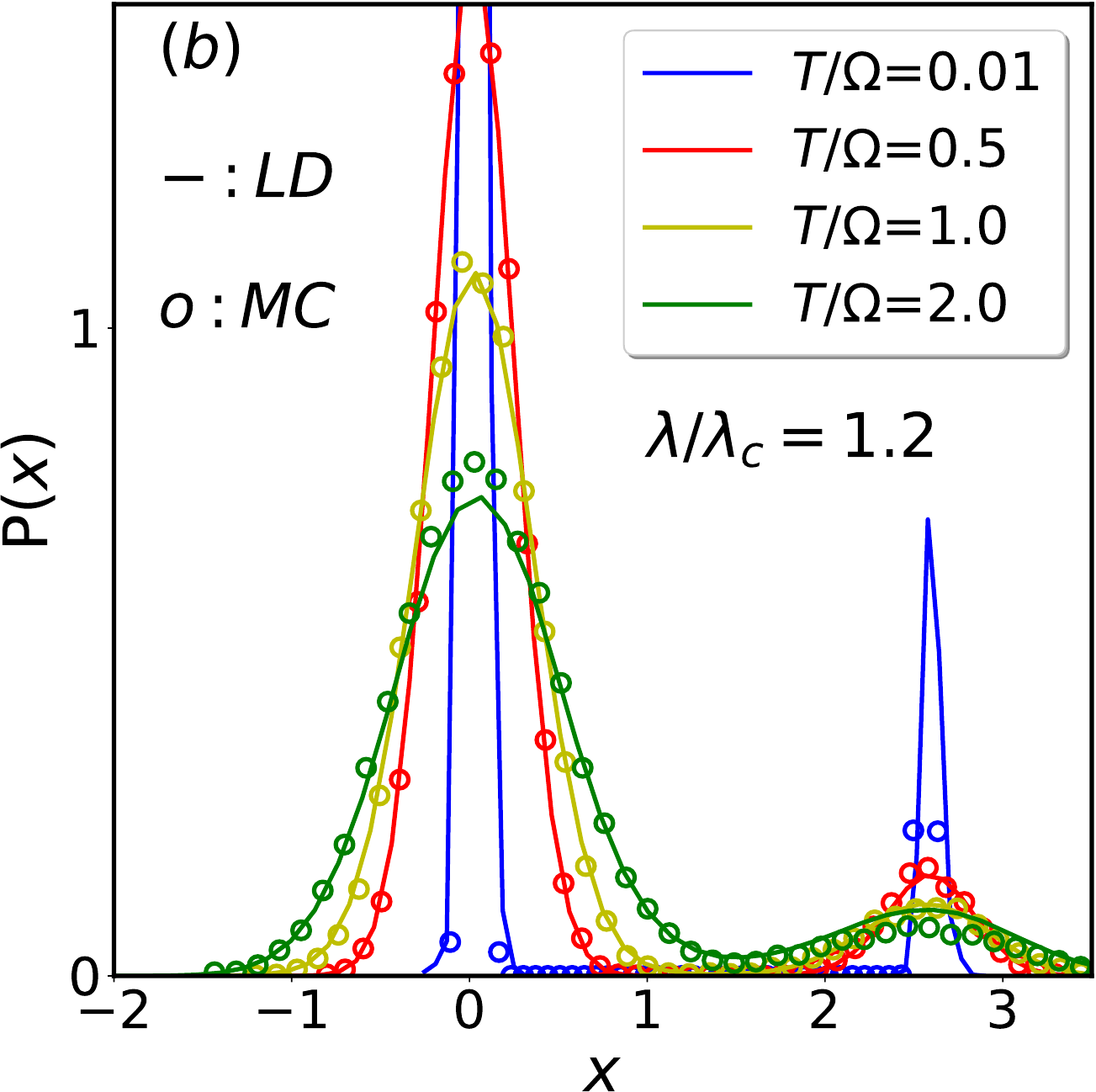}
~~~~~~~}
\centerline{
~~~~
\includegraphics[height=2.2cm,width=8.6cm]{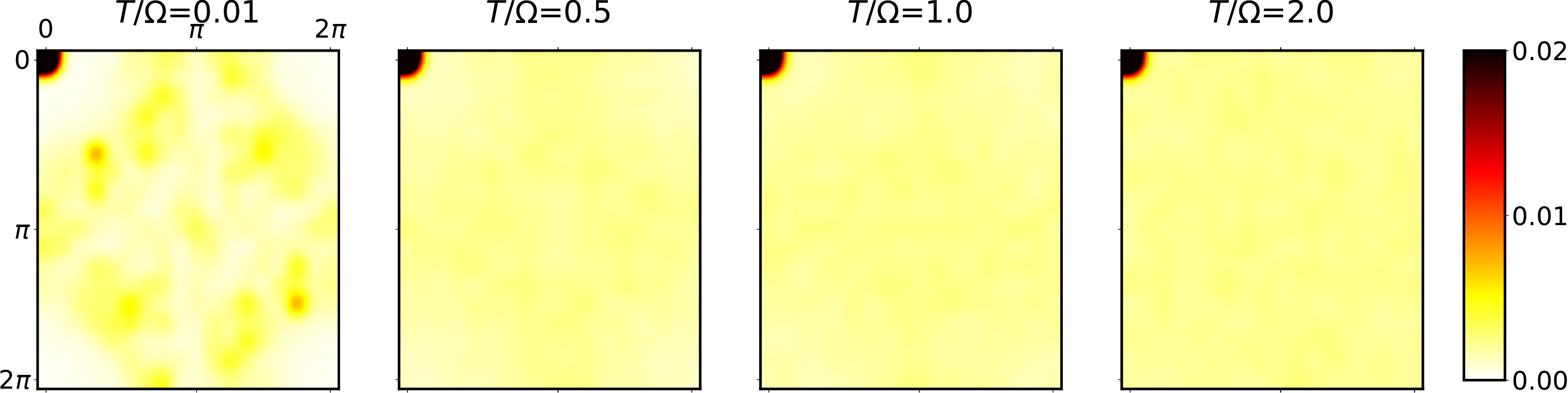}
}
\caption{Top:  The distribution $P(x)$ of lattice distortions 
at $n=0.1$ comparing Monte Carlo and Langevin dynamics results.
(a)~Weak coupling, $\lambda = 0.5 \lambda_c$
(b)~Strong coupling,  $\lambda = 1.2 \lambda_c$.
MC results match with LD at all $T$ and both couplings.
Weak coupling shows a sharp unimodal distribution
broadening with $T$. 
$P(x)$ at strong coupling is bimodal, the peak
at large $x$ corresponds to polaronic distortions and
has weight 0.1 (the electron density).
Bottom: Structure factor $S({\bf q},T)$ of distortions
at strong coupling. There are no features in ${\bf q}$,
except the ${\bf q}=(0,0)$ peak,
despite the presence of large distortions.}
\end{figure}

\begin{figure*}[t]
\centerline{~~~~~~~
\includegraphics[height=2.7cm,width=13cm]{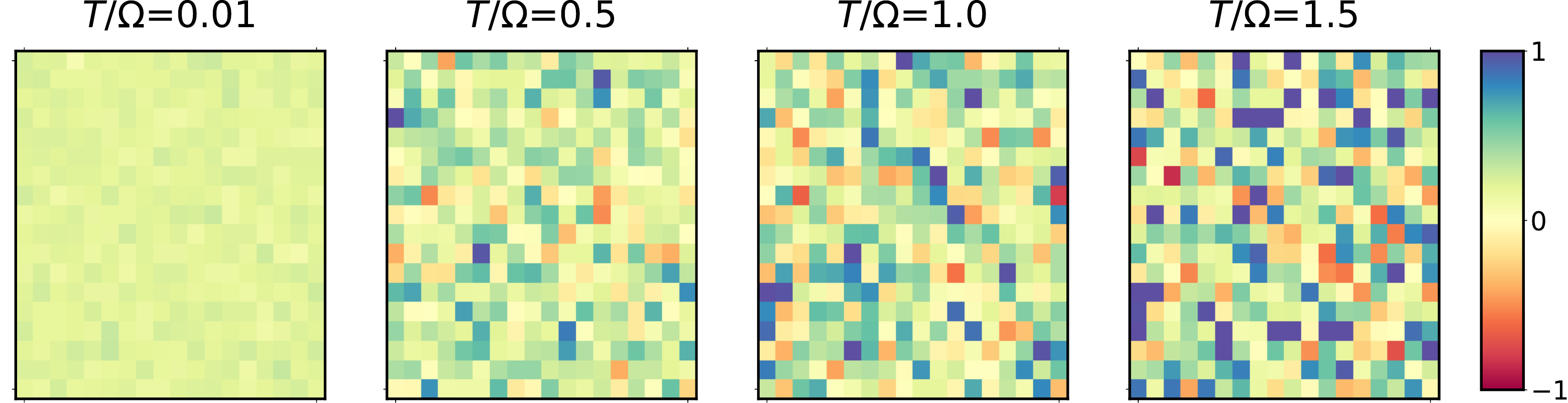}
}
\centerline{
~
\includegraphics[height=2.7cm,width=3.0cm]{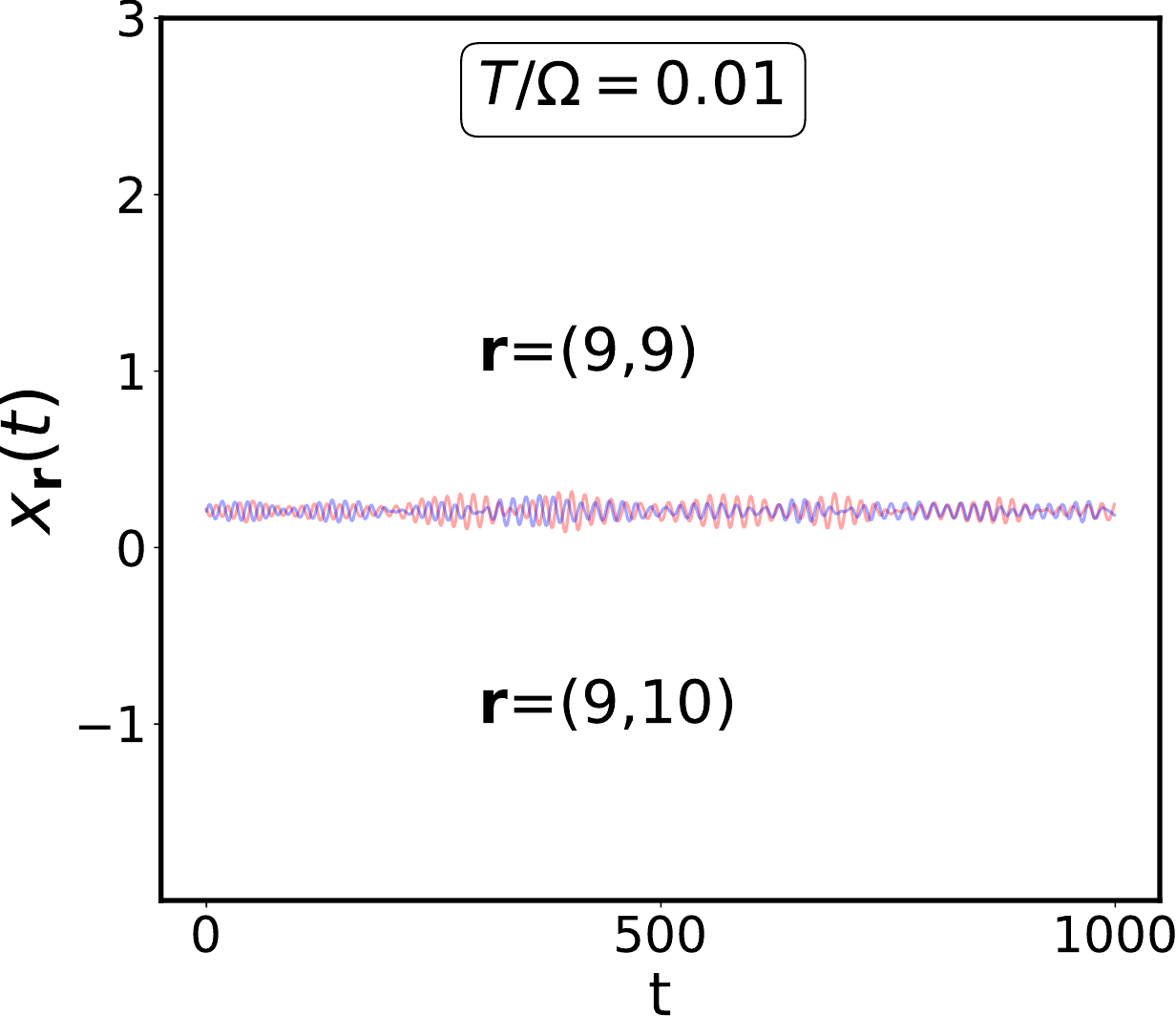}
\includegraphics[height=2.7cm,width=3.0cm]{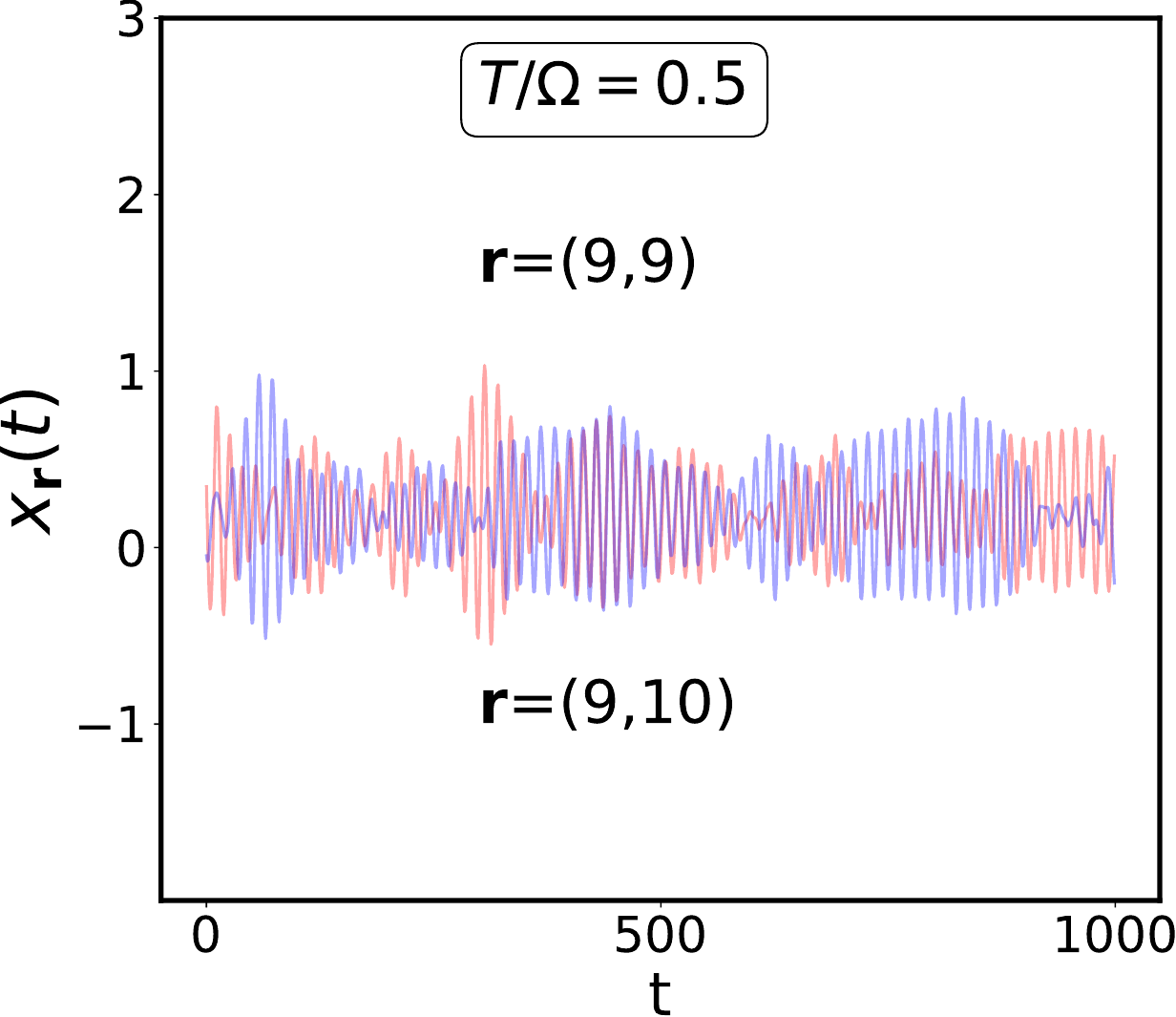}
\includegraphics[height=2.7cm,width=3.0cm]{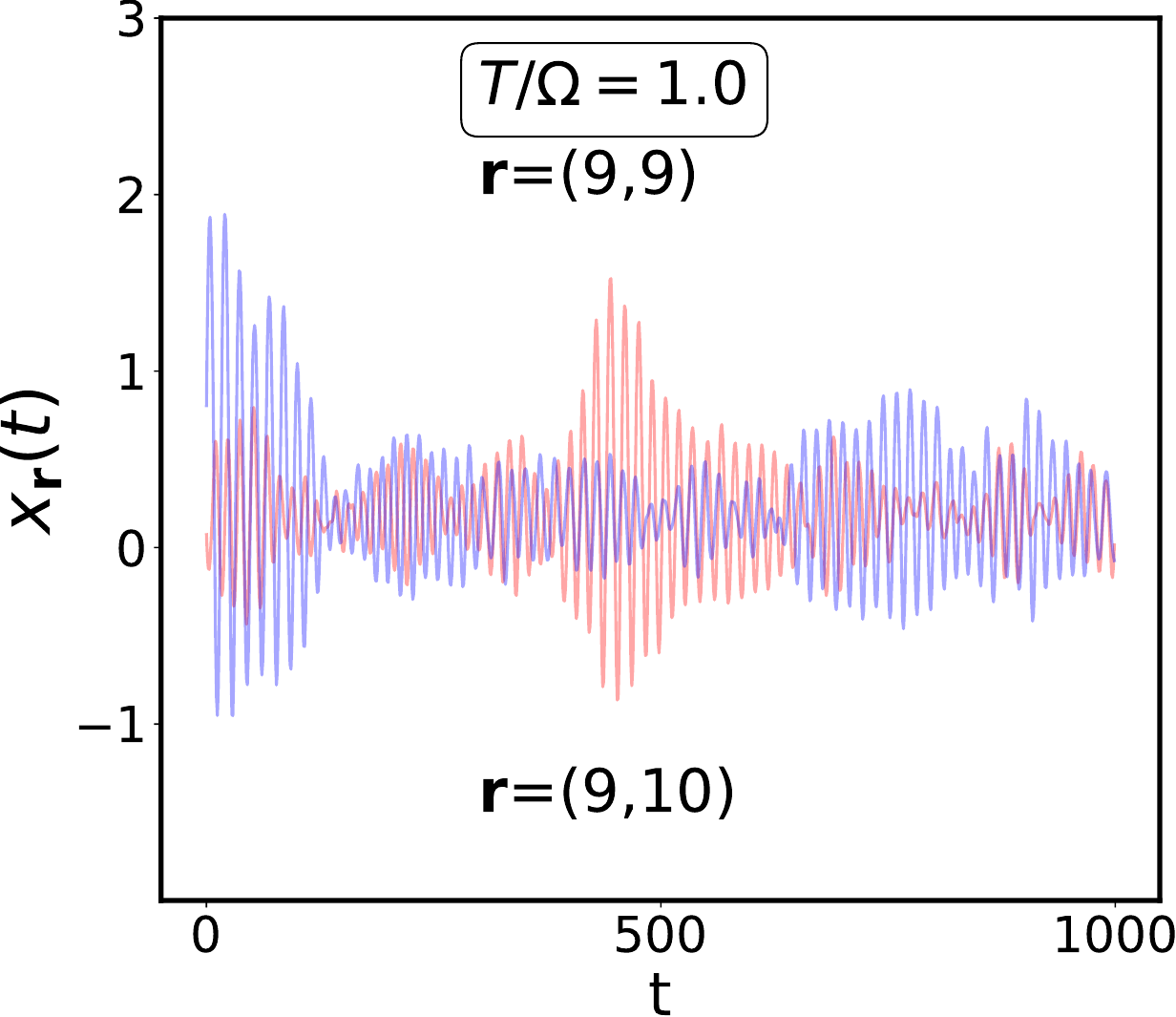}
\includegraphics[height=2.7cm,width=3.0cm]{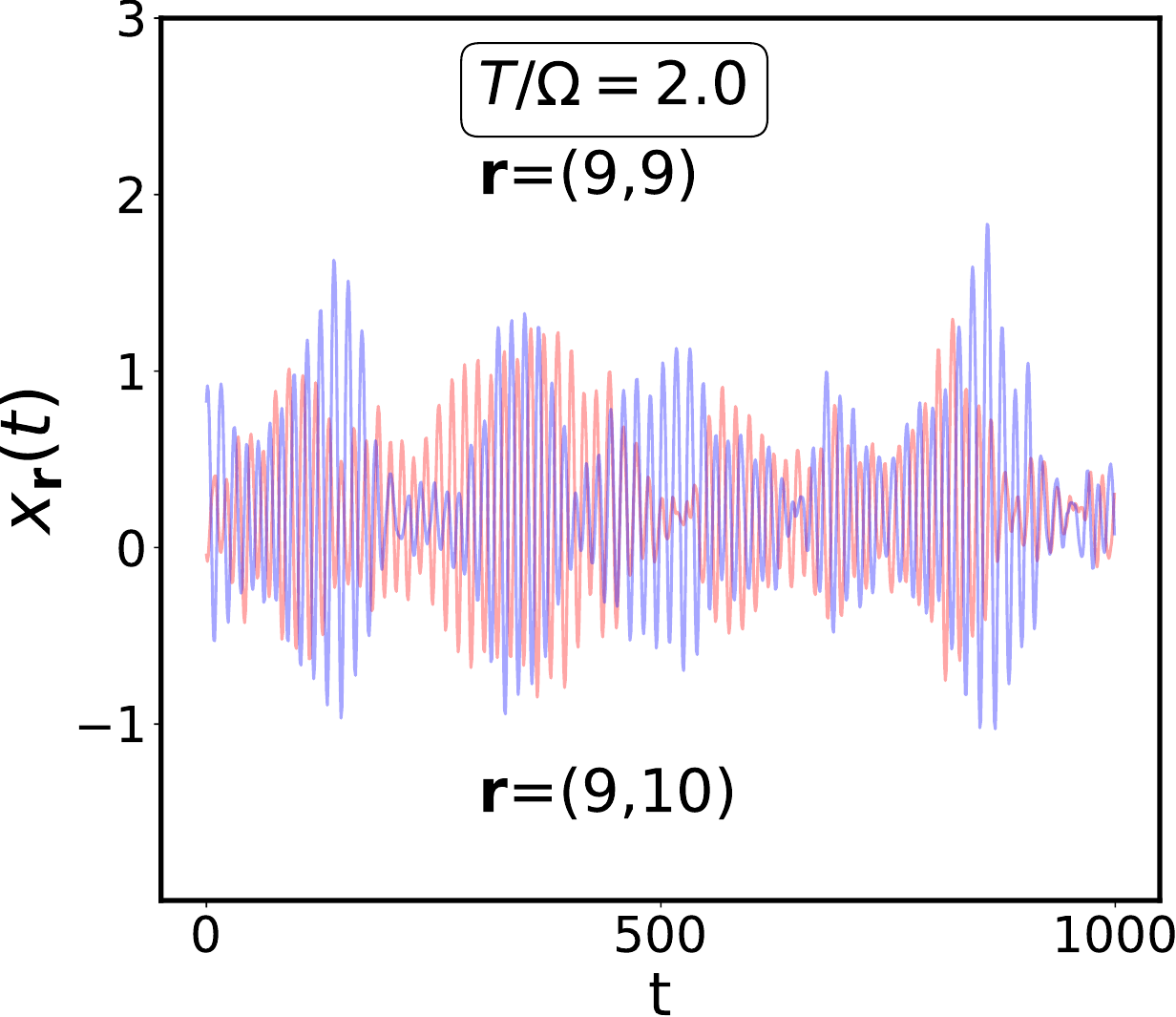}
~~~~~~~~~~~}
\centerline{
\includegraphics[height=2.7cm,width=14cm]{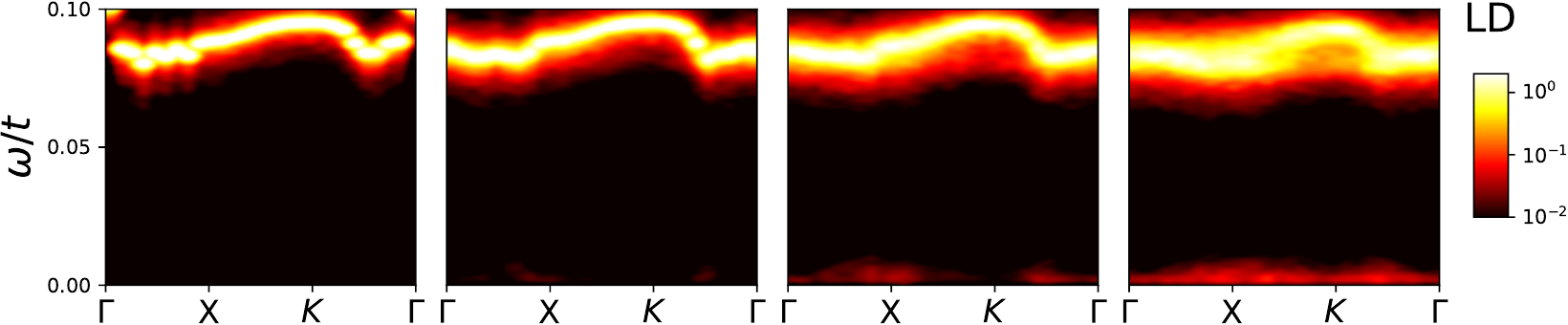}
}
\vspace{0.2cm}
\centerline{
\includegraphics[height=2.7cm,width=14cm]{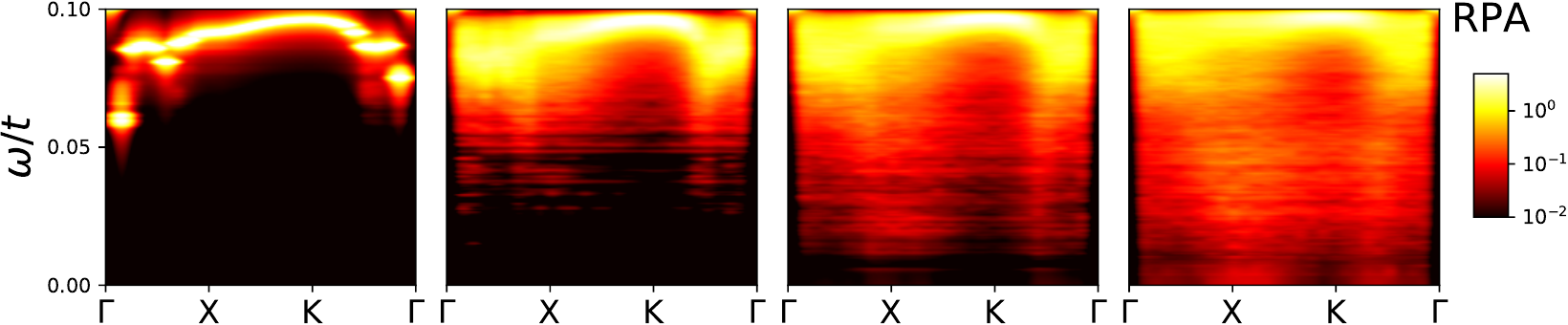}
}
\caption{
Dilute regime and weak coupling:  $n=0.1$ and $\lambda =0.5 \lambda_c$. 
Top row: instantaneous configuration $\{ x_i \}$.
Second row: real-time trajectory $x_i(t)$ at two sites. 
Third row: spectral map $\vert X({\bf q},\omega) \vert^2$ computed using LD.
Fourth row: RPA based spectrum for same parameters as in LD.
The Brillouin zone trajectory chosen is 
$(0,0) \rightarrow(\pi,0)\rightarrow (\pi,\pi)\rightarrow(0,0)$.  
The spatial maps show emergence of weak spontaneous inhomogeneity
with increasing $T$.
$x_i(t)$ shows harmonic fluctuations about a uniform state at low $T$, 
crossing over to larger amplitude vibrations accompanied by `burst' like 
events at higher $T$. The later are prominent for $T\sim\Omega$ and
generate leading to low frequency weight in $\vert X({\bf q},\omega) \vert^2$.
The RPA spectrum broadens rapidly with increasing $T$, due to the
dependence on instantaneous configurations, and also misses the low 
energy feature.
}
\end{figure*}

\subsection{Phonon dynamics}

\subsubsection{Weak coupling ($\lambda \ll  \lambda_c$)}

We first discuss the weak coupling ($\lambda/\lambda_{c}=0.5$) regime.
The top three rows of Fig.3 are results obtained using
Langevin dynamics. First, we plot the snapshots of the 
displacement field $\{x_{i}\}$. 
The real-time trajectories for
two nearest neighbour sites
are featured next, followed by LD based spectra in the third and 
RPA based spectra in the fourth rows respectively.
The first column represents the low temperature regime.
Here, the `background state' is translation invariant
and the electronic physics is that of a tight-binding model. 
The phonons execute harmonic oscillations for $T \ll \Omega$ and the
resulting dispersion arises out of the bare polarizability 
$\Pi_{0}({\bf q},\omega)$. The LD and RPA answers match reasonably
well. The latter is notionally more accurate as the 
full frequency dependence of $\Pi_{0}$ is kept.

\begin{figure}[b]
\centerline{
\includegraphics[width=2.85cm,height=3.6cm]{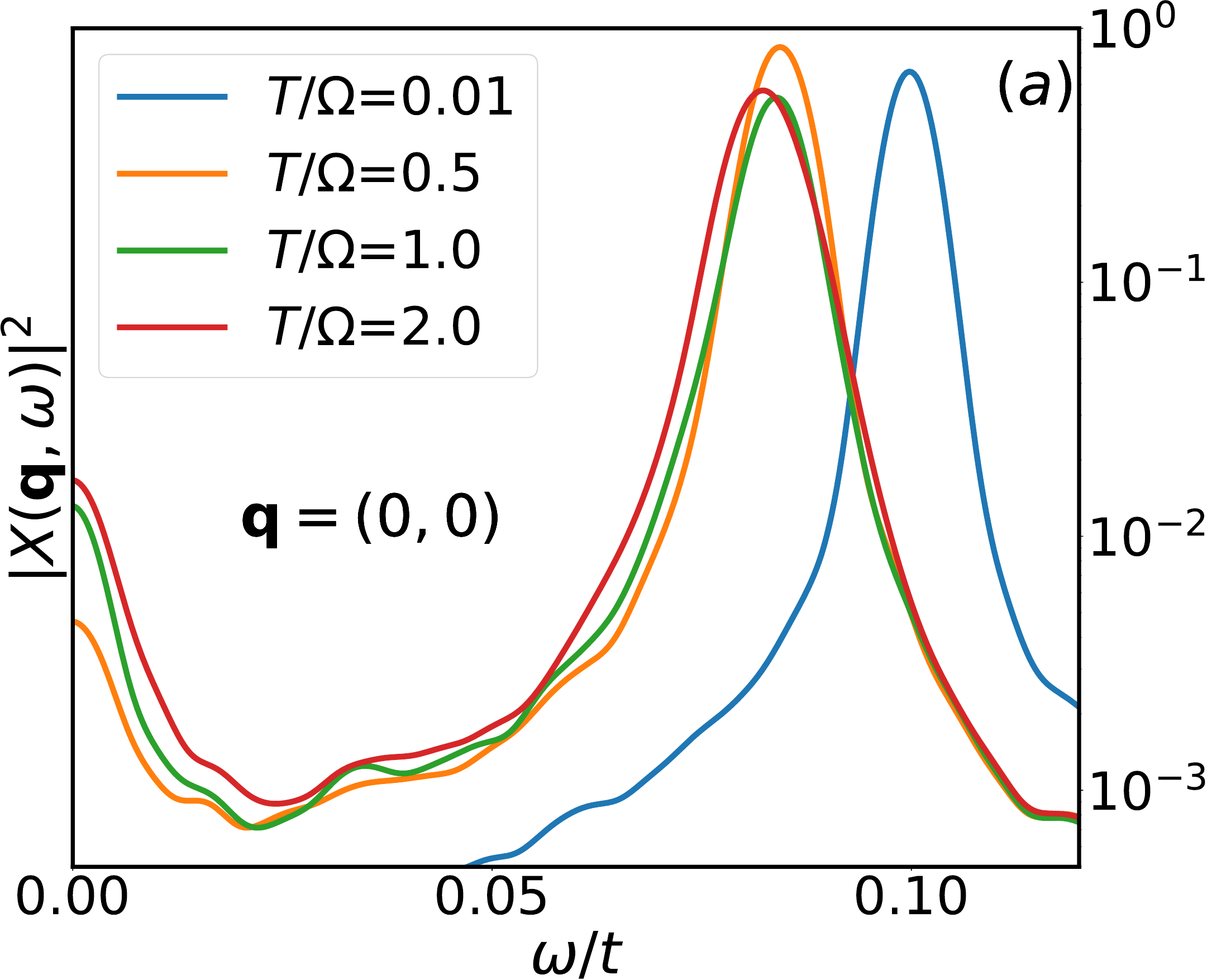}
\includegraphics[width=2.85cm,height=3.6cm]{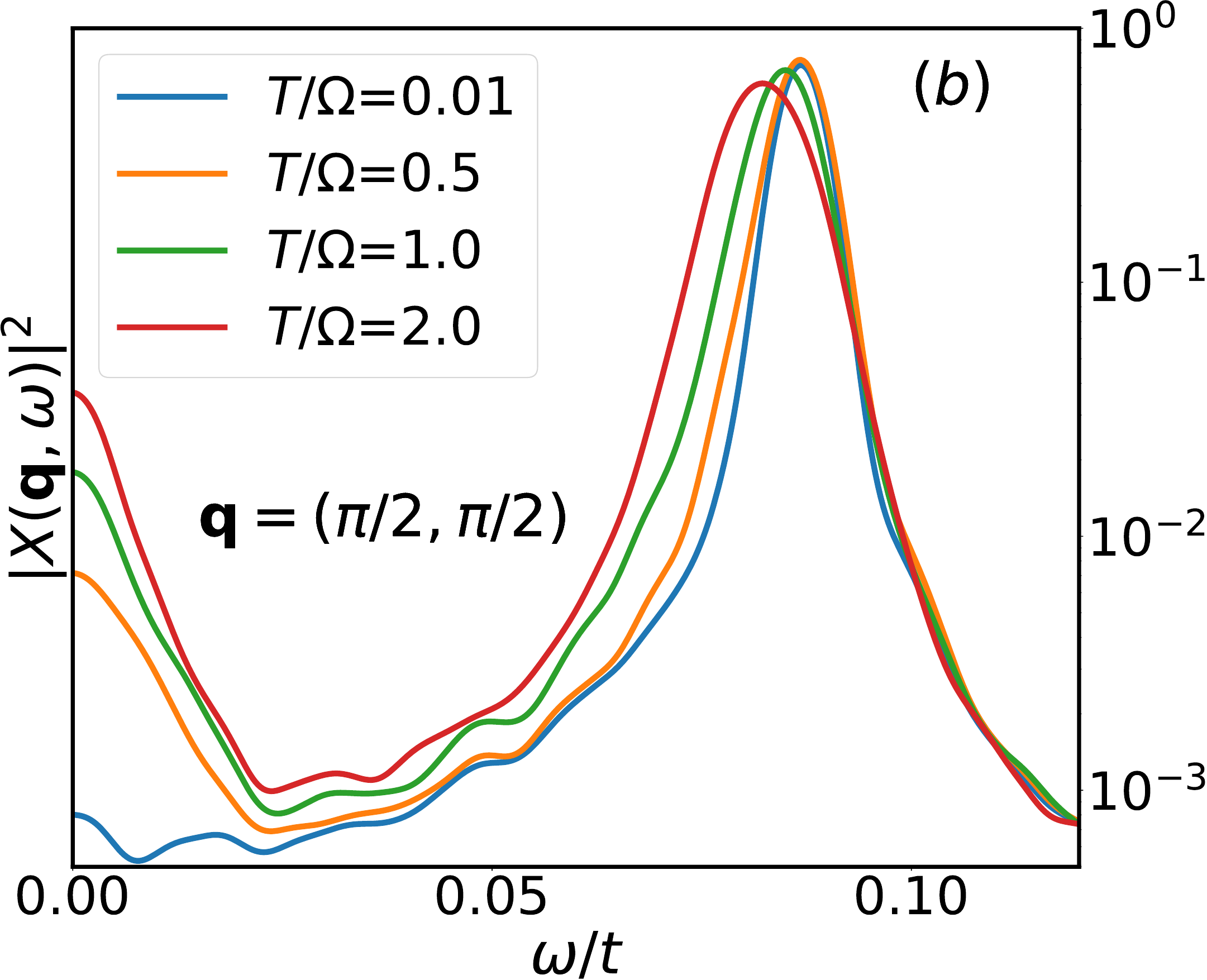}
\includegraphics[width=2.85cm,height=3.6cm]{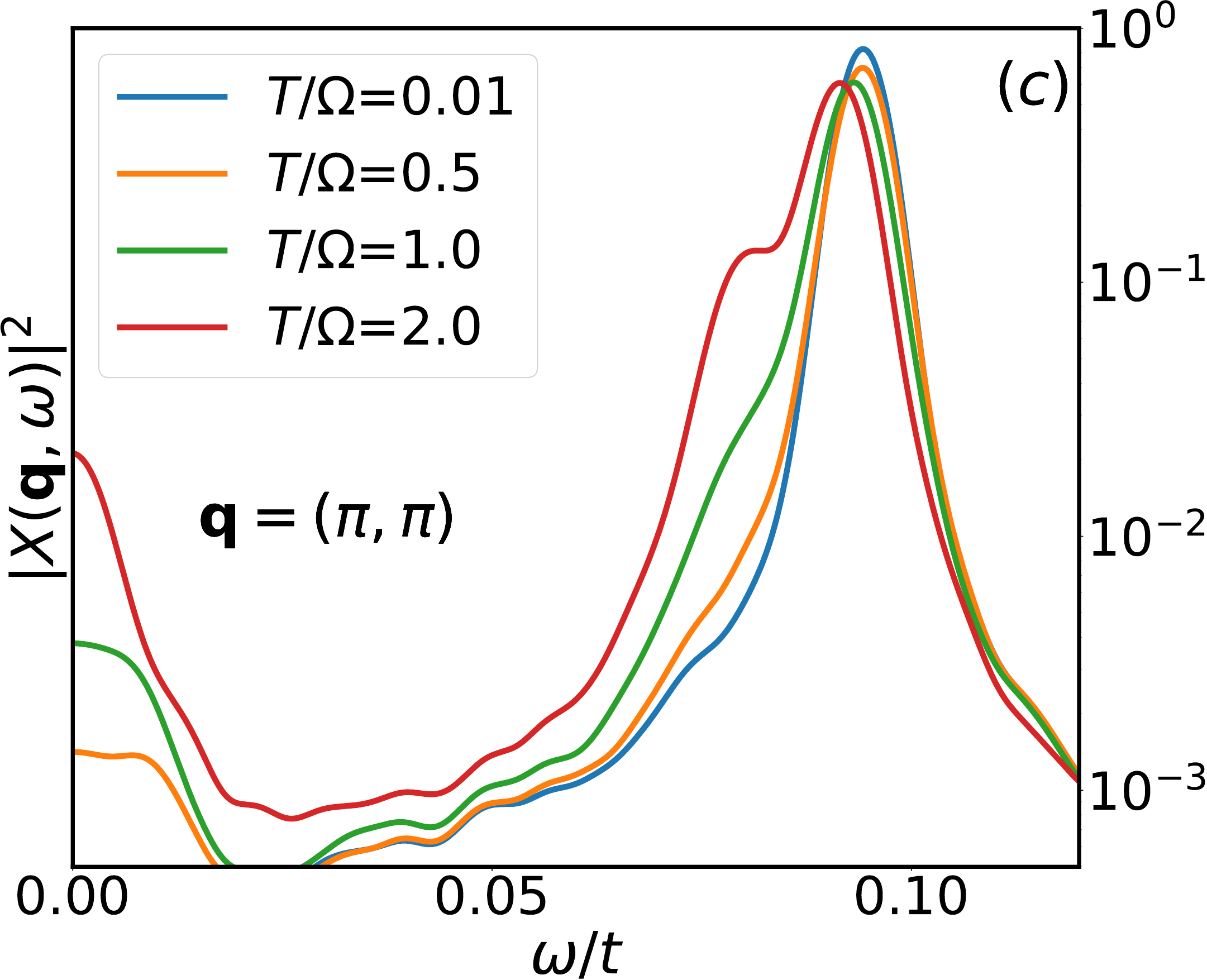} }
\caption{Lineshapes at $n=0.1$ and $\lambda = 0.5 \lambda_c$
and three characteristic momenta- ${\bf q}=(0,0)$,
$(\pi/2,\pi/2)$ and $(\pi,\pi)$.
Intensities are shown in a log-scale to emphasize the emergence of
low energy weight on heating up. The $(0,0)$ mode softens
`immediately' compared to the others.  Otherwise, the thermal trends
are similar for all wavevectors.  }
\end{figure}

\begin{figure*}[t]
\centerline{~~~~~~~
\includegraphics[height=2.7cm,width=13cm]{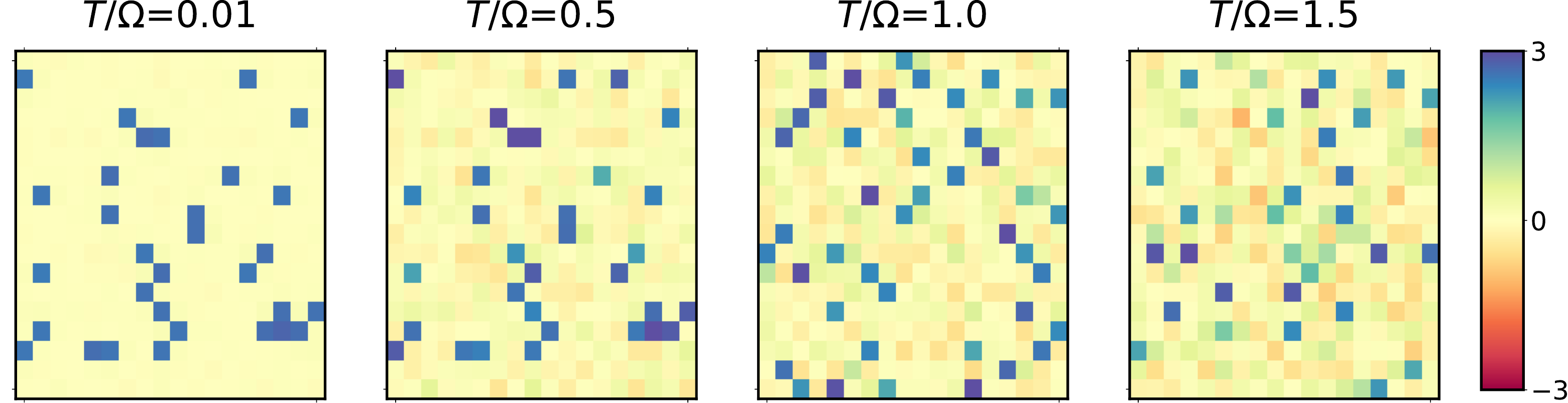}
}
\centerline{
~
\includegraphics[height=2.7cm,width=3.0cm]{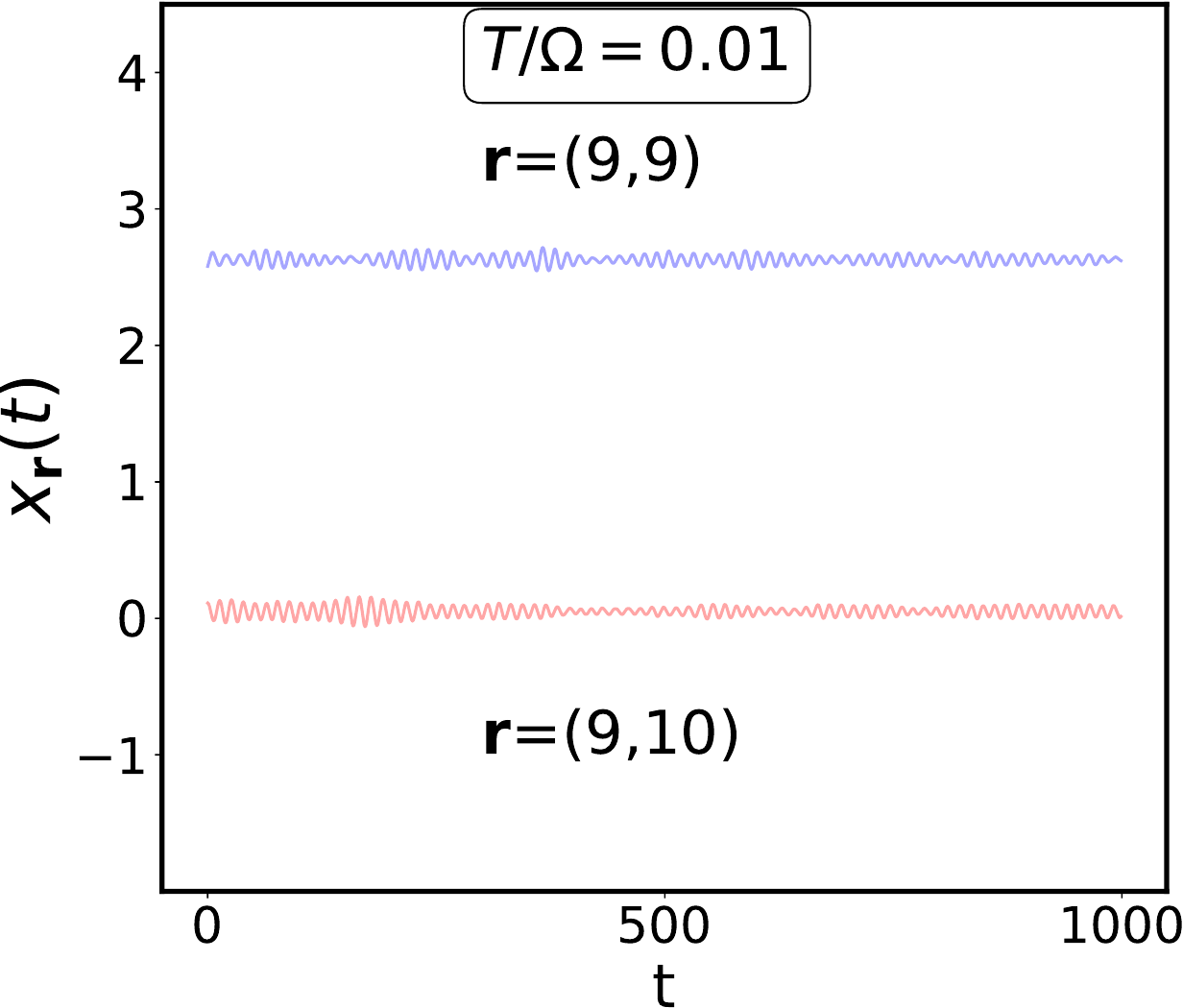}
\includegraphics[height=2.7cm,width=3.0cm]{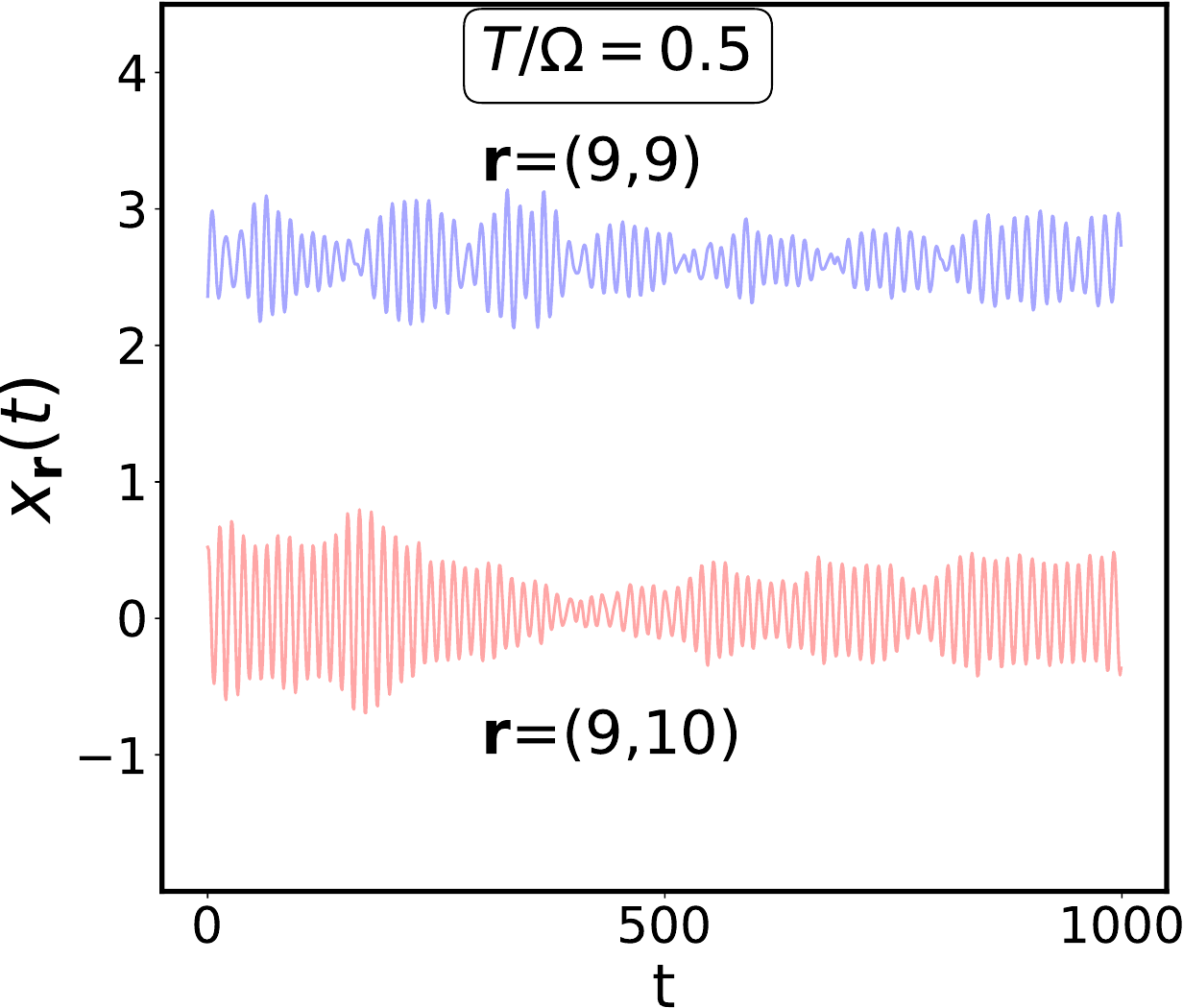}
\includegraphics[height=2.7cm,width=3.0cm]{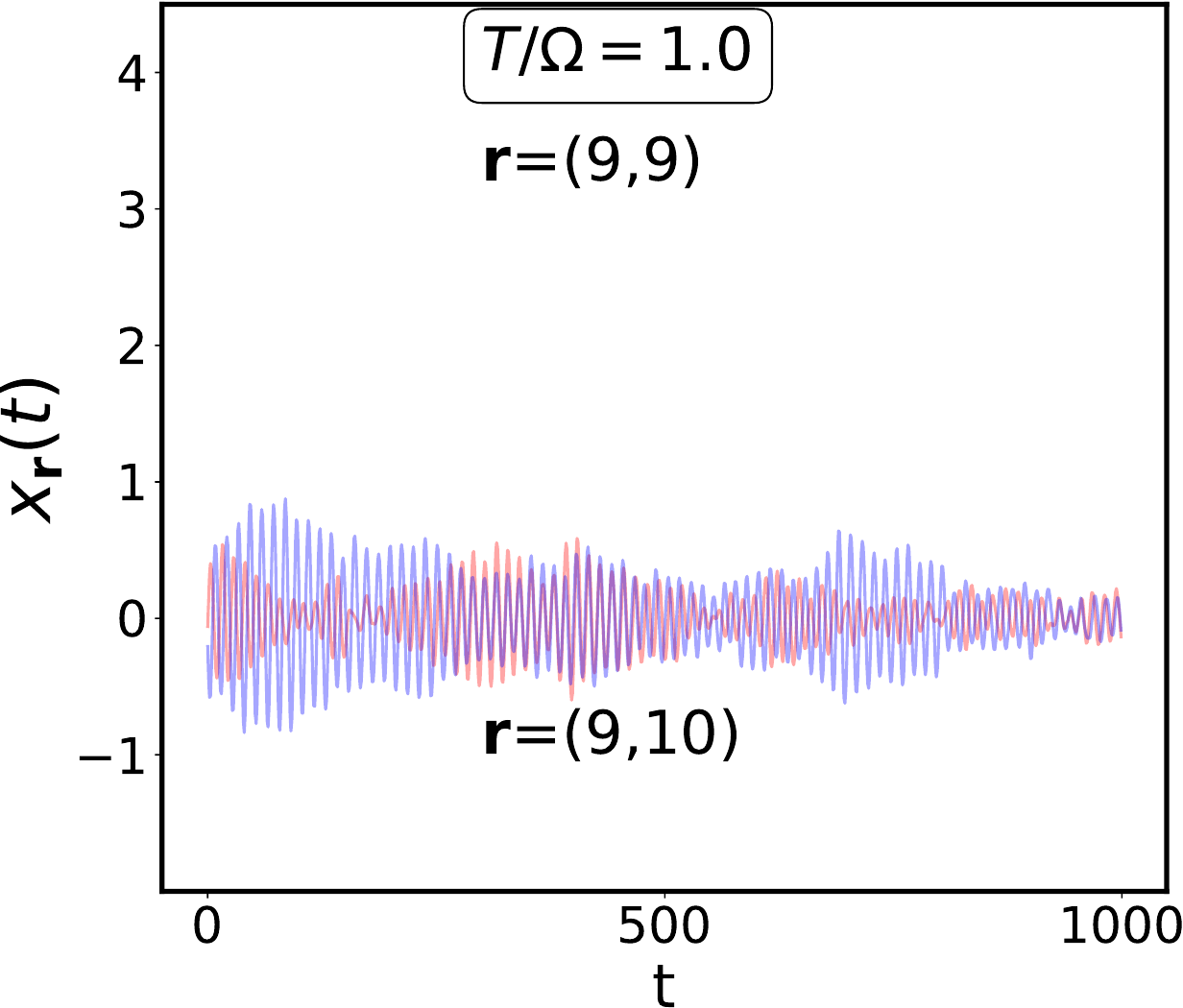}
\includegraphics[height=2.7cm,width=3.0cm]{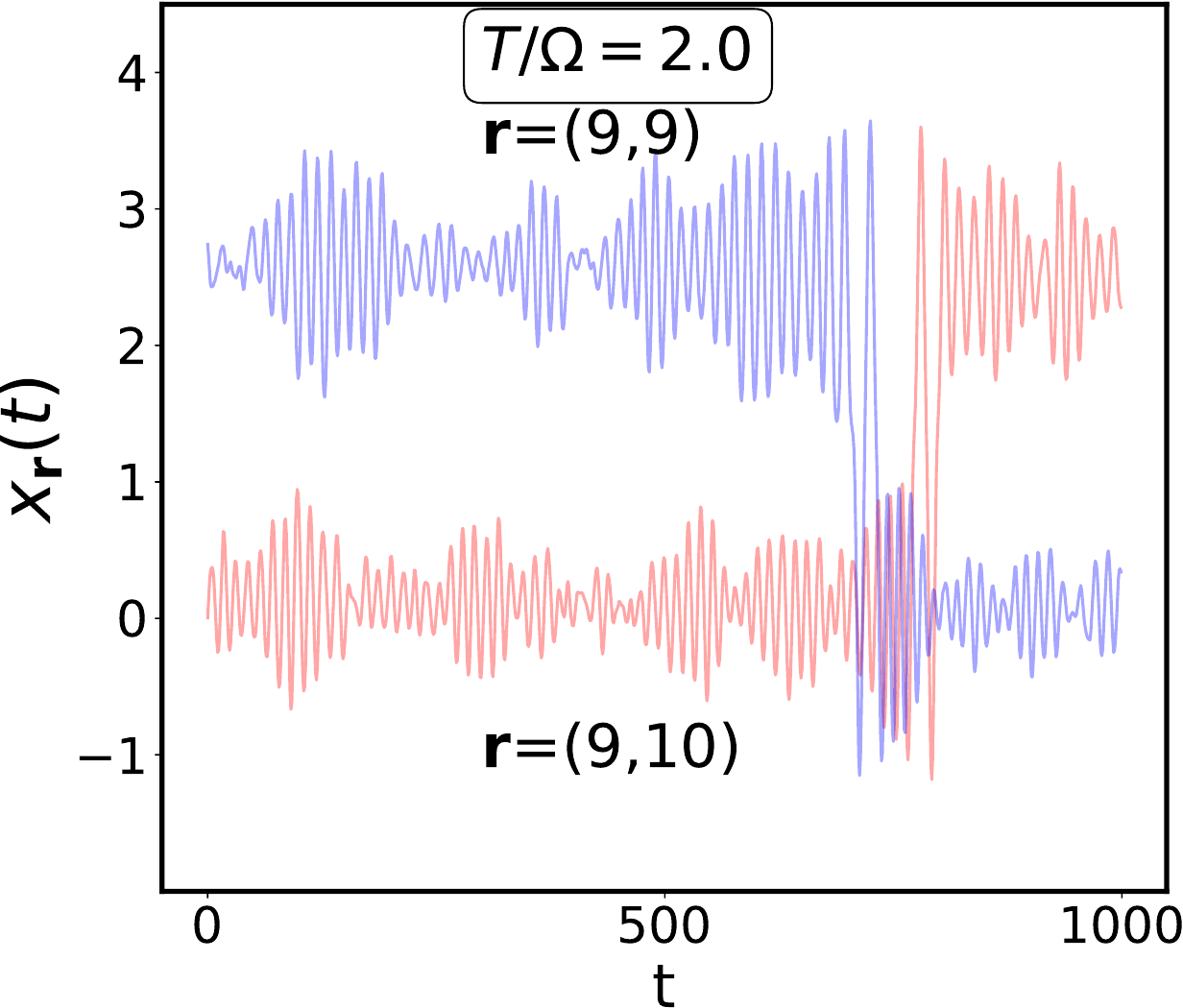}
~~~~~~~~~~~}
\centerline{
\includegraphics[height=2.7cm,width=14cm]{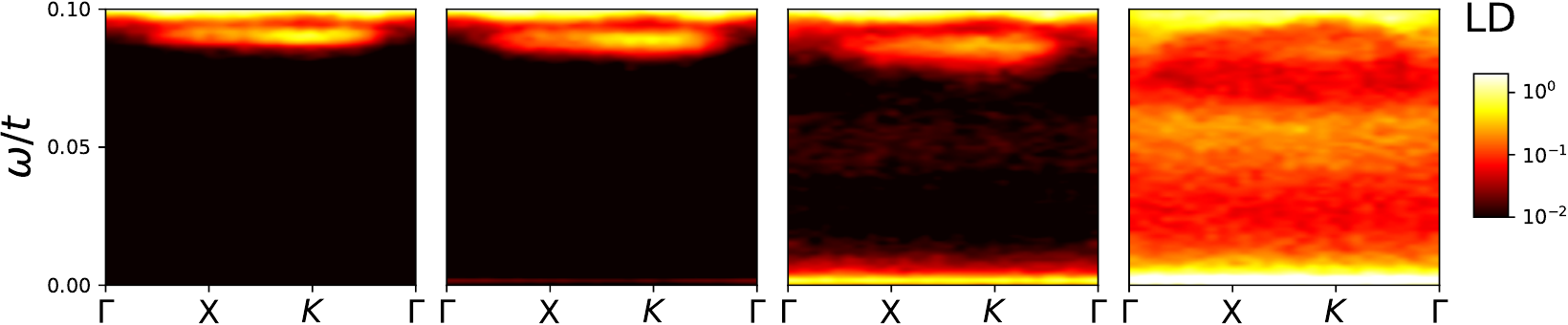} }
\caption{
Dilute regime and strong coupling:  $n=0.1$ and $\lambda =1.2 \lambda_c$.
Top row: instantaneous configuration $\{ x_i \}$, showing an essentially
uncorrelated patterns of large distortion at all $T$.
Second row: real-time trajectory $x_i(t)$ at two sites, highlighting
their different mean values at low $T$, `tunneling' events (column 3), 
and large oscillations.
Third row: spectral map $\vert X({\bf q},\omega) \vert^2$ computed using LD.
Brillouin zone trajectory as before is
$(0,0) \rightarrow(\pi,0)\rightarrow (\pi,\pi)\rightarrow(0,0)$.
The first two columns show an almost dispersionless `high energy'
feature.  The third column shows the emergence of low energy weight
due to tunneling events. The highest $T$ shows a broad spectrum 
due to large amplitude oscillations.
}
\end{figure*}
\begin{figure}[b]
\centerline{
\includegraphics[width=2.75cm,height=2.9cm]{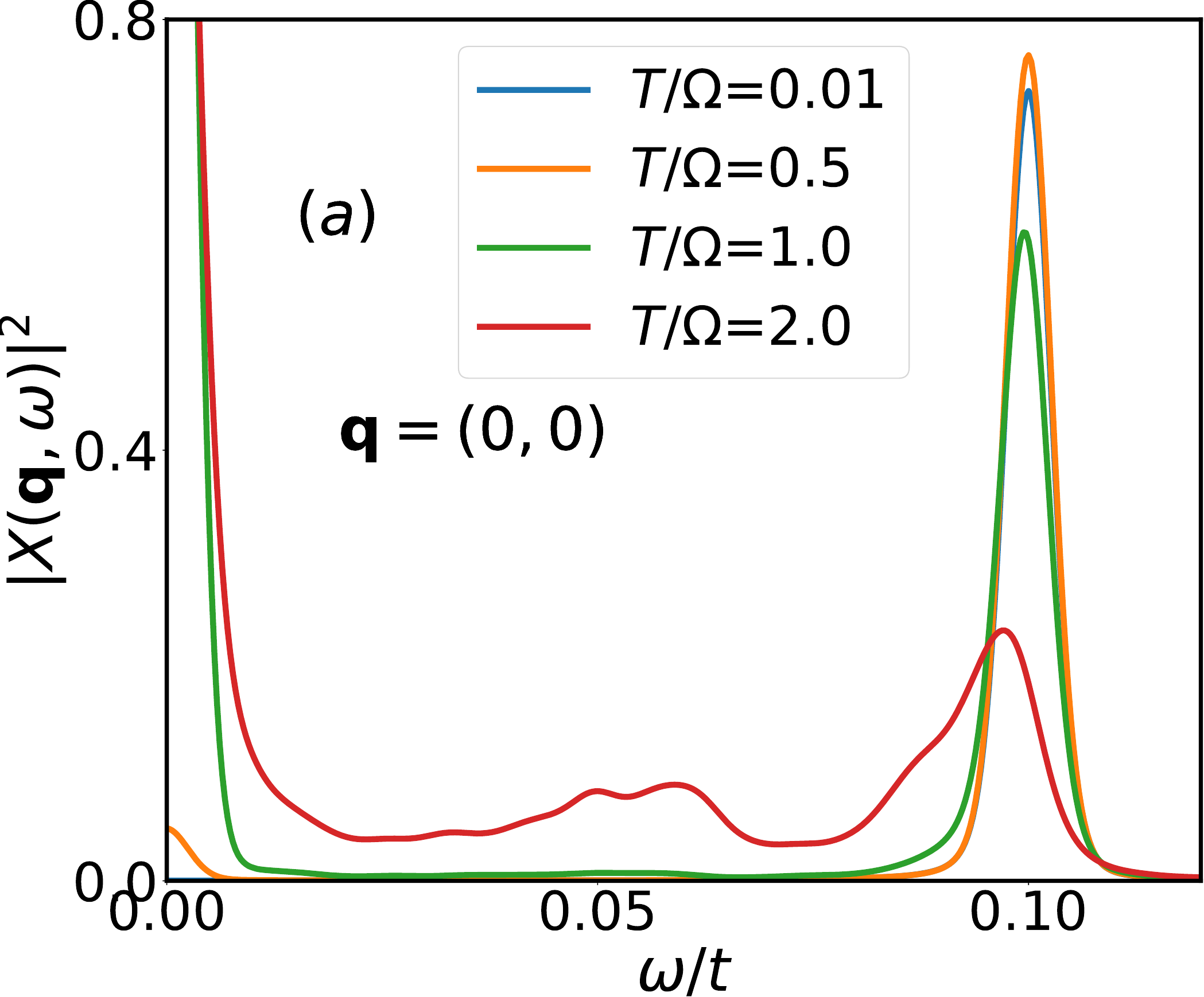}
\includegraphics[width=2.75cm,height=2.9cm]{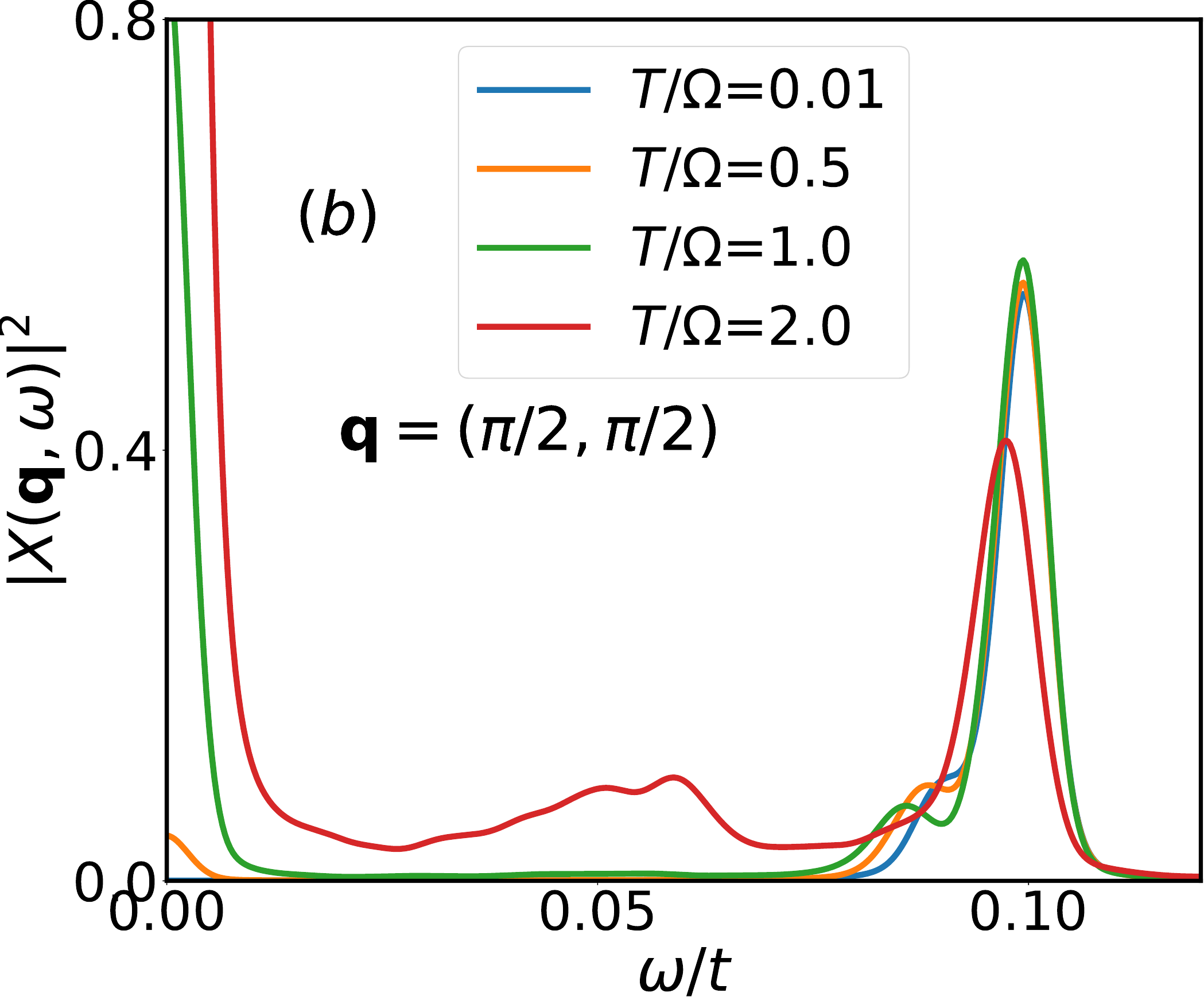}
\includegraphics[width=2.75cm,height=2.9cm]{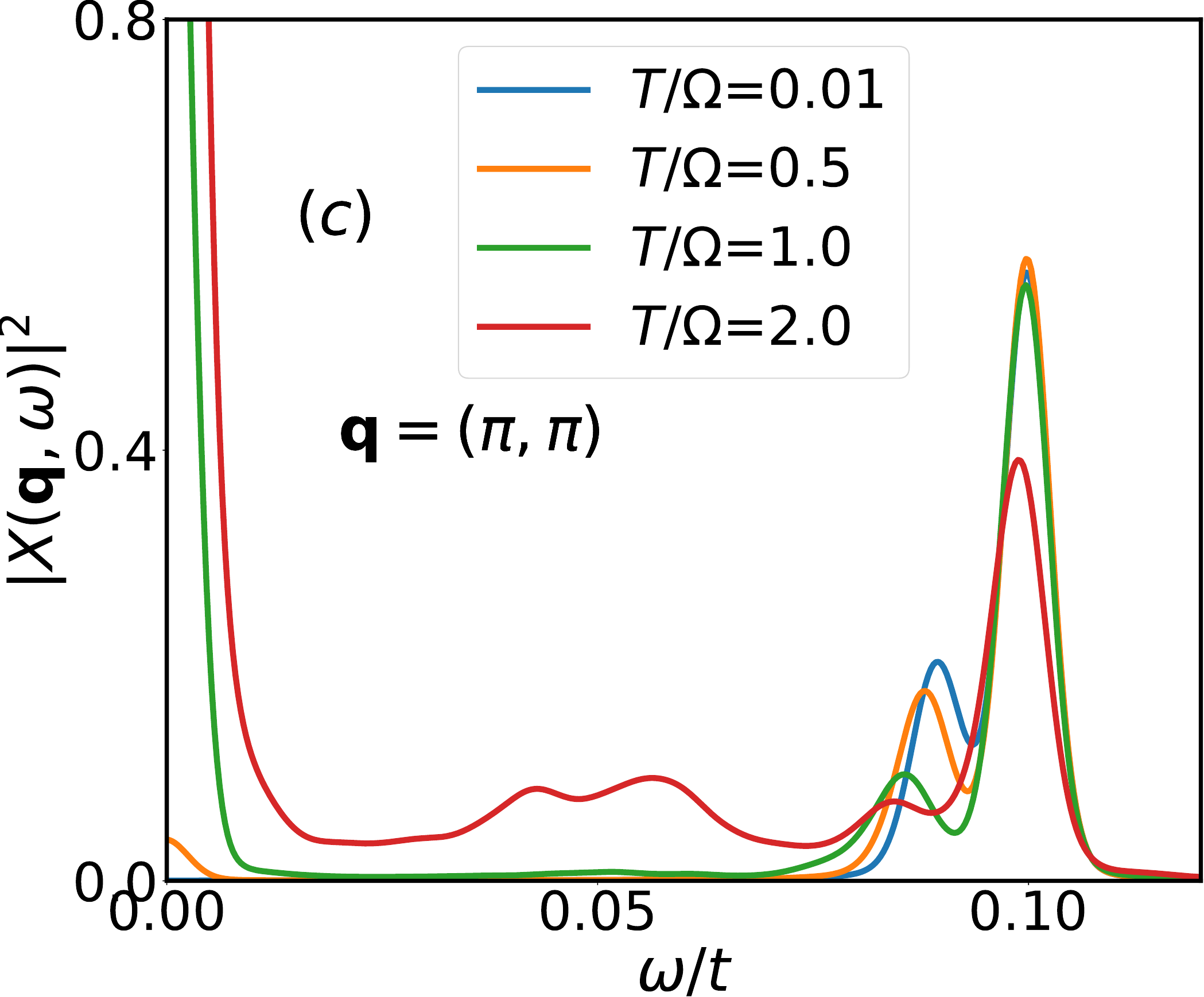}
}
\centerline{
\includegraphics[width=2.75cm,height=2.9cm]{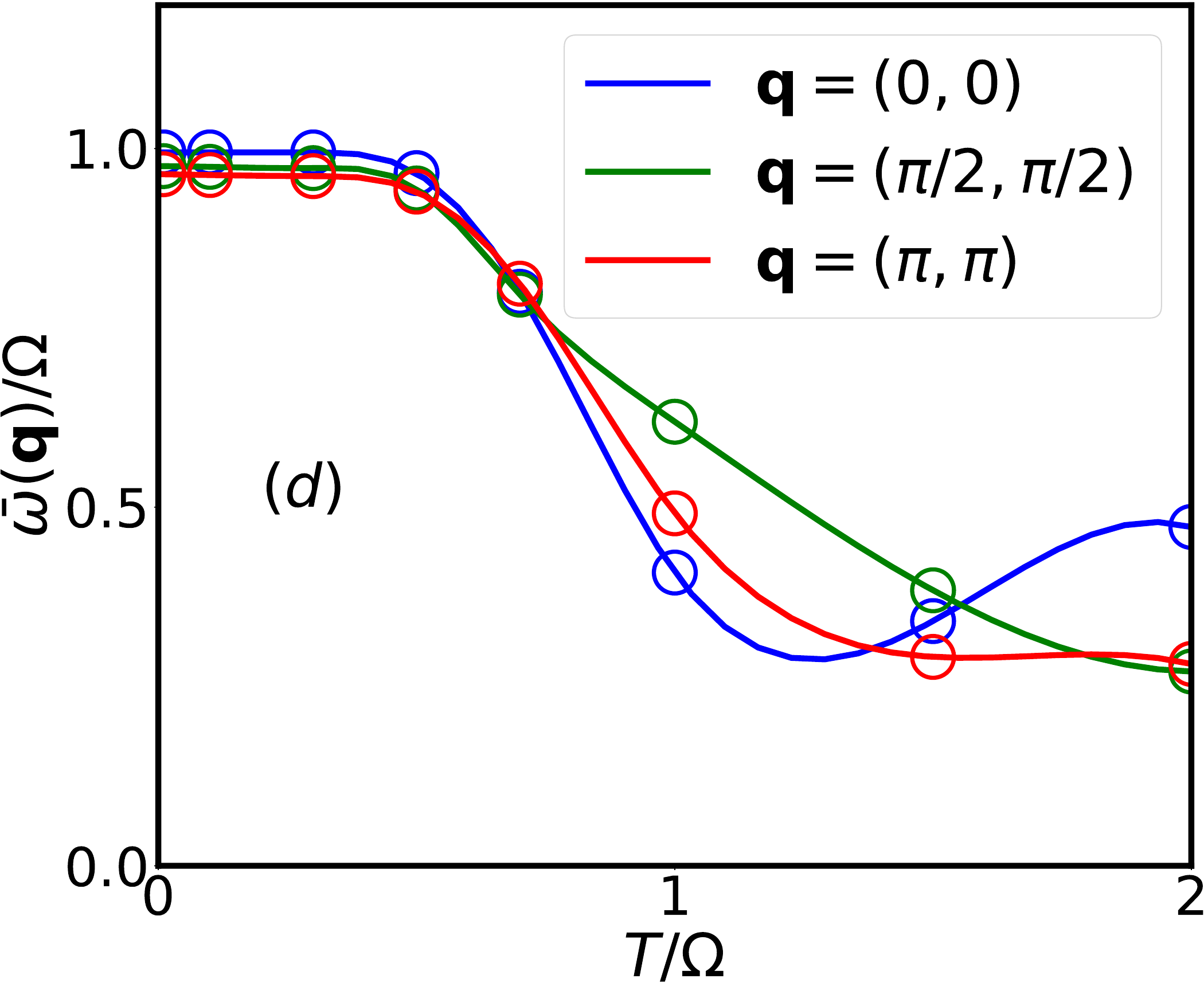}
\includegraphics[width=2.75cm,height=2.9cm]{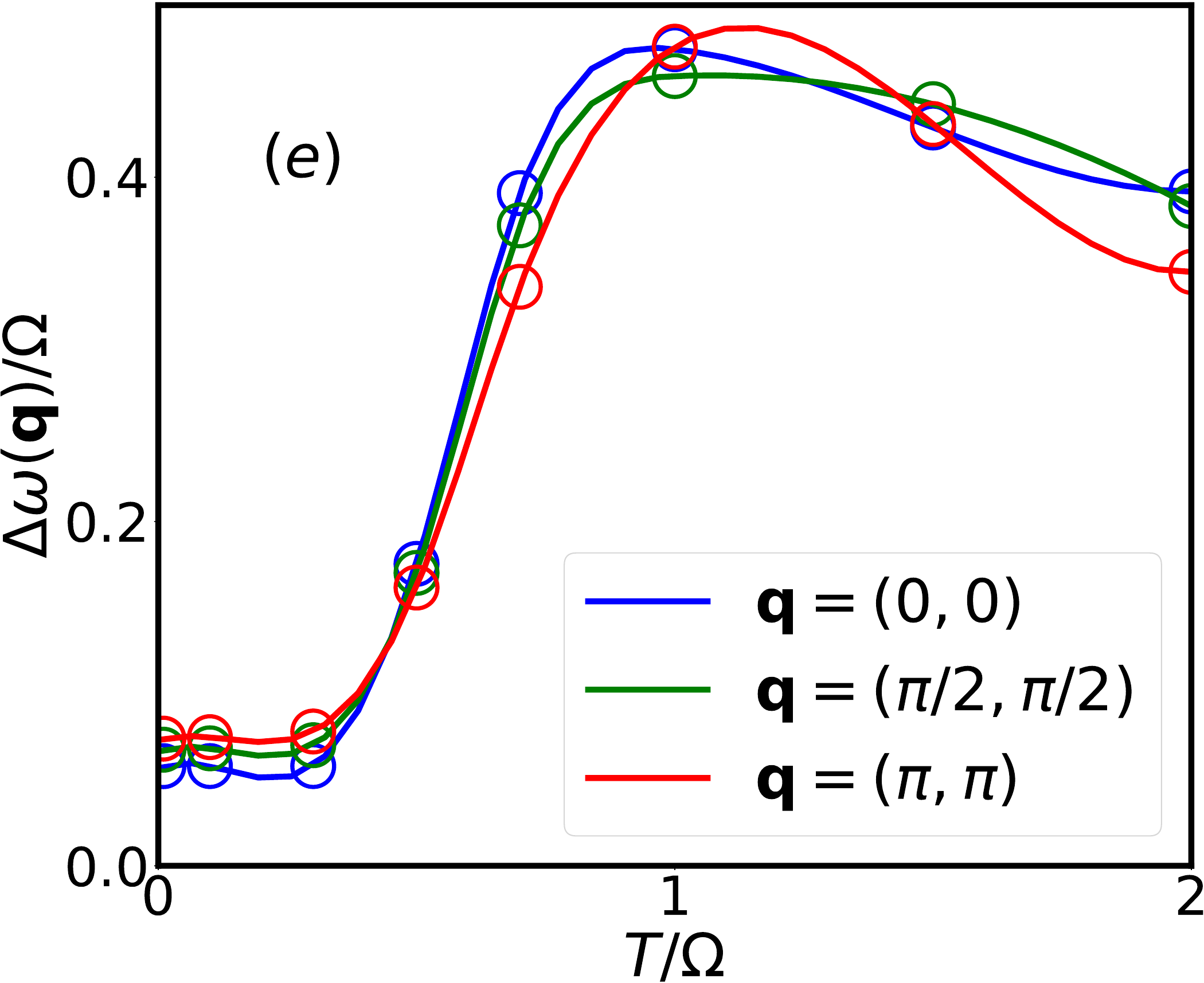}
\includegraphics[width=2.75cm,height=2.9cm]{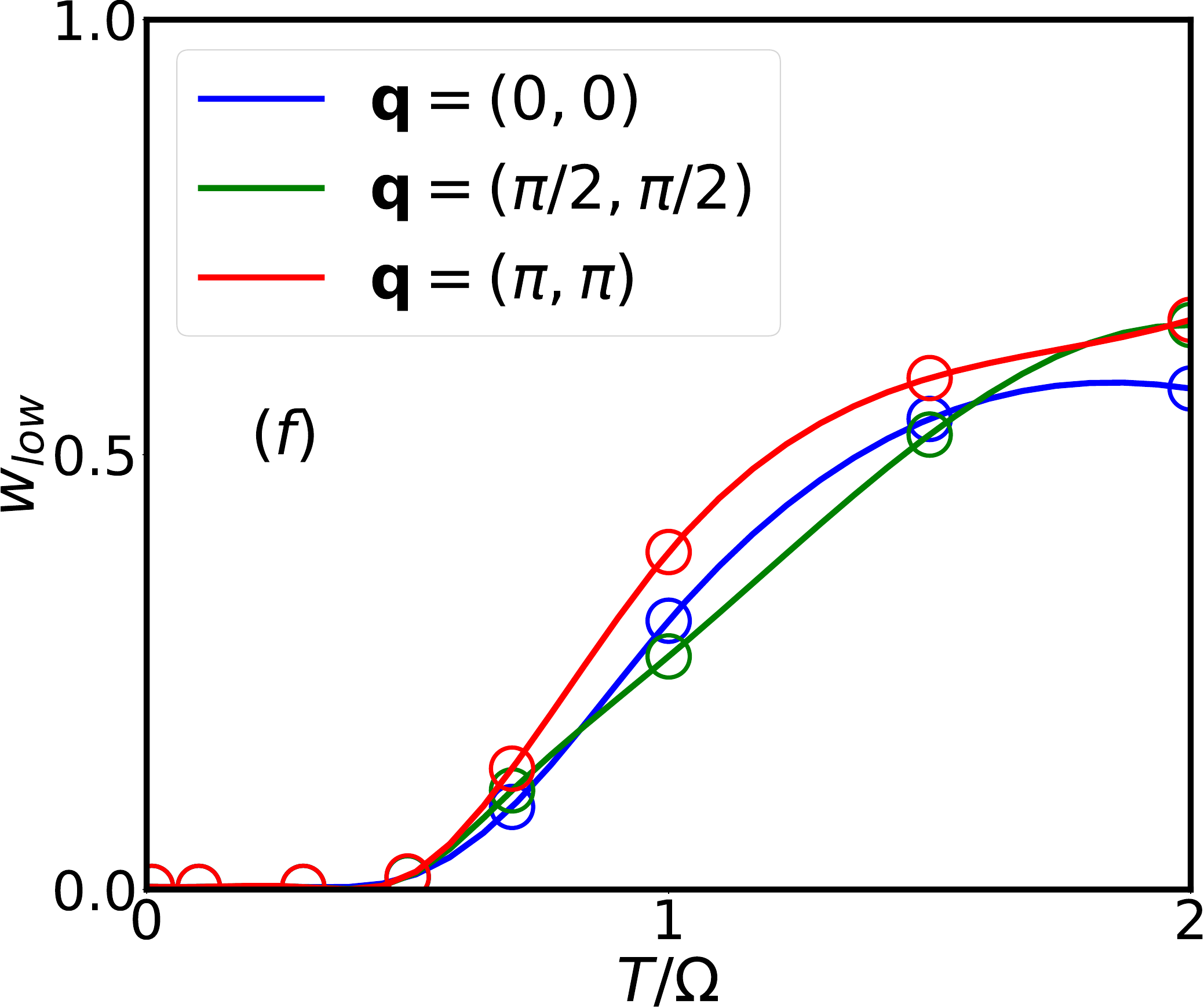}
}
\caption{Spectral features at $n=0.1$ and $\lambda = 1.2 \lambda_c$.
Top panel: Lineshapes from LD at three momenta-
${\bf q}=(0,0)$, $(\pi/2,\pi/2)$ and $(\pi,\pi)$.
The low temperature profile is sharp and unimodal
at $(0,0)$, but has a `peak splitting' at the other wavevectors.
Thermal behaviour is very similar for all three. The low energy
weight transfer is considerable for $T/\Omega\gtrsim1$.
Bottom: Mean frequency, $\bar{\omega}$, linewidth $\Delta\omega({\bf q})$,
 and low energy weight $w_{low}$ as function of temperature for
the three wavevectors.  The momentum selectivity is weak.
}
\end{figure}

On heating up to intermediate $T\sim0.5\Omega$, the amplitudes of 
oscillations increase and one also sees `burst' like 
events start to take place. However, at this stage, 
these are merged with large oscillations. The `high energy'
spectra in the LD case remain qualitatively unaltered, but 
a faint low energy ($\omega \ll \Omega$) weight starts to accumulate.
In the third column, we see results for $T\sim\Omega$, by 
the time the `bursts' become well separated in time and 
distinct from `bare oscillations'. 
However, spatial correlations are not 
apparent, as the `polaronic distortions'
are created far apart from each other.
The spectral signature, 
within LD, is clearly `two-peak' for most momenta. To 
emphasize the emergence of low frequency weight, we plot
the respective intensities in a logarithmic scale. 
Finally, at high temperature, the frequency of bursts
increase, along with oscillation amplitudes. 
This strengthens the low energy weight and broadens
the high energy band appreciably. We comment that
RPA fails to capture the `large amplitude' behaviour
of trajectories and consequently gives a gradually 
broadened spectrum on increasing temperature. This
arises from the growing `thermal disorder' of the 
background states, about which the spectrum of
small fluctuations are calculated.

Figs.4(a)-(c) exhibits the thermal dependence of lineshapes at three 
momentum values- ${\bf q}=(0,0),(\pi/2,\pi/2),(\pi,\pi)$.
The basic trends are similar for all three. The softening
is accompanied by a low energy weight transfer, which is 
not significantly momentum selective, as is physically expected.

\begin{figure}[b]
\centerline{
\includegraphics[height=3.7cm,width=4cm]{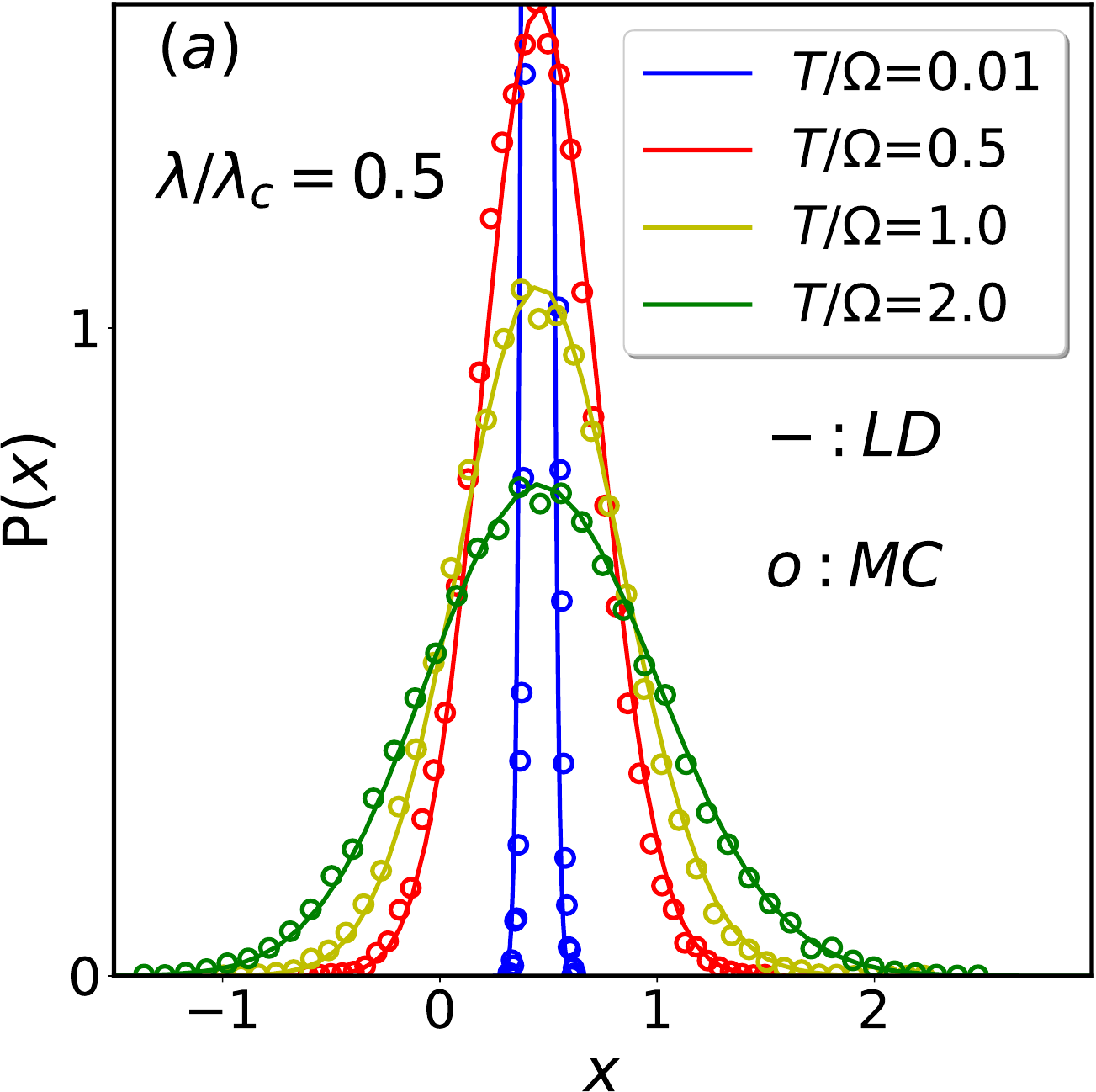}
\includegraphics[height=3.7cm,width=4cm]{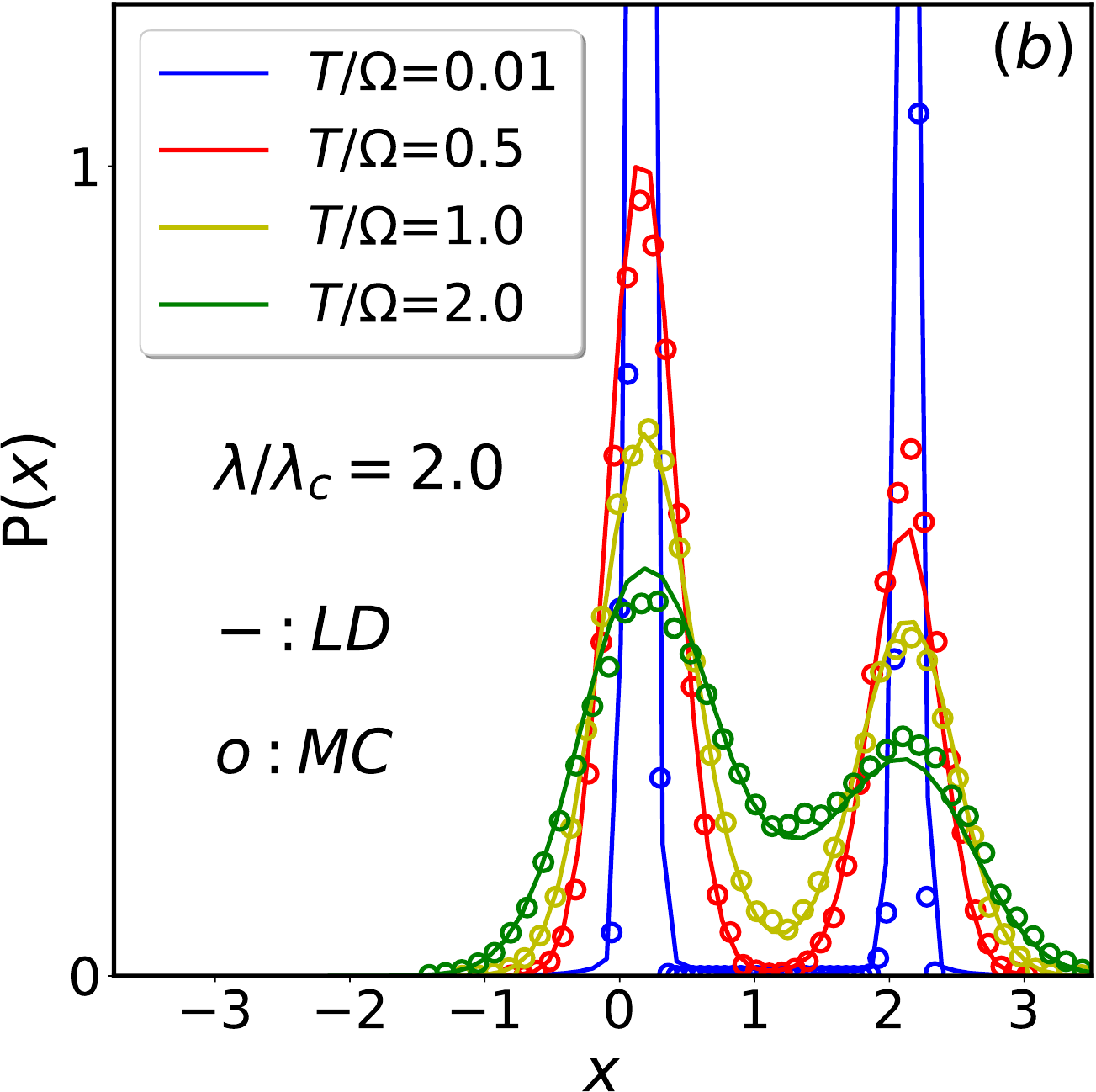}
~~~~~~~}
\centerline{~~~~
\includegraphics[height=2.2cm,width=8.5cm]{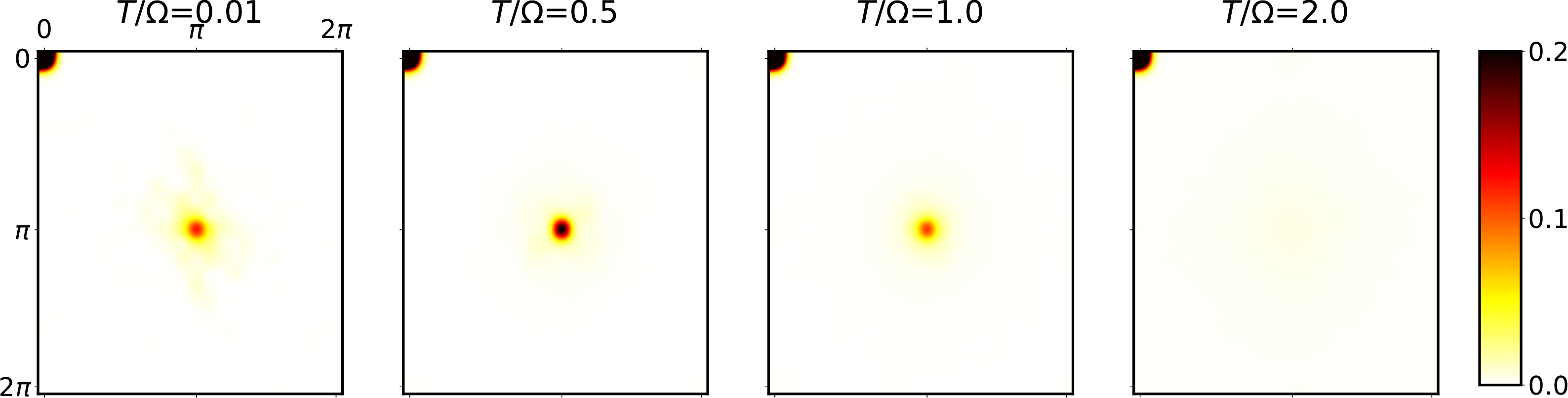}
}
\caption{Static indicators at $n=0.4$.  Top panels:
$P(x)$ at different $T$, comparing MC and LD. (a) -
weak coupling, $\lambda/\lambda_{c}=0.5$, (b) - strong coupling,
$\lambda/\lambda_{c}=2.0$.  Bottom: Structure factor $S({\bf q})$ at
strong coupling shows a peak feature at $(\pi,\pi)$ at
low $T$, signifying significant charge correlations.
}
\end{figure}

\begin{figure*}[t]
\centerline{~~~~~~~
\includegraphics[height=2.7cm,width=13cm]{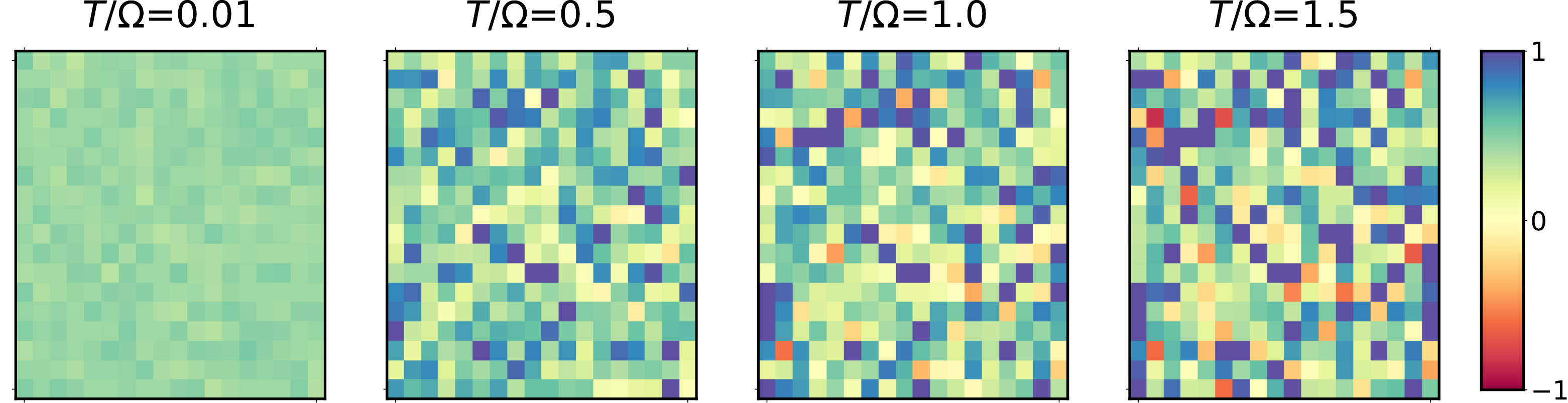}
}
\centerline{
~
\includegraphics[height=2.7cm,width=3.0cm]{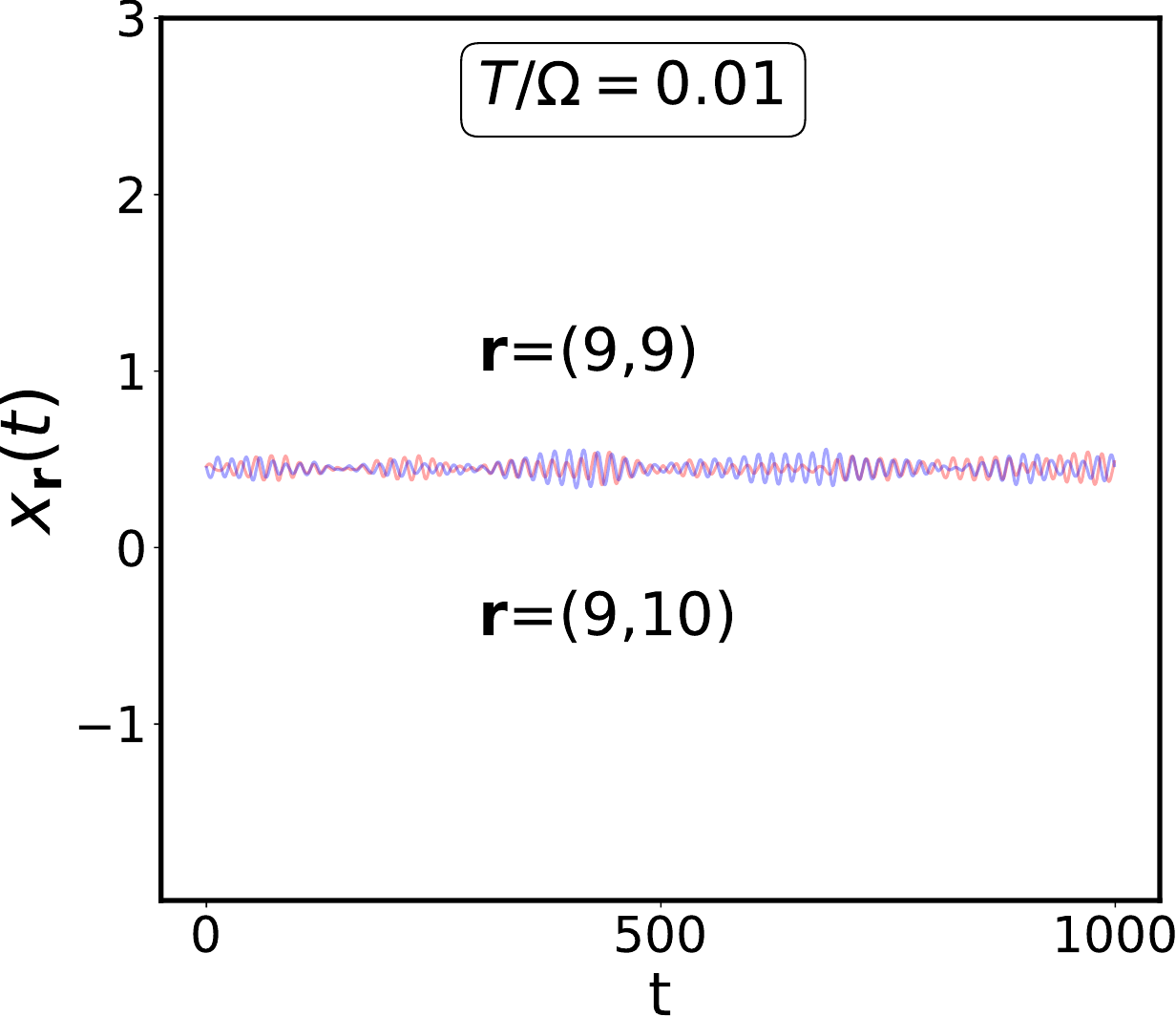}
\includegraphics[height=2.7cm,width=3.0cm]{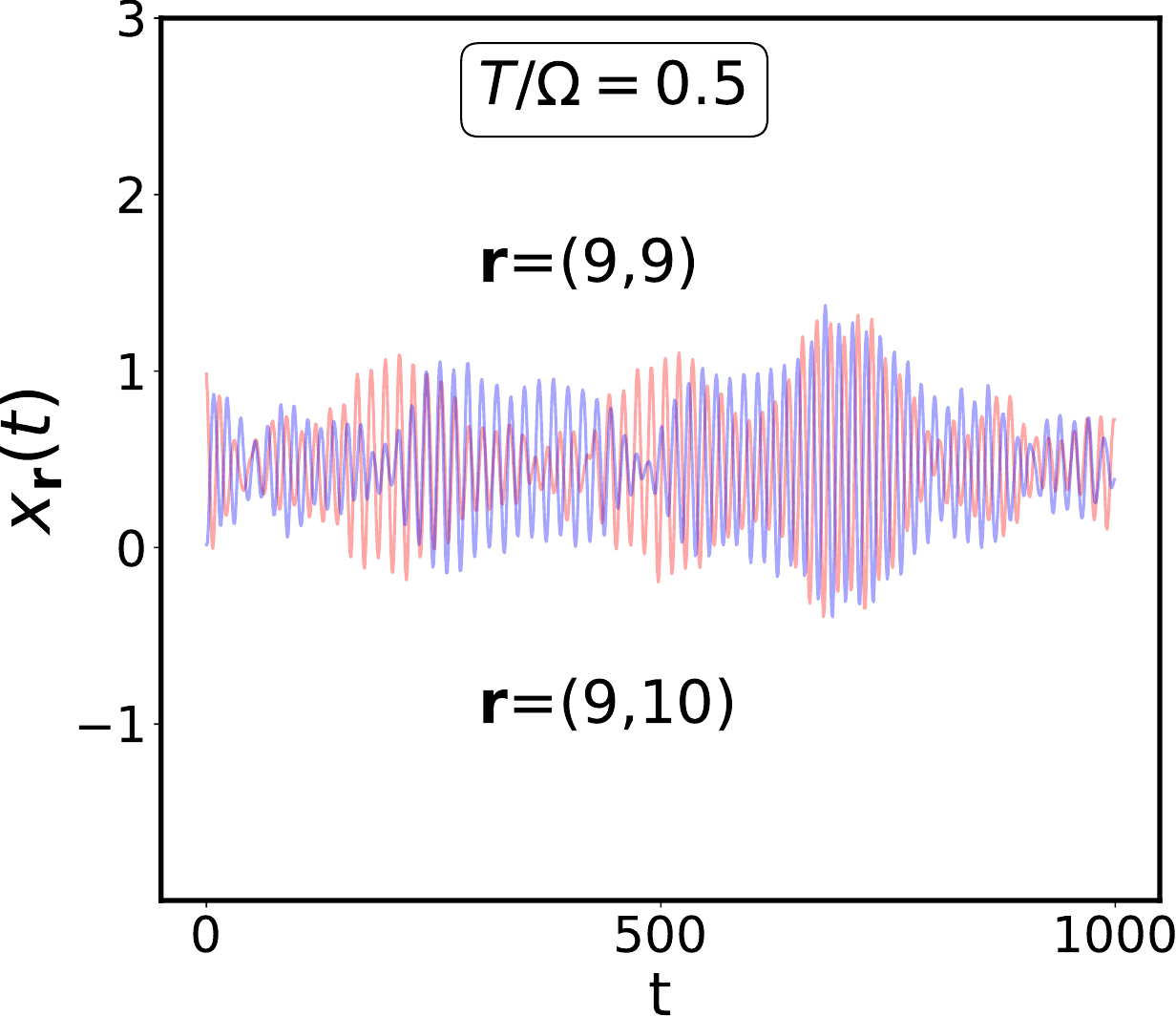}
\includegraphics[height=2.7cm,width=3.0cm]{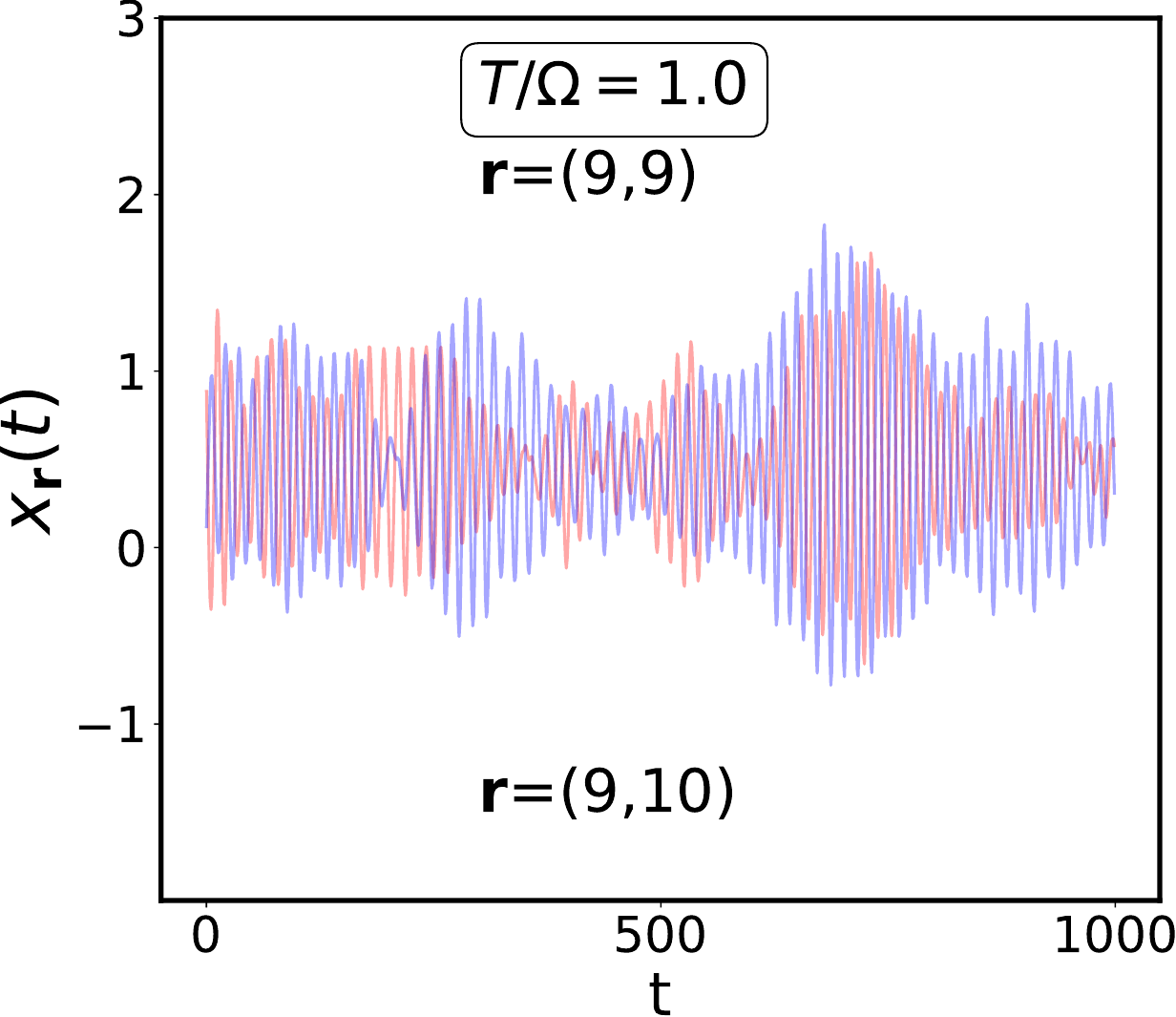}
\includegraphics[height=2.7cm,width=3.0cm]{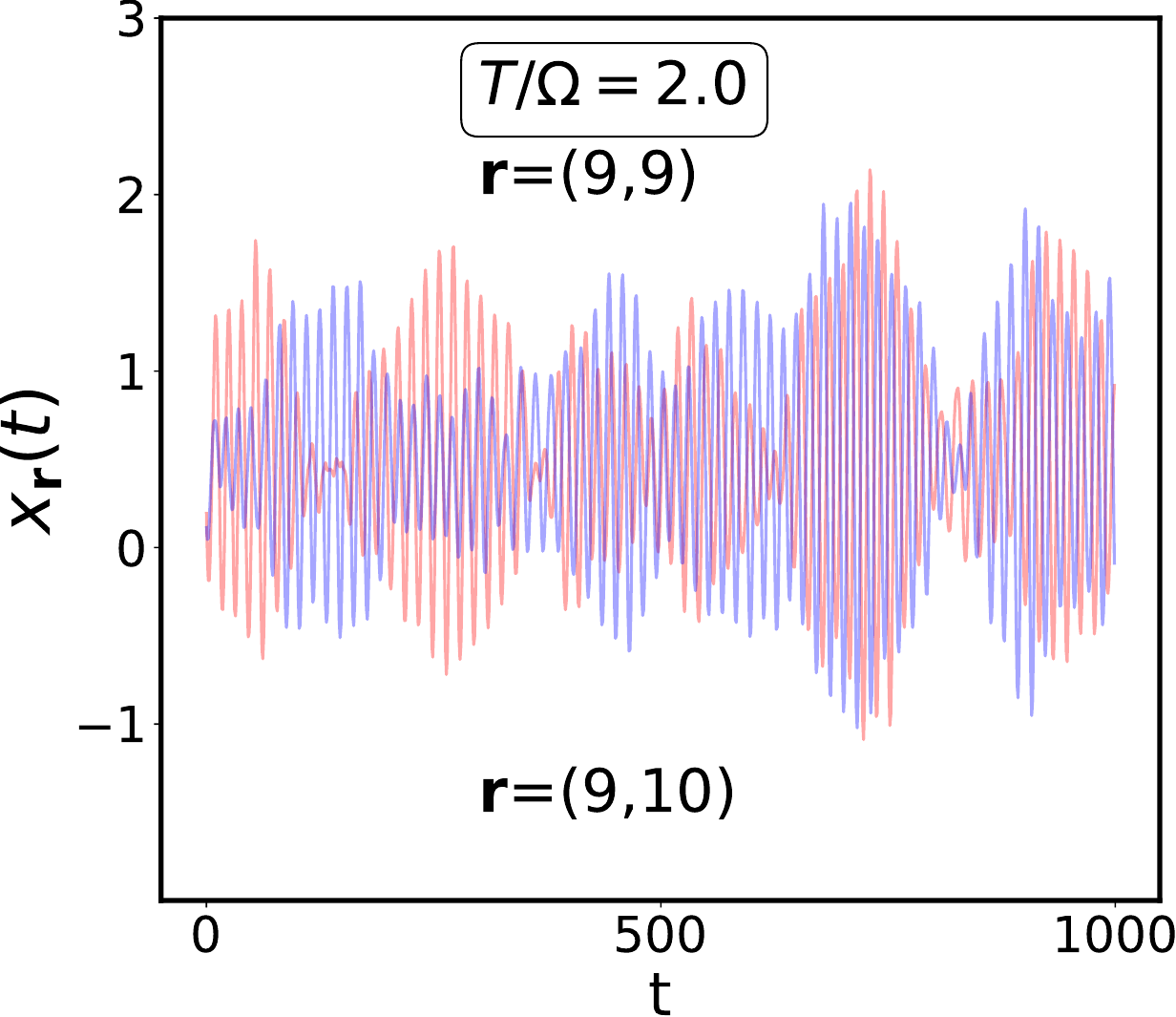}
~~~~~~~~~~~}
\centerline{
\includegraphics[height=2.7cm,width=14cm]{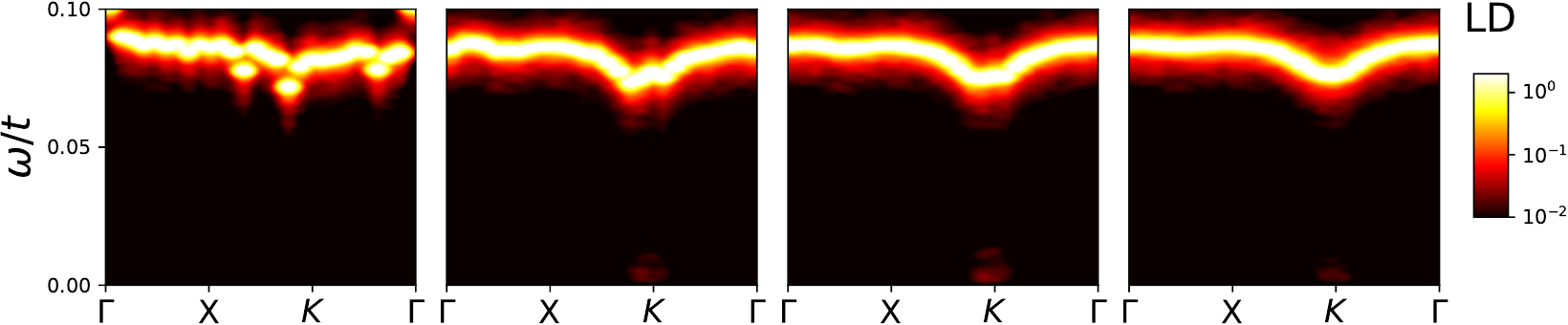}
}
\centerline{
}
\caption{
Charge correlated regime and weak coupling:  $n=0.4$ and $\lambda =0.5 
\lambda_c$.
Top row: instantaneous configuration $\{ x_i \}$, showing an
undistorted $T=0$ configuration evolving into a weakly distorted
and short range correlated pattern with increasing $T$.
Second row: real-time trajectory $x_i(t)$ at two sites, highlighting
essentially similar oscillations at neighbouring sites, with an
amplitude that grows with $T$.
Third row: spectral map $\vert X({\bf q},\omega) \vert^2$ computed using LD.
Brillouin zone trajectory 
$(0,0) \rightarrow(\pi,0)\rightarrow (\pi,\pi)\rightarrow(0,0)$.
The dispersion is essentially $T$ independent with a faint
trace of low energy weight that shows up at intermediate $T$
for ${\bf q} \sim (\pi,\pi)$.
Overall, in this $n-\lambda$ regime we mainly see a perturbative
to large oscillation crossover in the dynamics.
}
\end{figure*}

\subsubsection{Strong coupling ($\lambda  > \lambda_c$)}

Fig.5 focusses on the strong coupling regime ($\lambda/\lambda_{c}=1.2$).
As before, the top row depicts snapshots of the displacement
field. The trajectories on nearest-neighbour
sites feature different mean values ($\sim0$ and $\sim g/K$), 
signifying that one of them has a trapped electron. Harmonic 
oscillations are seen in the first column, which grow in amplitude
at intermediate $T$ ($\sim0.5\Omega$). The phonon spectrum is 
basically composed of the bare band and a split-off weight
near ${\bf q}=(\pi,\pi)$. No `dispersion' is discernable, 
signfying spatially localized phonon modes. The RPA answer 
agrees well with LD, which means there are no qualitatively new 
effects brought in by large fluctuations upto this temperature.

\begin{figure}[b]
\centerline{
\includegraphics[width=2.9cm,height=3.5cm]{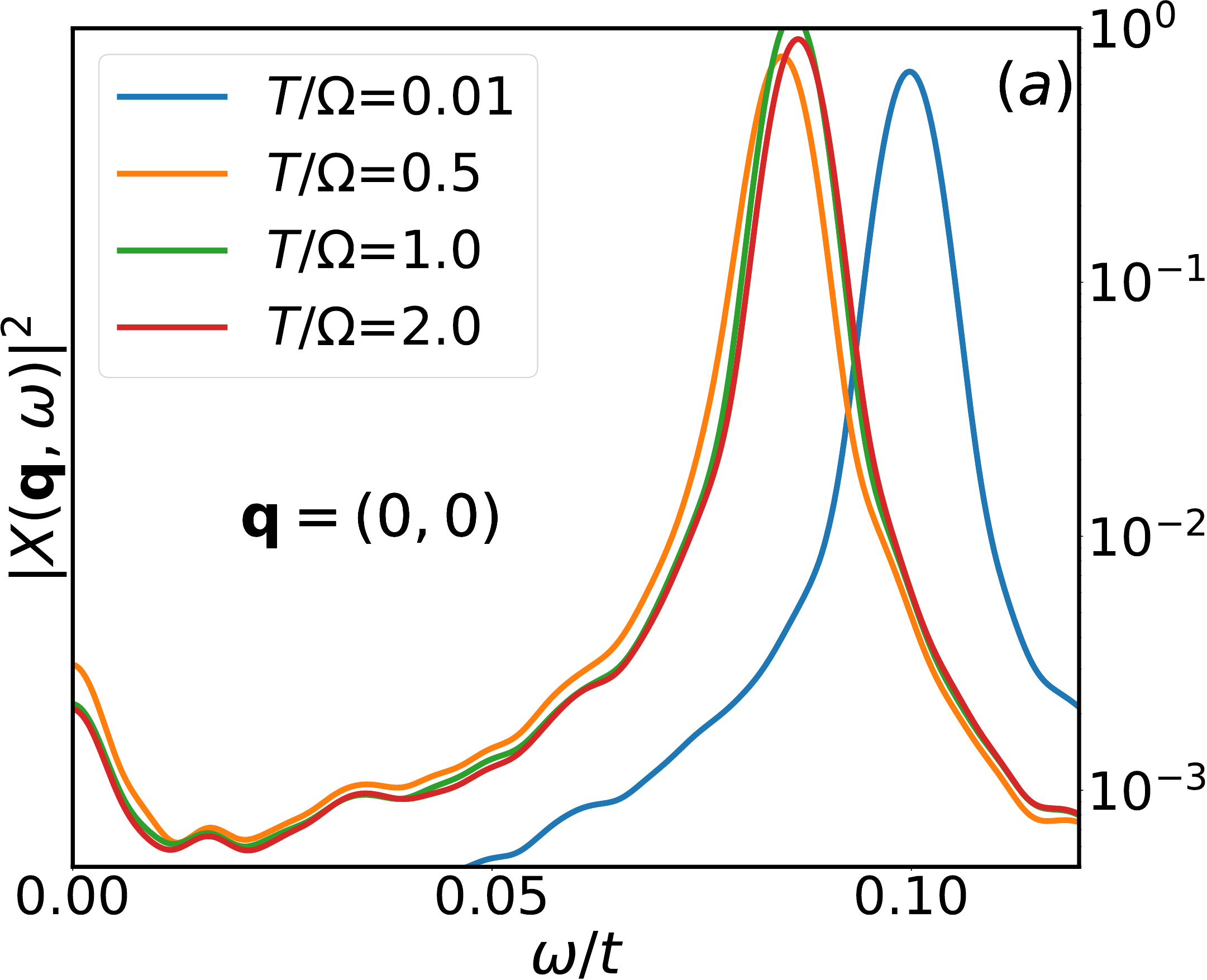}
\includegraphics[width=2.9cm,height=3.5cm]{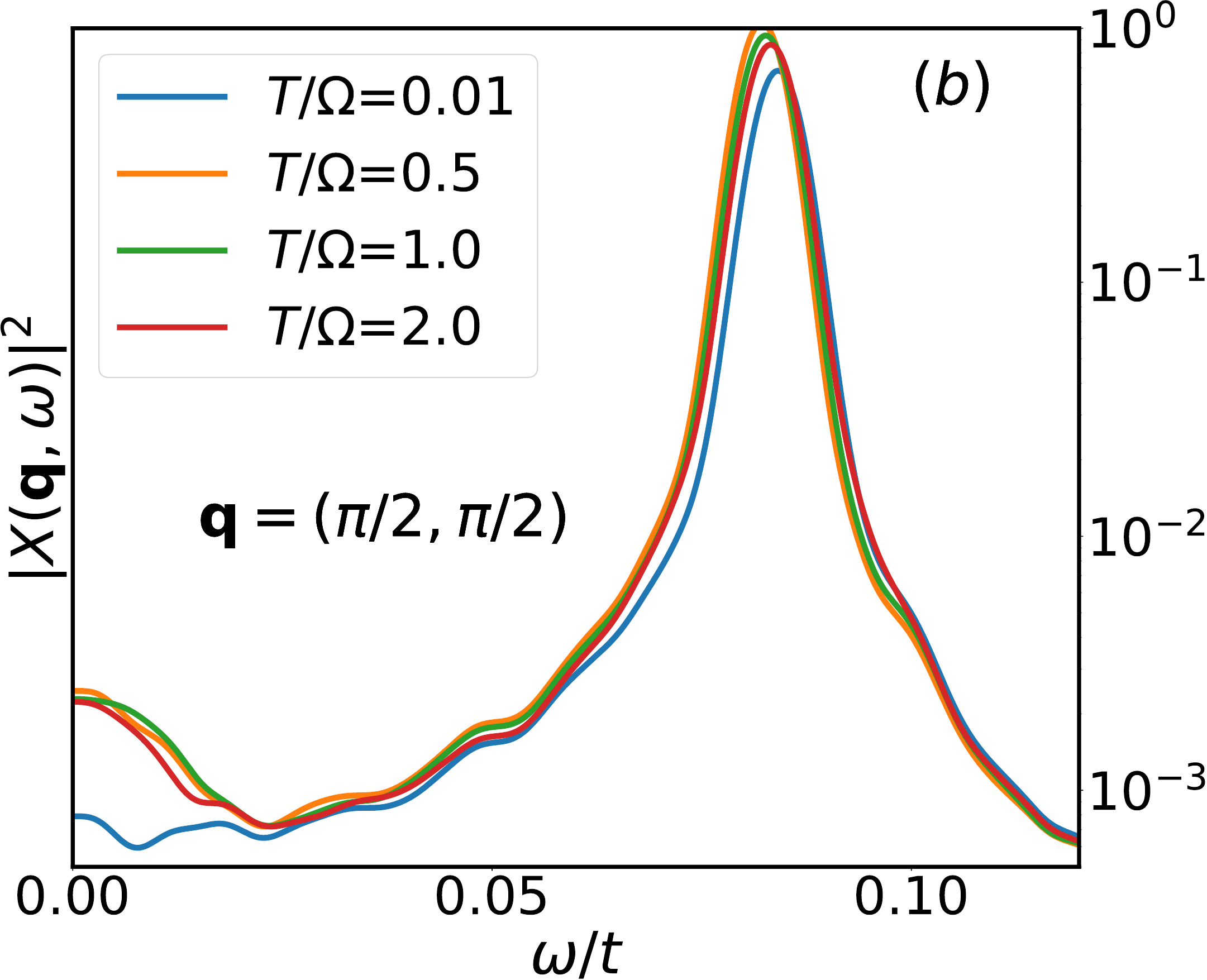}
\includegraphics[width=2.9cm,height=3.5cm]{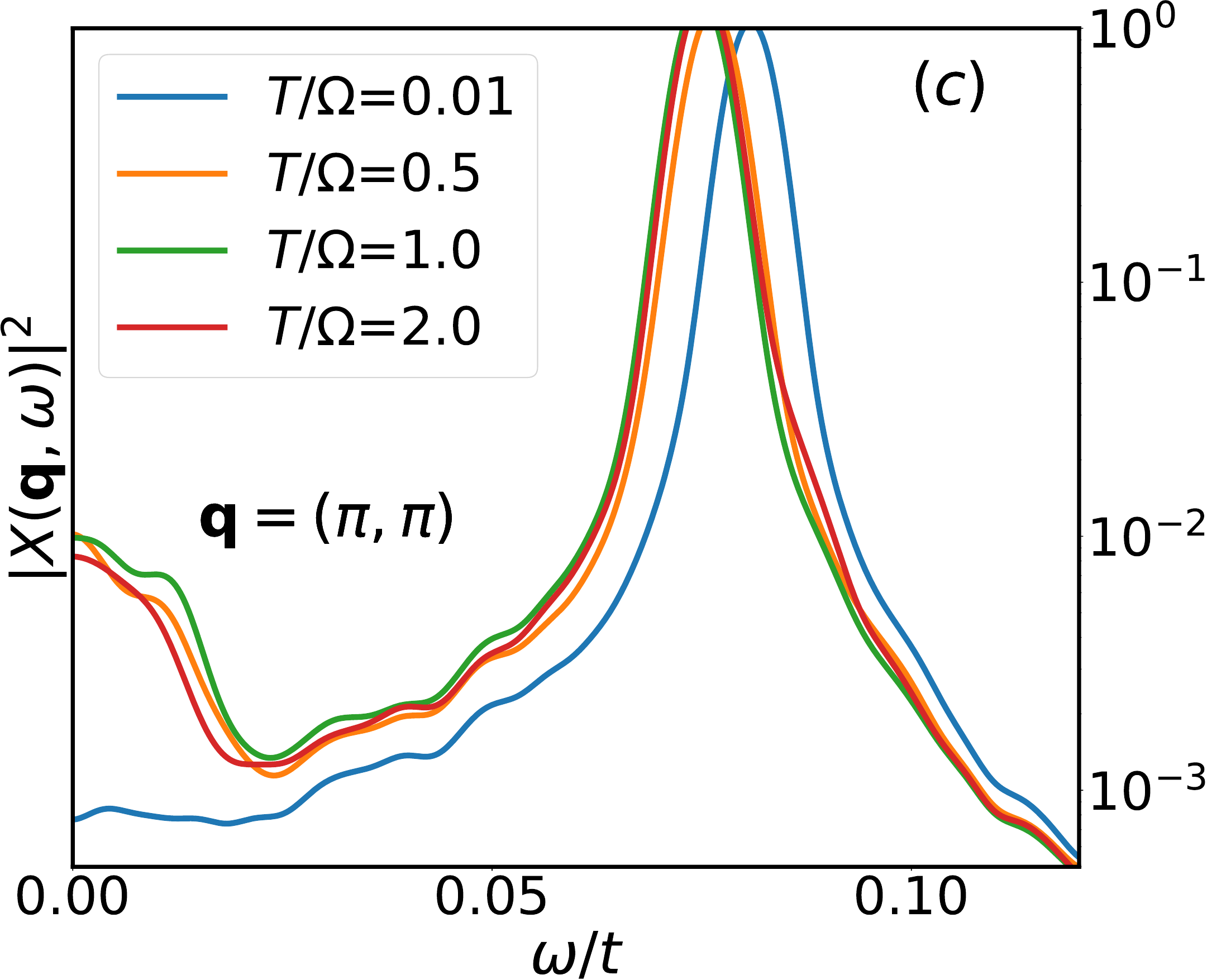}
}
\caption{Lineshapes at $n=0.4$ and $\lambda = 0.5 \lambda_c$
and three characteristic momenta-
${\bf q}=(0,0)$, $(\pi/2,\pi/2)$ and $(\pi,\pi)$. 
Intensities are
shown in a log-scale to emphasize the emergence of
low energy weight in heating. For this density, the
$(\pi,\pi)$ mode shows much more prominent weight
transfer at low energies. The high-energy parts
have similar thermal behaviour at all ${\bf q}$'s.
}
\end{figure}

However, heating up further ($T\sim\Omega$) dislodges
polarons on neighbouring empty spaces. Since this is a 
dilute system, there's a lot of room for polarons to move.
This creates a prominent, `momentum independent' low-energy
band, which is completely missed out by the RPA approach.
The physical reason for this weight to crop up is local
`flip' moves, which change densities by $0(1)$ on neighbouring
sites. We see a reappearance of the polaron for
higher $T$ (last column), and a clean occurence of the `flip'
amidst large oscillations. The low energy weight is also 
remarkably increased.

\begin{figure*}[t]
\centerline{~~~~~~~
\includegraphics[height=2.7cm,width=13cm]{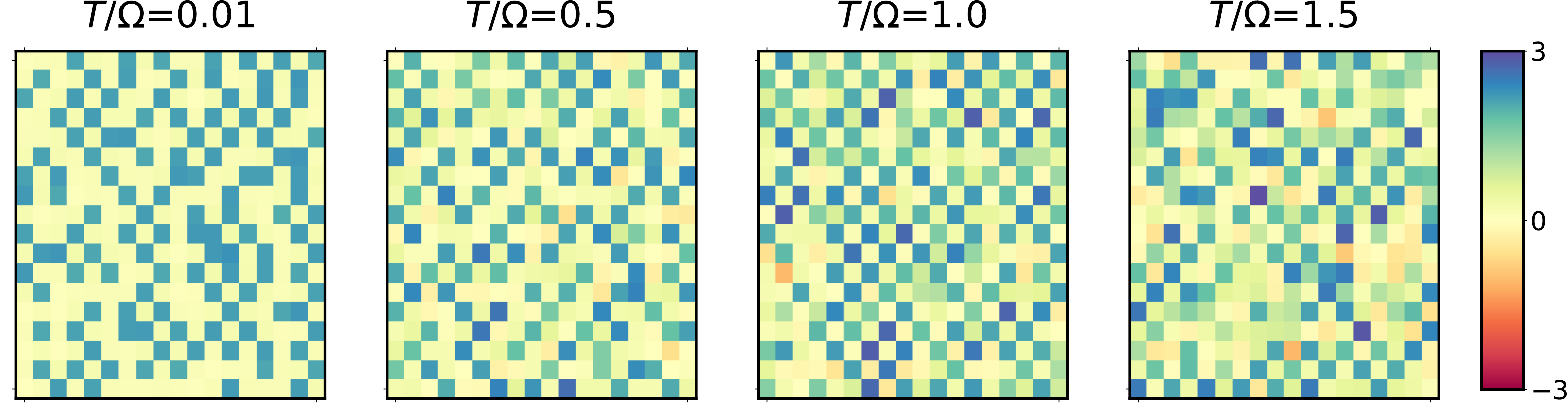}
}
\centerline{
~
\includegraphics[height=2.7cm,width=3.0cm]{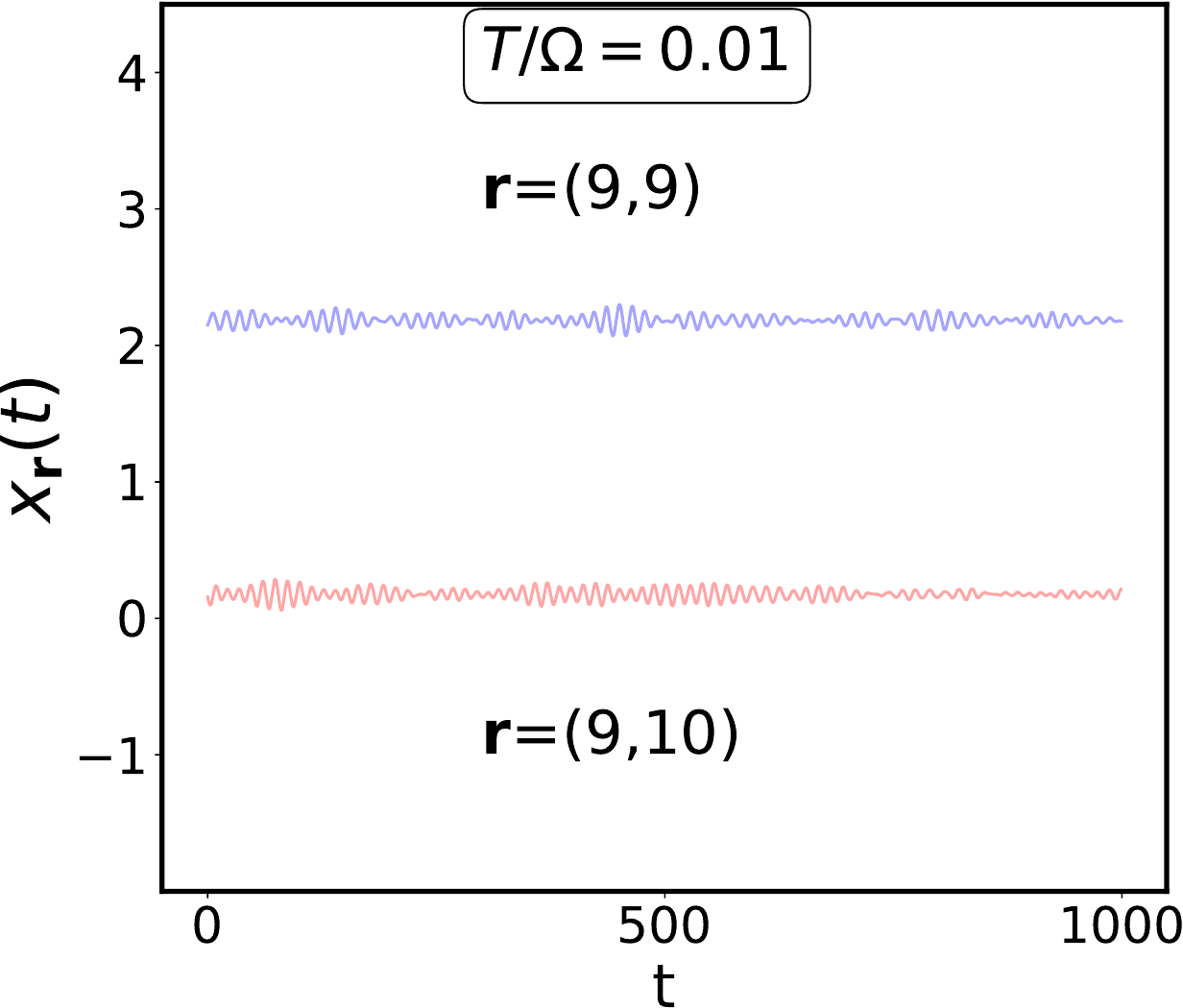}
\includegraphics[height=2.7cm,width=3.0cm]{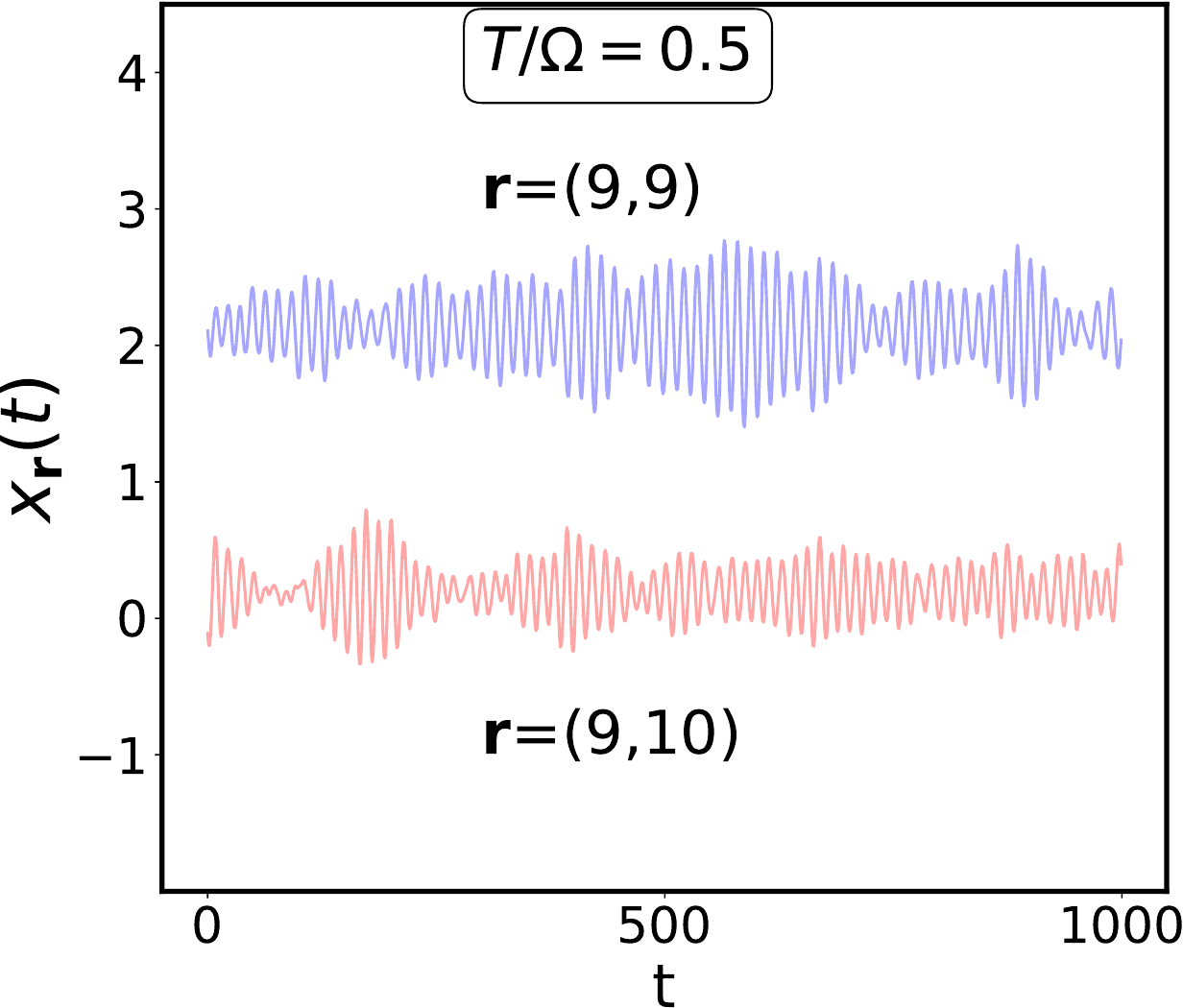}
\includegraphics[height=2.7cm,width=3.0cm]{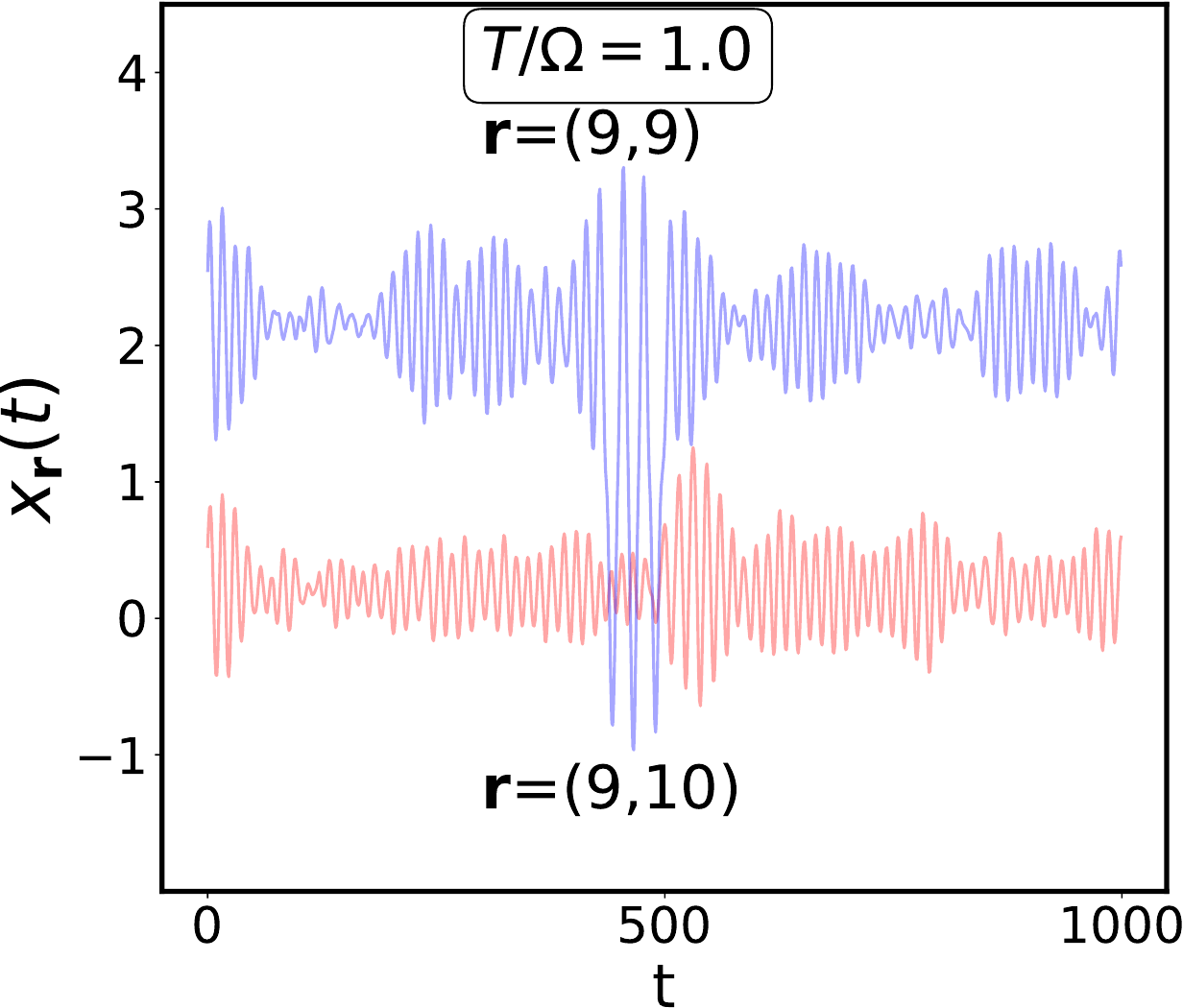}
\includegraphics[height=2.7cm,width=3.0cm]{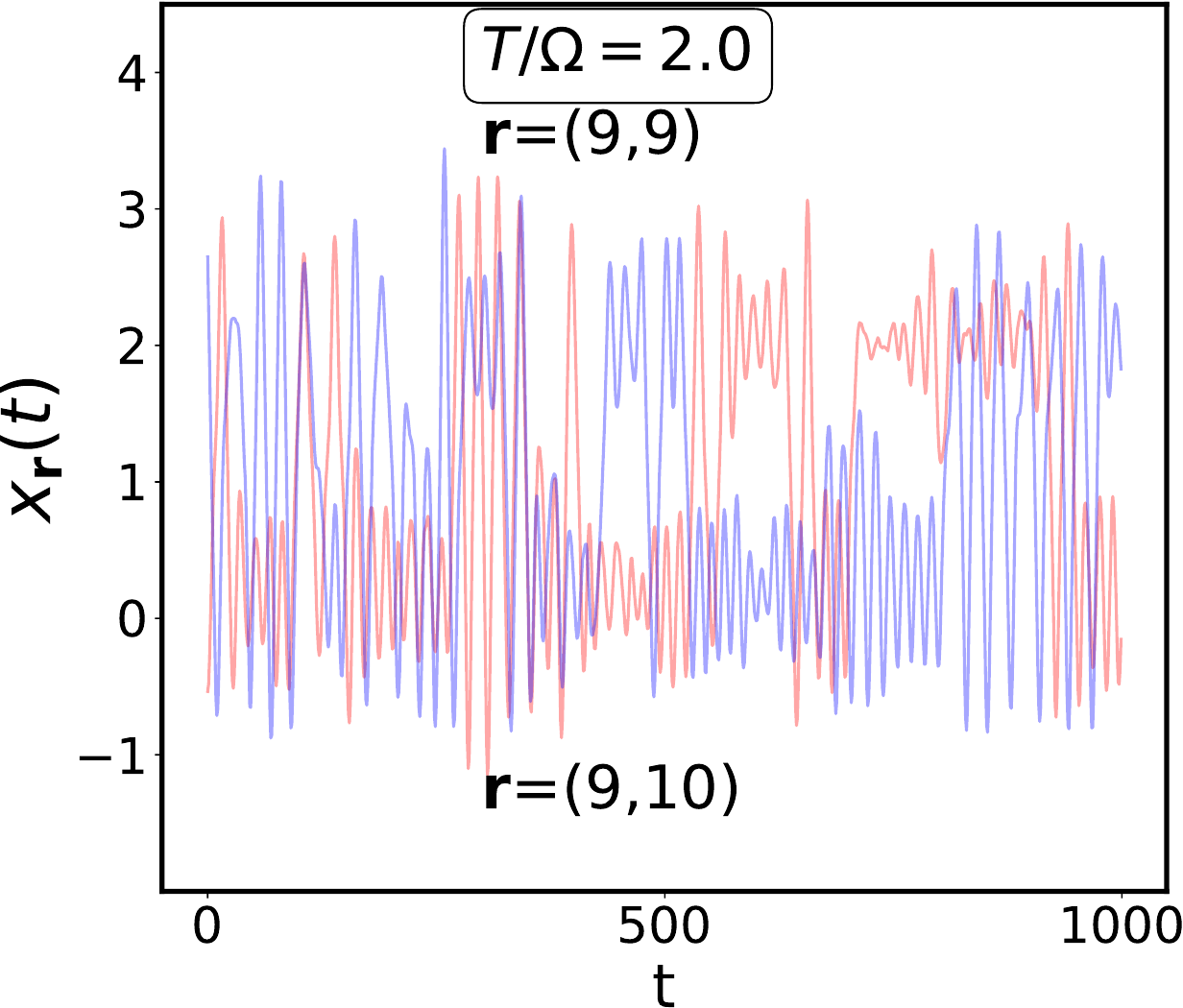}
~~~~~~~~~~~}
\centerline{
\includegraphics[height=2.7cm,width=14cm]{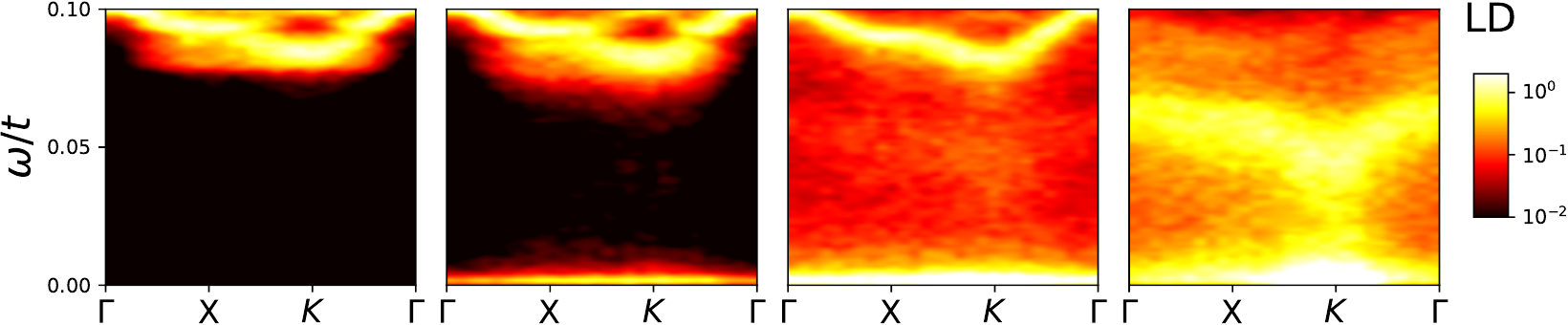}
}
\caption{
Charge correlated regime, strong coupling: $n=0.4$ and $\lambda = 1.2 
\lambda_c$.
Top row: Spatial snapshot, 2nd row: real-time trajectories, 3rd row:
spectral distribution $|X({\bf q},\omega)|^2$   obtained via LD.
The BZ trajectory chosen is $(0,0) \rightarrow(\pi,0)
\rightarrow(\pi,\pi)\rightarrow(0,0)$.
Compared to the dilute case, here the room for polaron
tunneling is less. This reduces the probability of `dislodging'.
At higher $T \gtrsim \Omega$  one observes
correlated tunneling moves. The momentum selectivity of low
frequency weight transfer is increased on heating.
}
\end{figure*}

Figs.6(a)-(c) shows lineshapes at three characteristic
momenta, as in the weak coupling case. The lowest $T$ lineshape 
is bimodal for ${\bf q}=(\pi,\pi)$. Thermal trends are very similar
at all momenta. The low energy weight transfer is small till 
intermediate $T$ ($\sim0.5\Omega$), but rises dramatically thereafter.
The bottom panel focusses on mean ($\bar{\omega}({\bf q})$), 
width ($\Delta\omega({\bf q})$) and fraction of total
spectral weight at low frequency ($w_{low}$) in Figs.6(d)-(f) 
respectively for the same momentum points. 
The window for computing $w_{low}$ is chosen to be $0-0.4\Omega$.
Again, momentum independence
in all these trends are underscored. 
We see appreciable softening
near $T\sim\Omega$ and a sharp rise in dampings 
beyond $0.5\Omega$.

\section{The charge correlated regime}

\subsection{Static properties}

The statics again is characterized by $P(x,T)$ and $S({\bf q},T)$.
The former is featured for weak and strong coupling 
regimes in Figs.7(a)-(b) (top panel) and has 
qualitatively similar features as in the dilute case. 
The critical coupling $\lambda_{c}=0.4$
at this density. The large distortion peak has 
more weight compared to the dilute regime.

The structure factor $S({\bf q},T)$ 
(shown in Fig.7, bottom panel) at low $T$ 
shows a mild peak at ${\bf q}=(\pi,\pi)$, 
signifying `charge order'. We comment that 
this is an artifact of neglecting
quantum fluctuations However, above a
charge correlation scale $T_{CC}\sim\Omega$, 
this peak is significantly subdued and the
system becomes thermally disordered.

\begin{figure}[b]
\centerline{
\includegraphics[width=2.75cm,height=3.3cm]{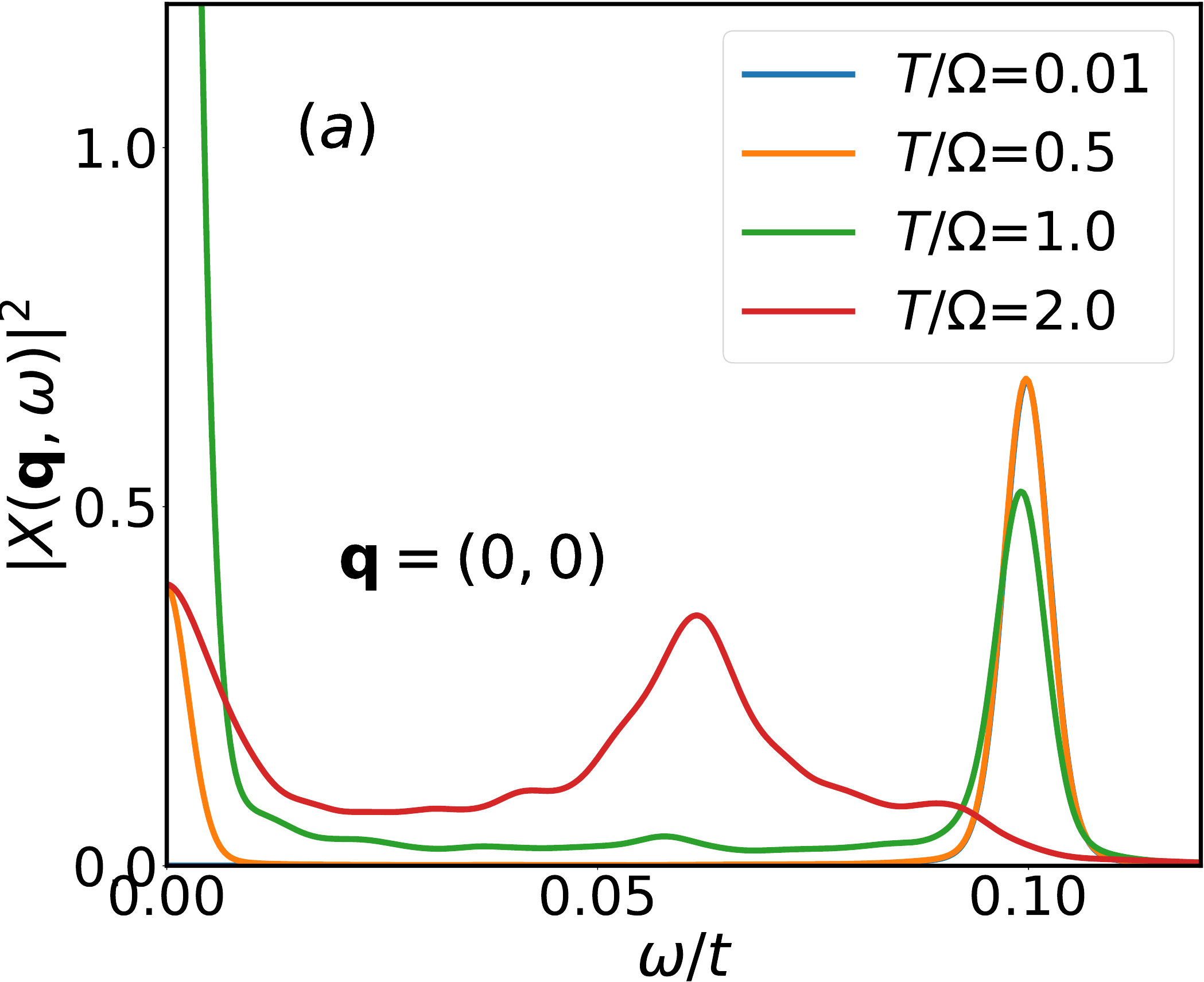}
\includegraphics[width=2.75cm,height=3.3cm]{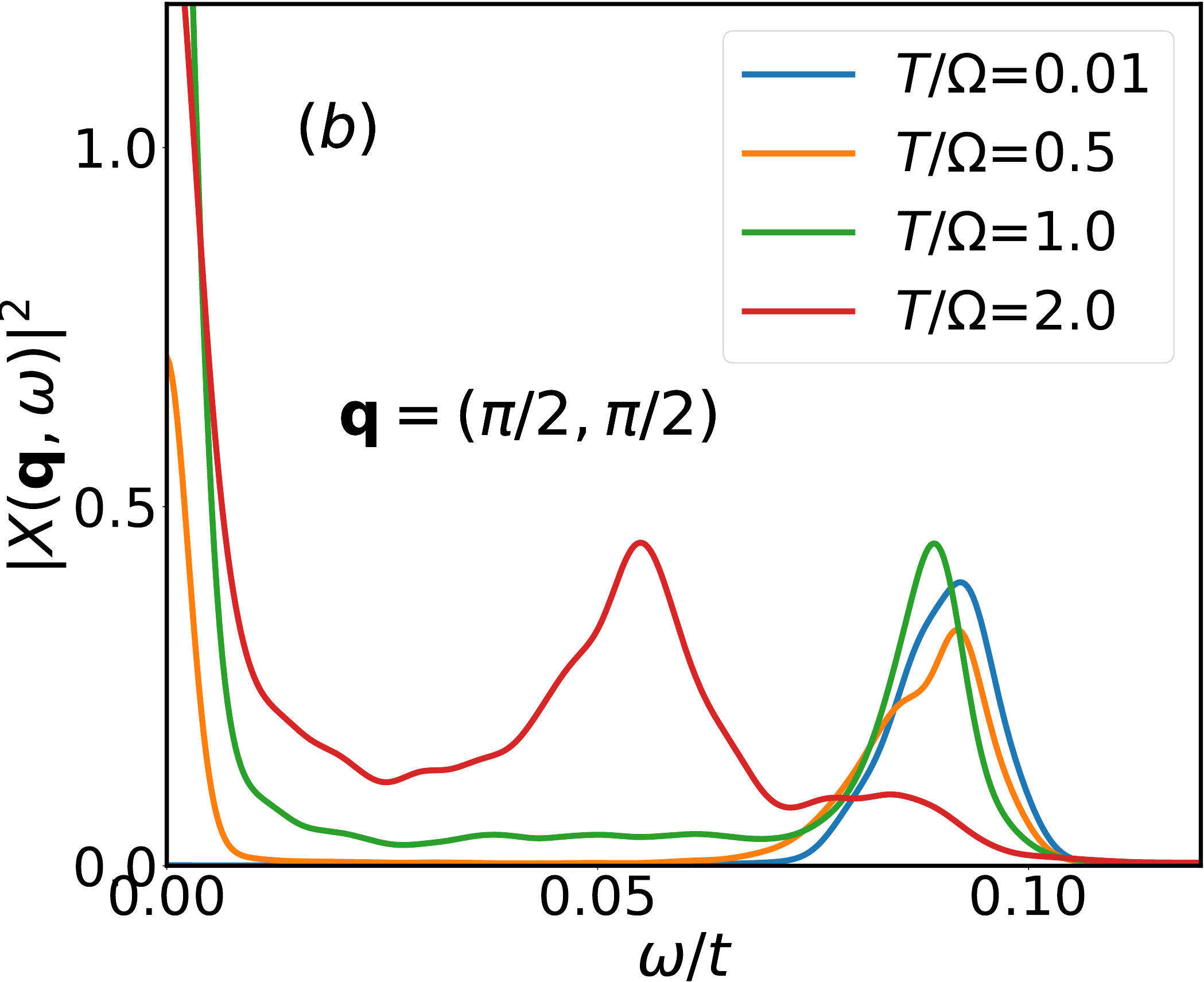}
\includegraphics[width=2.75cm,height=3.3cm]{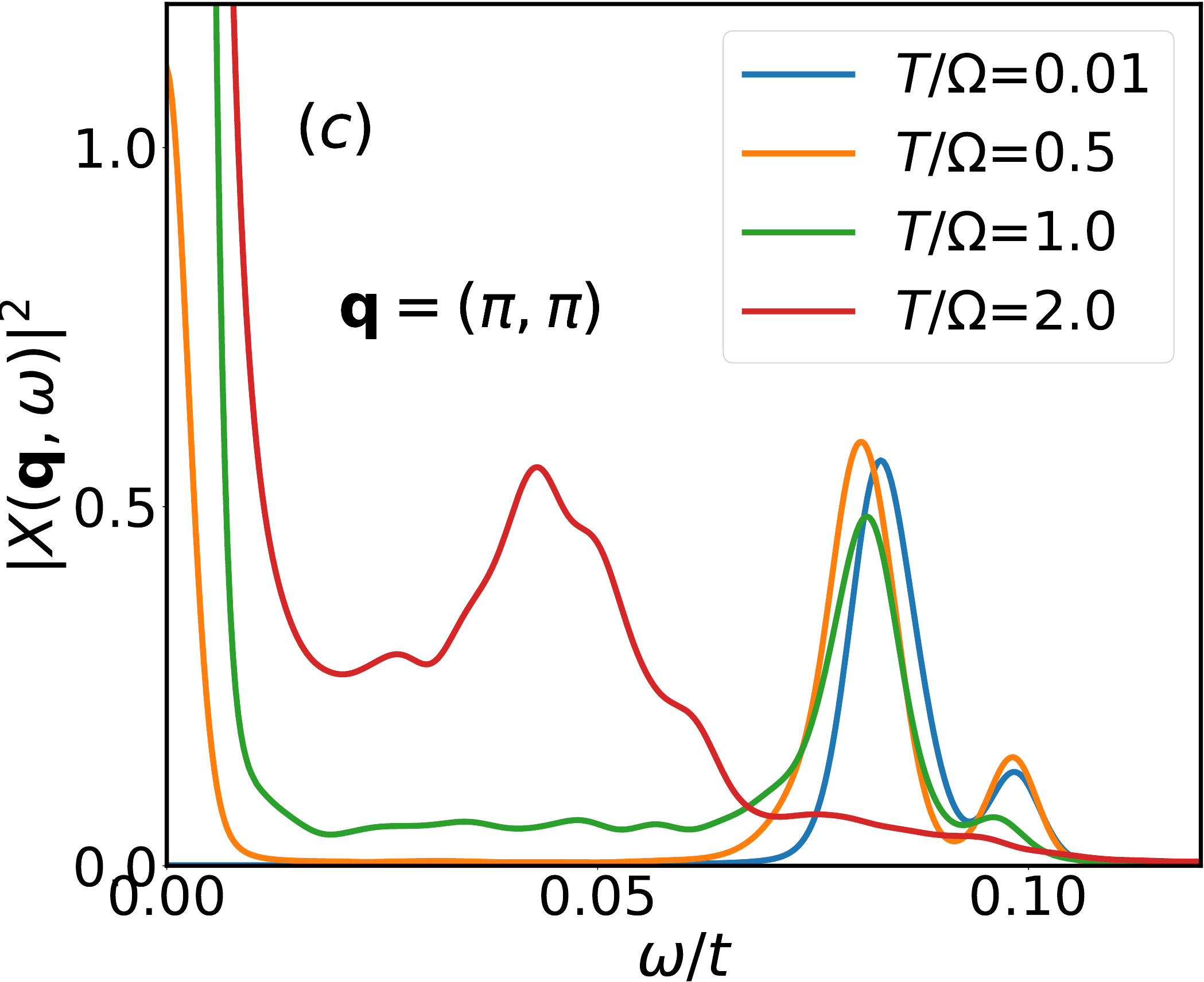}
}
\centerline{
\includegraphics[width=2.75cm,height=3.3cm]{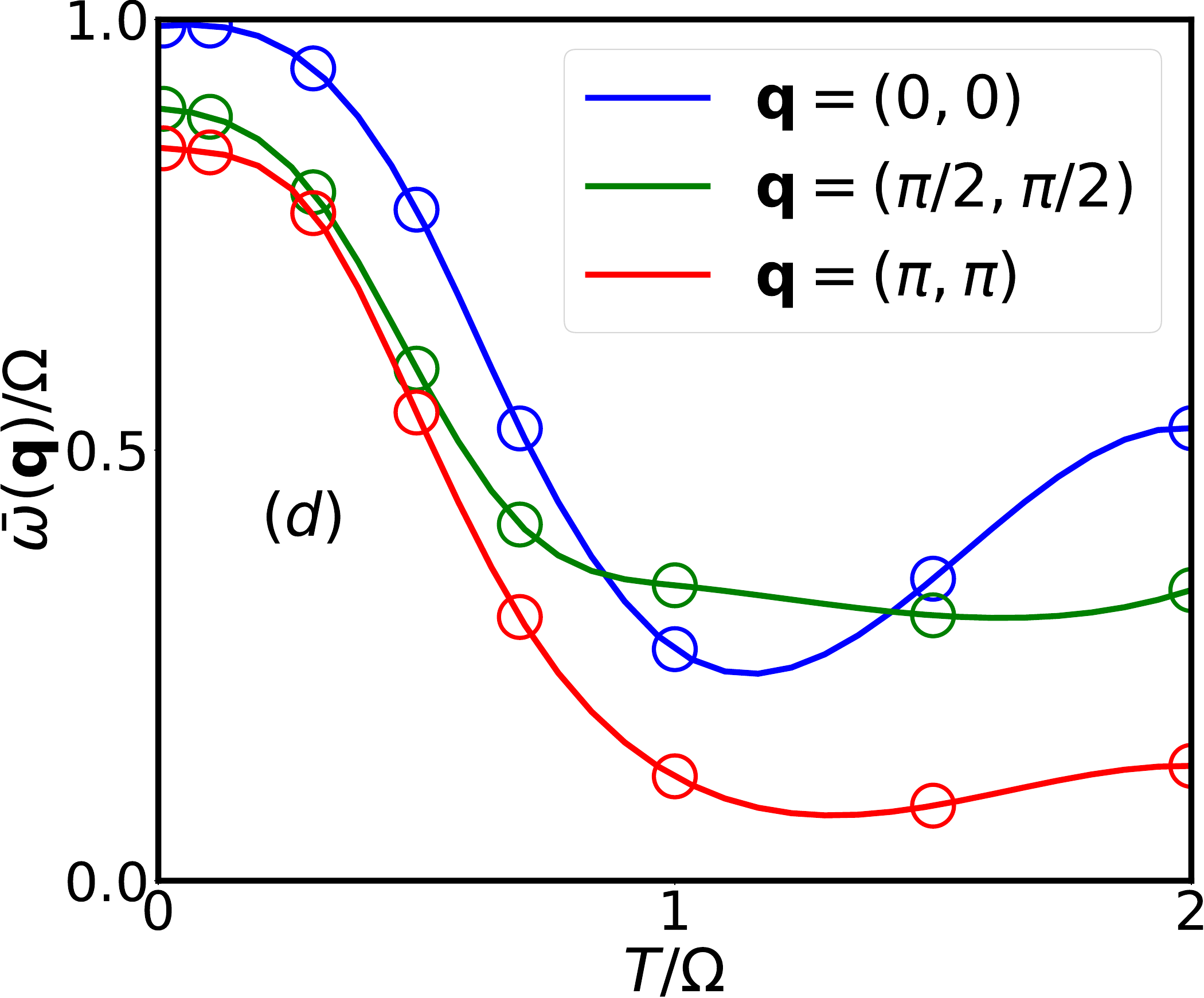}
\includegraphics[width=2.75cm,height=3.3cm]{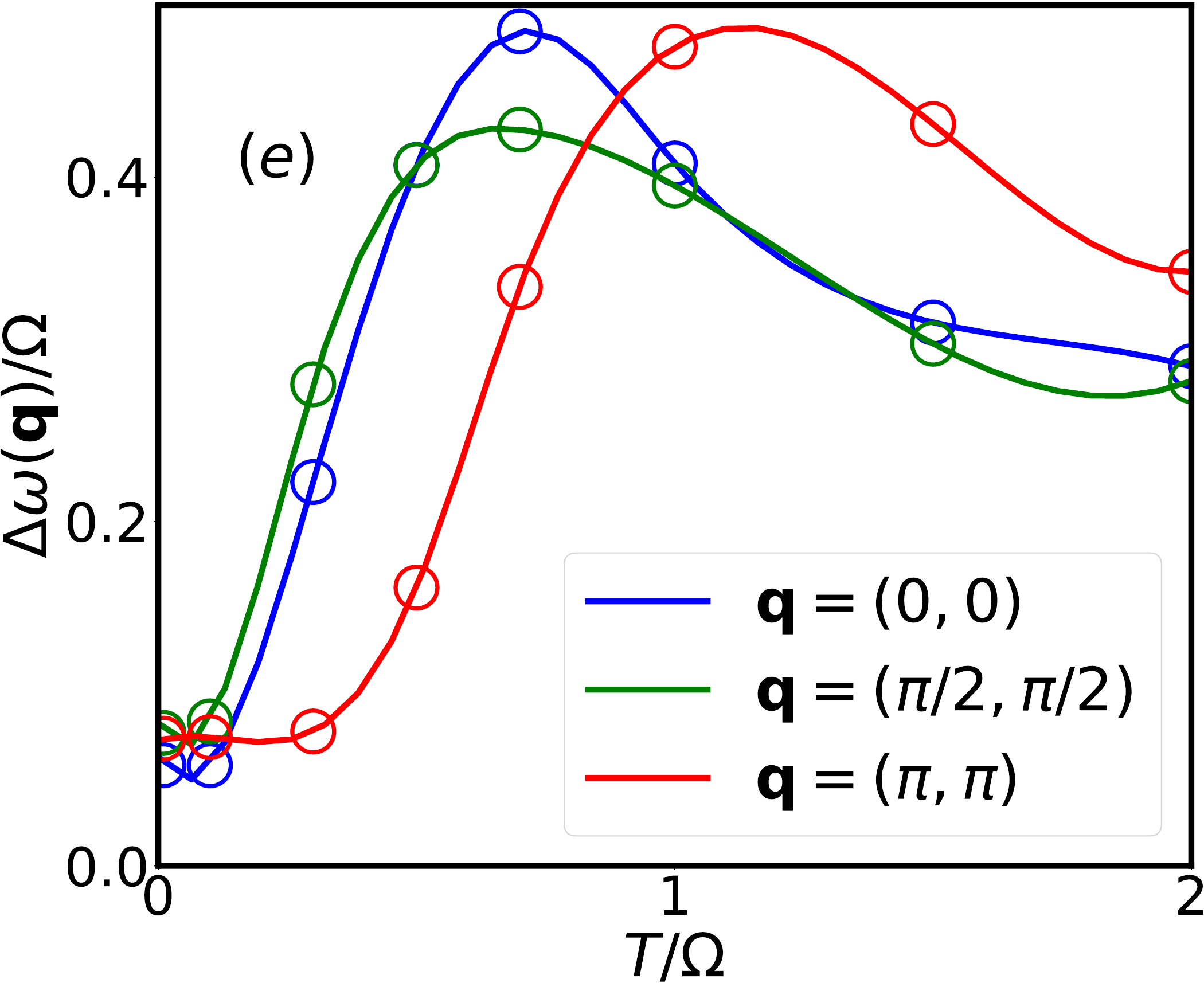}
\includegraphics[width=2.75cm,height=3.3cm]{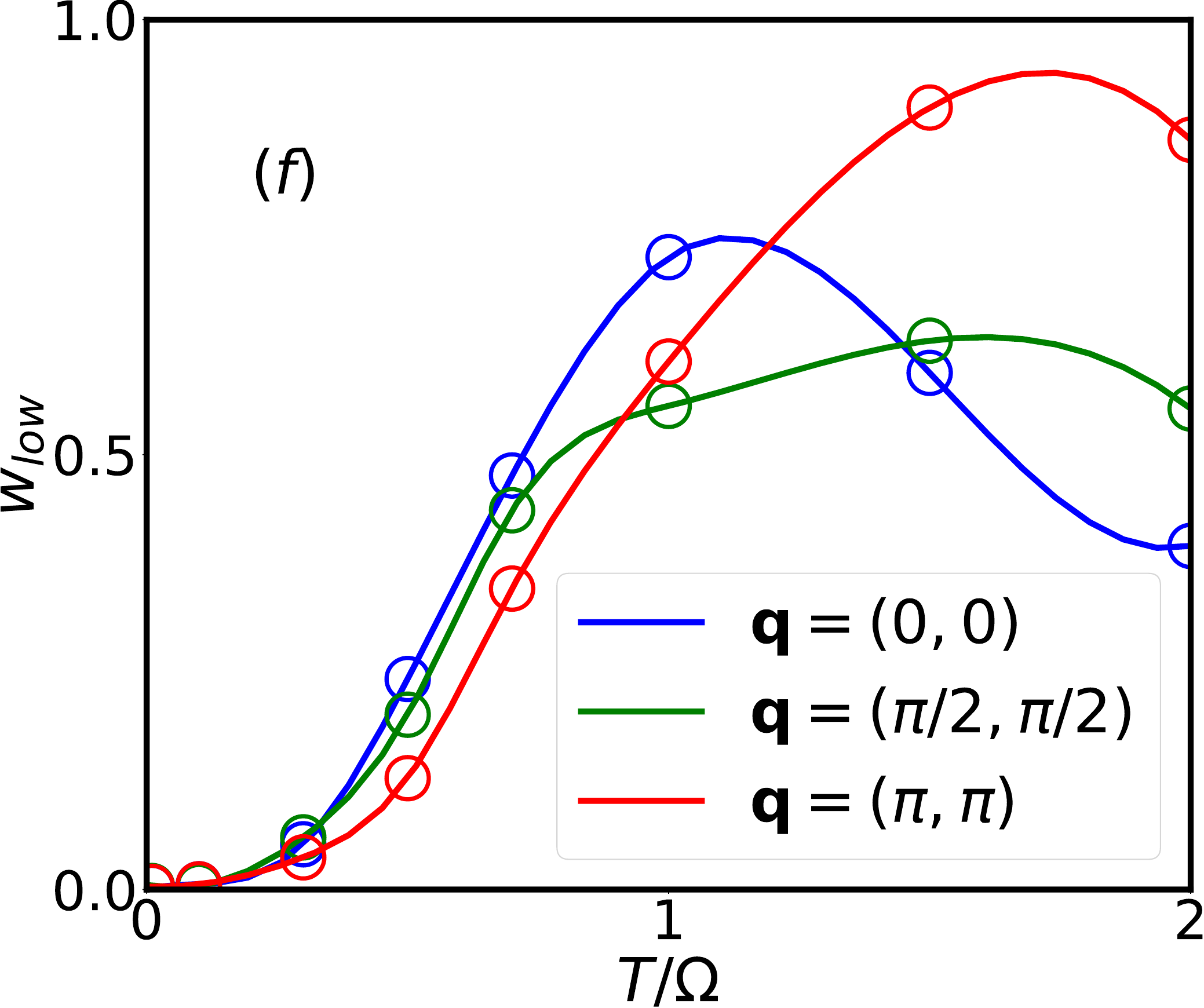}
}
\caption{Spectral features at $n=0.4$ and $\lambda = 1.2 \lambda_c$.
Top: Lineshapes obtained using LD for three momenta-
${\bf q}=(0,0)$, $(\pi/2,\pi/2)$ and $(\pi,\pi)$.
The low $T$ result has a `peak splitting'
at $(\pi,\pi)$.  Thermal behaviour is clearly momentum selective.
Bottom: Mean frequency, $\bar{\omega}$, linewidth, $\Delta
\omega({\bf q})$,  and low energy weight,  $w_{low}$
for the same three wavevectors.  The $(\pi,\pi)$ mode shows enhanced
softening. Low-energy weight rises similarly for all ${\bf q}$'s.
}
\end{figure}

\subsection{Phonon dynamics}

\subsubsection{Weak coupling ($\lambda \ll  \lambda_c$)}

Fig.8 summarizes our findings on the full phonon spectrum
in the weak coupling ($\lambda/\lambda_c=0.5$) scenario.
The top row features real space snapshots, followed
by real-time trajectories in the second row.
The overall behaviour is similar to the dilute case, with
the main difference being a correlation between nearest
neighbours in the `burst' like events. 
The impact of this is directly seen as a faint 
low-frequency weight at ${\bf q}=(\pi,\pi)$, visible
from $T\sim0.5\Omega$. 
The high-energy band is much more coherent
within the LD approach
compared to RPA. This happens as the former 
captures crucial anharmonic fluctuations, 
while the latter is limited to gaussian fluctuations on 
`thermally disordered' backgrounds. Once again, 
we've used a log-scale 
to emphasize the low frequency weights.
We also comment that the momentum selective small 
weight transfer gets quantitatively more prominent 
on moving closer to the crossover ($\lambda\sim \lambda_{c}$). 

In Figs.9(a)-(c), the detailed lineshapes in this regime reveal
that the low energy weight transfer is most prominent 
at ${\bf q}=(\pi,\pi)$, the wavevector relevant for 
short-range correlations. The high energy thermal trends
are similar for all momenta. However, the $(0,0)$ mode
softens considerably on mild heating, closely followed
by the $(\pi,\pi)$ mode.

\subsubsection{Strong coupling ($\lambda  > \lambda_c$)}

Fig.10 highlights the strong coupling 
($\lambda/\lambda_{c}=2.0$) dynamics in the high density regime. 
The snapshots clearly depict a dense polaronic
system at low temperature.
The nearest neighbour trajectories reveal the usual
harmonic-anharmonic crossover in the first two
columns. However, rare `flip' moves are present 
even for $T\sim0.5\Omega$, generating some low-frequency
weight quantified in detail later. Moving close to
the bare phonon scale ($T\sim\Omega$), interesting
`stray' flips are seen, which accentuate the low-frequency
weight transfer. The high energy band retains its low $T$
shape. At even higher temperatures, this band itself moves to
lower frequencies. The spatial correlations amongst
flip moves strengthen, leading to increased momentum
selectivity. On further heating, the correlations weaken.
The RPA method, as before, gives progressive
dampings on heating and fails to capture the 
rich thermal behaviour. 

Fig.11 summarizes the detailed phonon properties. First, in 11(a)-(c),
the lineshapes (top panel) at three characteristic wavevectors
$(0,0)$, $(\pi/2,\pi/2)$ and $(\pi,\pi)$ reveal that there's 
indeed more pronounced momentum selectivity compared to the
dilute case. The low temperature lineshapes are unimodal for
the first two while the $(\pi,\pi)$ mode features a bifurcation.
On heating up, even at intermediate $T\sim0.5\Omega$, there's 
a $\sim20\%$ weight transfer near zero frequency. This 
is related to rare thermal tunneling moves. The bottom panel
(Figs.11(d)-(f)) shows prominent softening at $(\pi,\pi)$ 
compared to other momenta, although broadenings show an opposite trend. 
Low energy weight rises similarly for all wavevectors.

\section{Commensurate filling and charge order}

In this section, we summarize the phonon properties for 
charge ordered ($n=0.5$) systems. In this case, $\lambda_{c}=0$.
The system has an order-disorder thermal transition at 
$T\sim2\Omega$ for $\lambda=0.5$, that falls as $\sim1/\lambda$ 
at strong coupling. 

\subsection{Static properties}

The top row of Fig.12 shows the $P(x)$ distributions (a) 
and structure factor $S(\pi,\pi)$ (b)
at this density for various temperatures. The results obtained using
Langevin dynamics and Monte Carlo agree very well over all thermal
regimes. The order-disorder transition at this coupling takes place
around $T\sim0.1t$. The distribution of distortions is bimodal and 
symmetric about $x=1.0$ at low $T$. Thermal effects broaden the peaks
and almost merge them for $T\sim2T_c$.

\subsection{Phonon dynamics}

The bottom two rows of Fig.12 feature spectral maps for $n=0.5$. 
The top panel shows spectra obtained using the LD approach while
the bottom one exhibits RPA based spectra. We see that the perturbative
theory fails to capture the low-energy weight transfer at $(\pi,\pi)$
near $T\sim\Omega\sim T_c$. These are evidently due to non-Gaussian
`correlated flip' moves in the lattice. The high temperature dynamics
is that of a `disordered polaron liquid', also captured by the LD 
approach and not by the RPA, which just gives a very broad spectrum.

\begin{figure}[t]
\centerline{
\includegraphics[height=3.6cm,width=4cm]{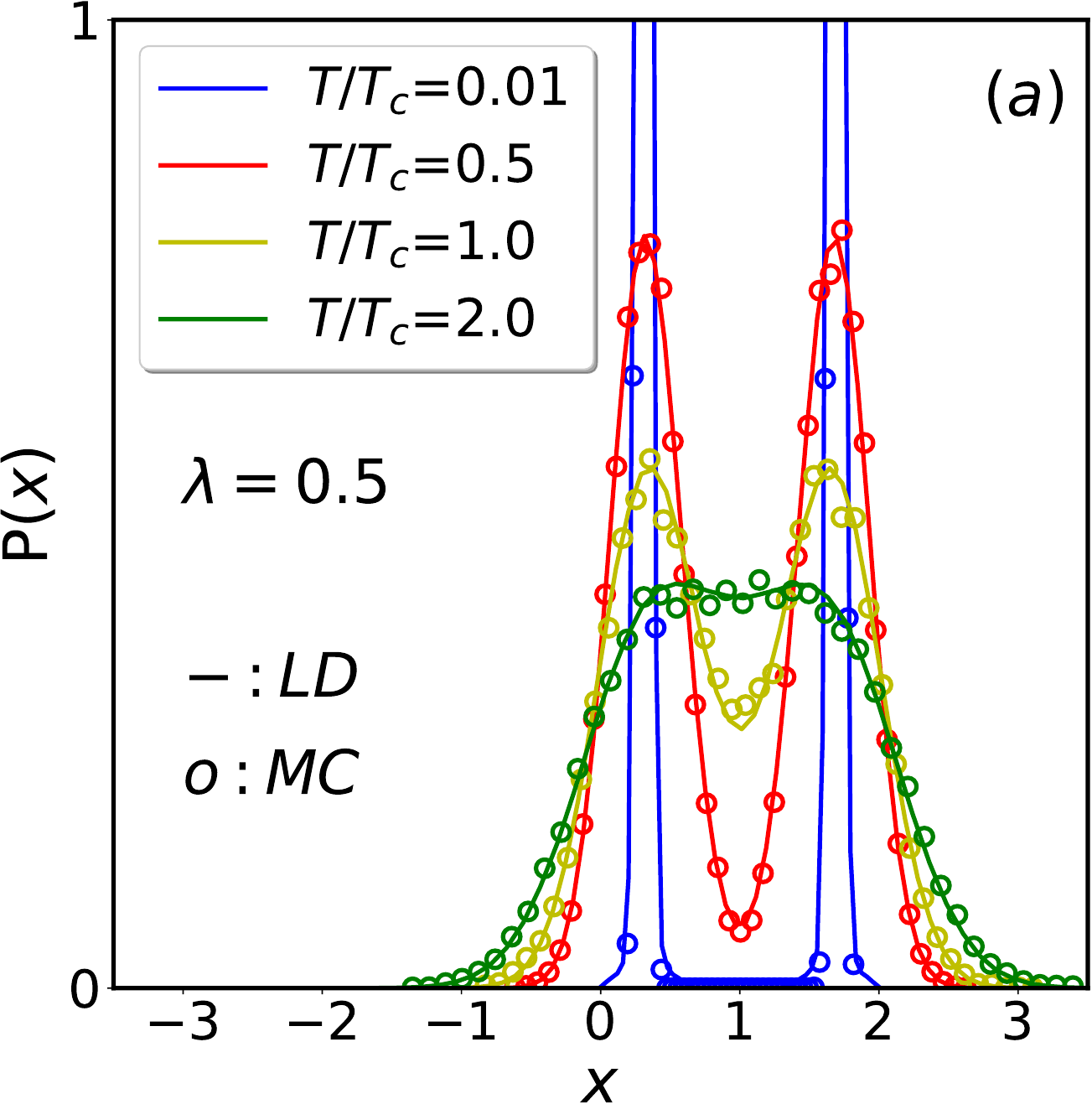}
\includegraphics[height=3.6cm,width=4cm]{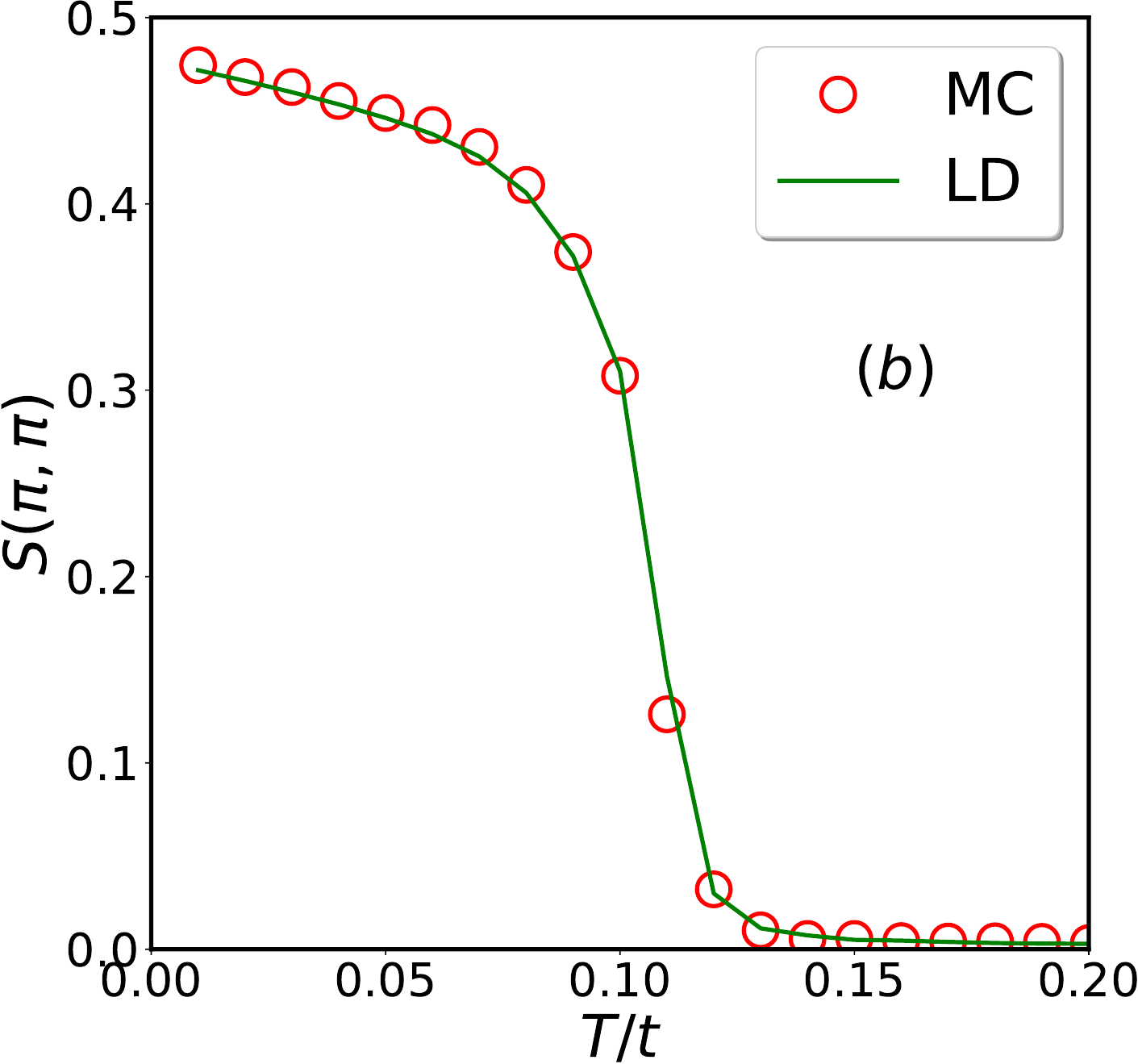}
}
\centerline{~~~~
\includegraphics[height=2.2cm,width=8.7cm]{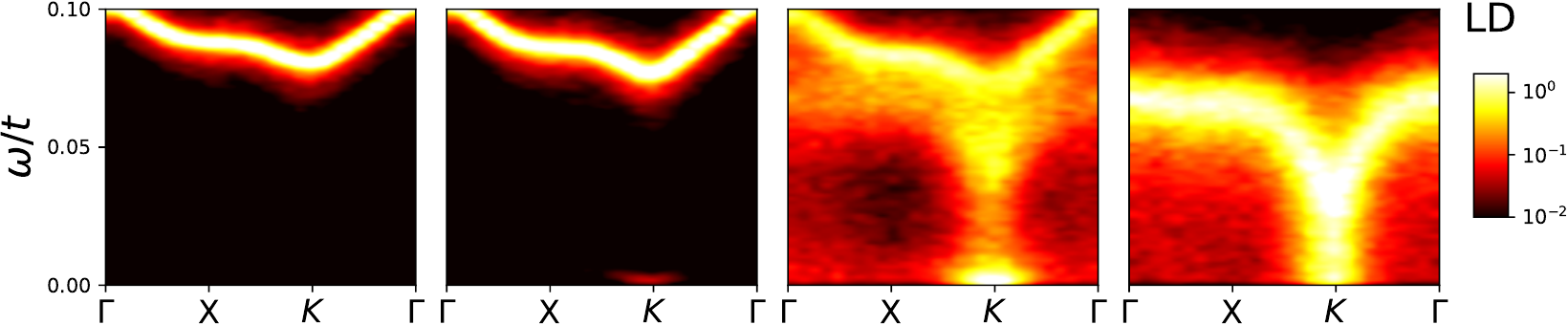}
}
\vspace{0.2cm}
\centerline{~~~~
\includegraphics[height=2.2cm,width=8.7cm]{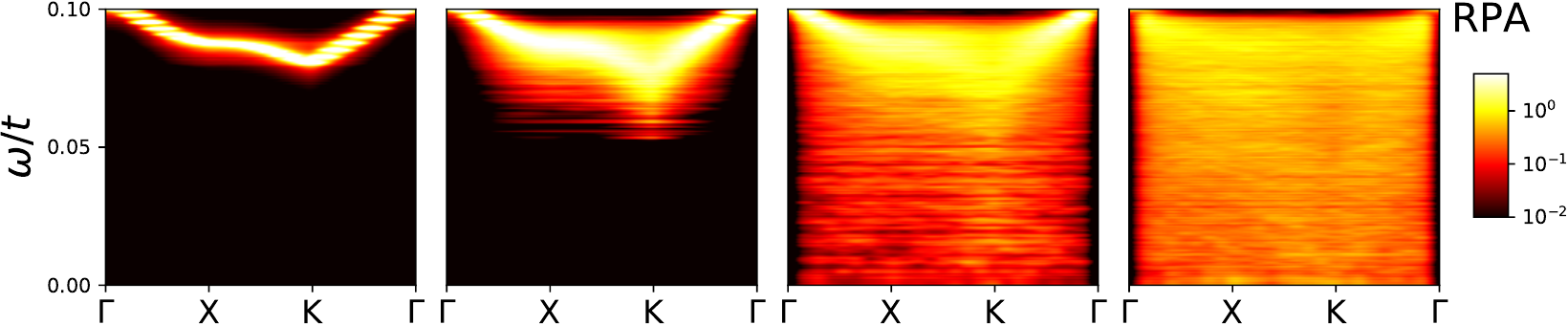}
}
\caption{Features at $n=0.5$, increasing $T$ on a CO state.
Top row: (a)~Distribution of distortions, $P(x)$, and
(b)~structure factor $S(\pi,\pi)$. The former shows symmetric
bimodal profile at all temperatures.
The structure factor is shown only for the ordering wavevector, 
and shows an order-disorder transition for $T_{c}\sim0.1t$.
Bottom: Phonon spectra computed using LD (upper row) 
and RPA (lower row).
The dramatic low-energy weight transfer near criticality
is captured by the former, while RPA just shows an 
increasingly broad spectrum due to the disordered
thermal backgrounds. Beyond $T_c$, LD reveals
dynamics in the thermal `polaron liquid'.
}
\end{figure}

\section{Discussion}

\subsection{RPA versus the Langevin approach}

We've used two complementary strategies to investigate the phonon
physics in the Holstein model, namely traditional RPA and Langevin
dynamics. Both of them are viable in the physically relevant adiabatic 
regime, where phonon variables are slow. The former is restricted
to small amplitude quantum fluctuations about finite temperature
static `backgrounds', while the latter can access large amplitude 
classical dynamical fluctuations. 

In the strong coupling scenario, large distortions are already
`preformed' in the lattice. Within RPA, we only probe small 
amplitude fluctuations about the `thermally disordered' backgrounds.
These don't faithfully represent the true dynamics of the system, 
which involves re-organization of the polarons on neighbouring
sites. These are rare, non-Gaussian moves. The Langevin dynamics 
method approximately captures these events and hence gets interesting
low-energy signatures, which RPA fails to access. In other words, 
while the latter approach does contain large distortions at the 
static level, their dynamics `as a whole' is missed out. One 
only calculates the impact of small excursions on top of the 
`distorted' states. Fig.13 illustrates the difference in the
family of fluctuations accessed by the respective methods.

\begin{figure}[t]
\centerline{
\includegraphics[width=8.5cm,height=6.2cm]{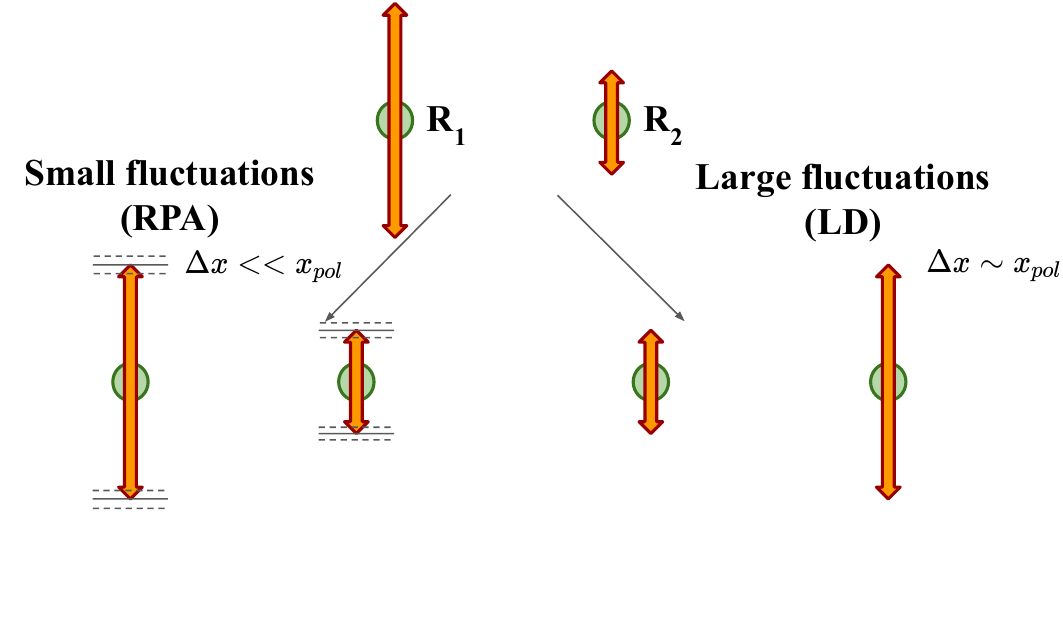}
}
\caption{Schematic cartoons of the RPA and LD schemes.
In the former, one 
explores small amplitude fluctuations about `large' or `small'
background distortions. 
Within LD, the novelty is to
capture `swapping' moves of large and small distortions
on adjacent sites.
}
\end{figure}

\begin{figure}[b]
\centerline{
\includegraphics[width=4cm,height=3.5cm]{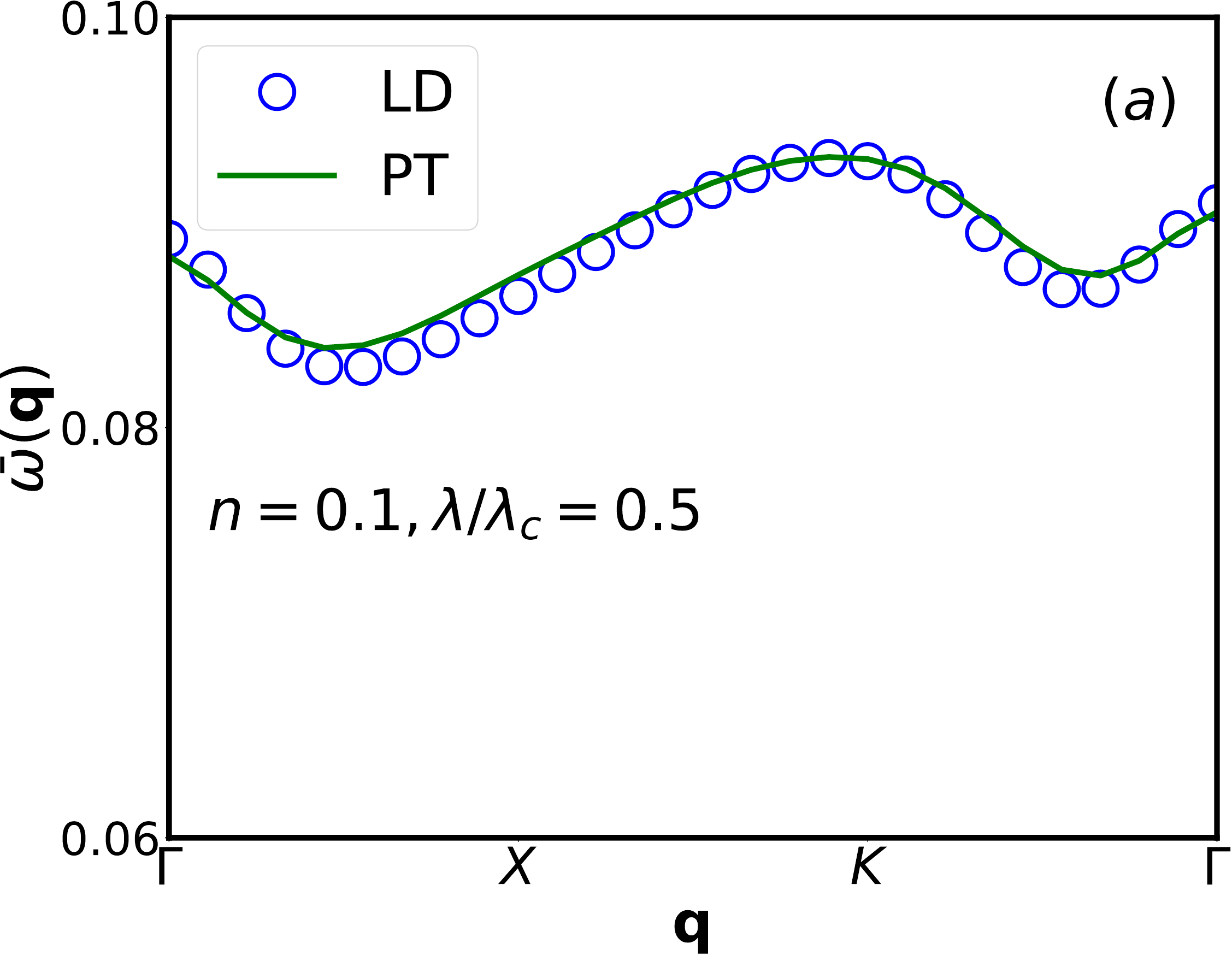}
\includegraphics[width=4cm,height=3.5cm]{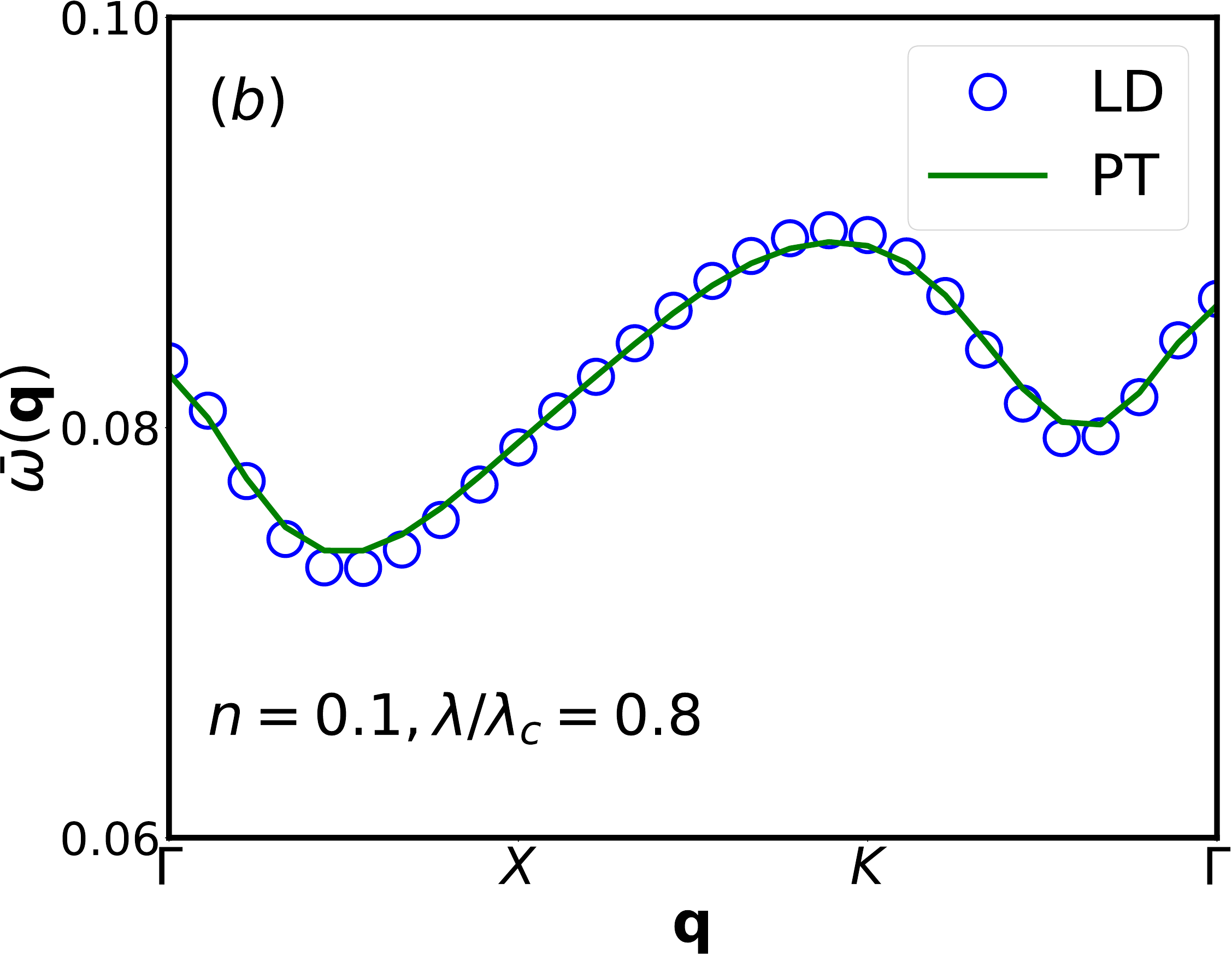}
}
\centerline{
\includegraphics[width=4cm,height=3.5cm]{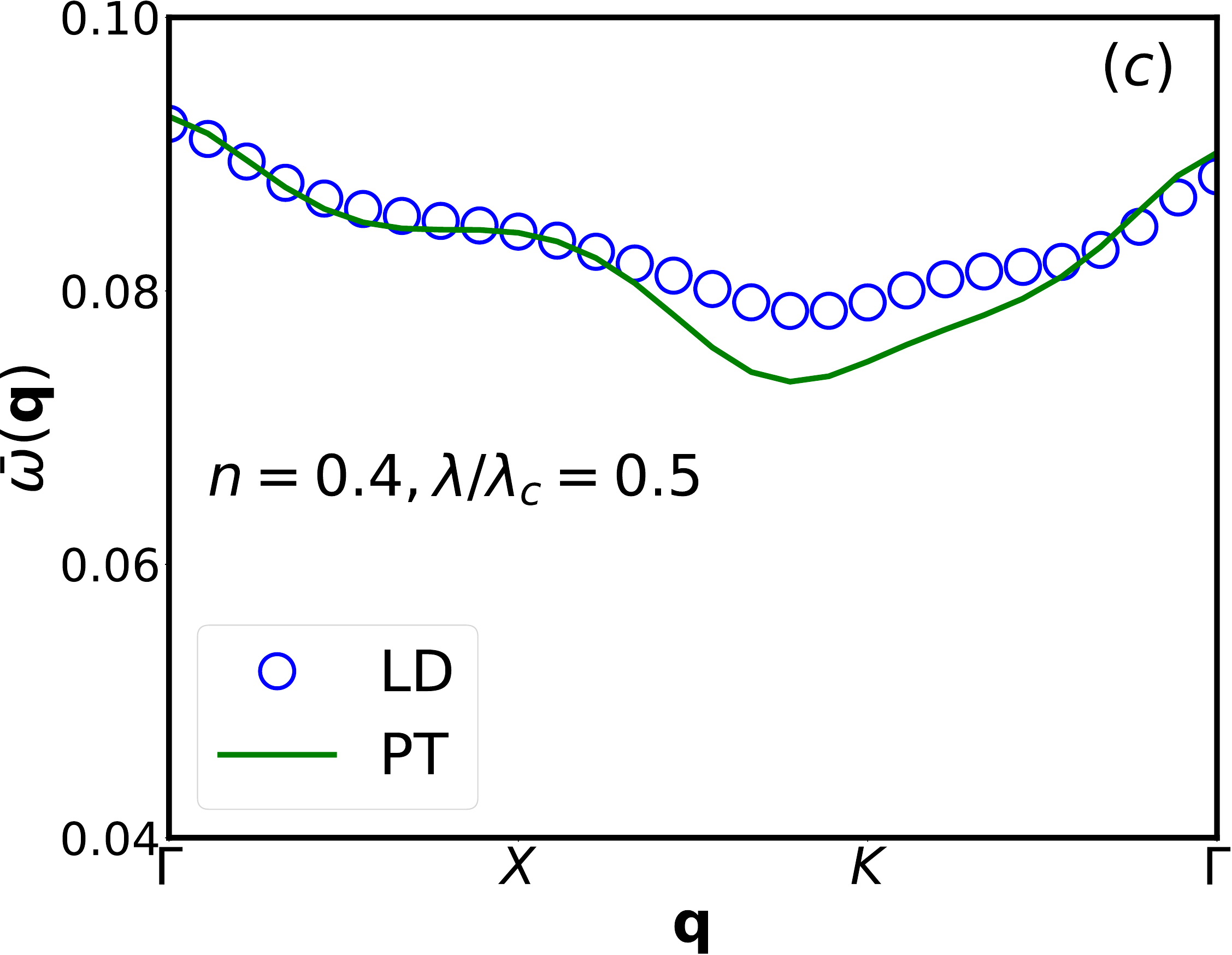}
\includegraphics[width=4cm,height=3.5cm]{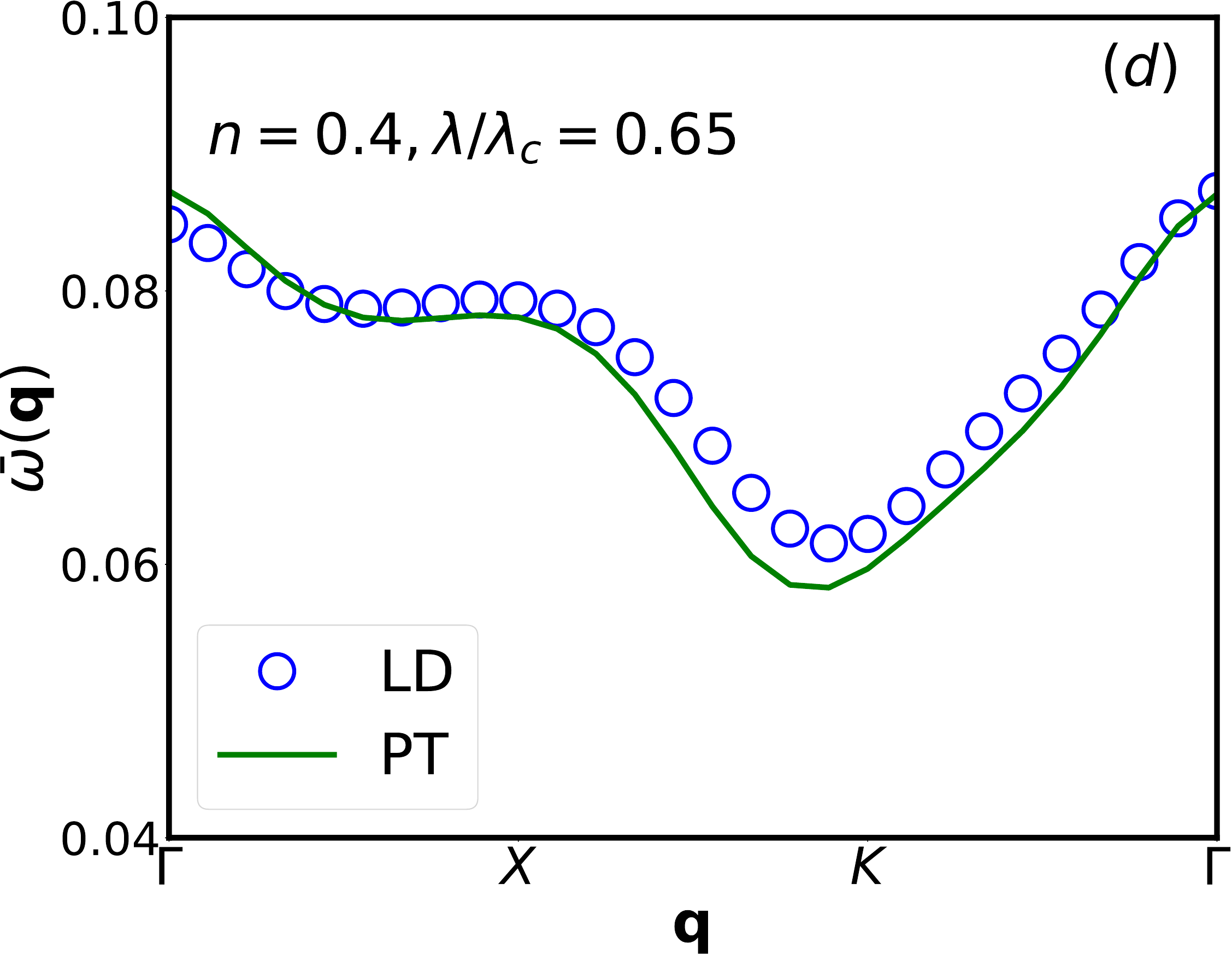}
}
\caption{Phonon dispersions ($\bar{\omega}({\bf q})$) obtained using
Langevin dynamics (LD) and standard perturbation theory (PT),
implemented in momentum space, in the weak coupling scenario.
We show four typical parameter points. The top row features
results in the dilute regime, while the bottom depicts the
correlated case. We observe a quantitative agreement between
the low $T$ LD answer and the perturbative one in all cases.  }
\end{figure}

\begin{figure}[t]
\centerline{
\includegraphics[width=8.8cm,height=3.0cm]{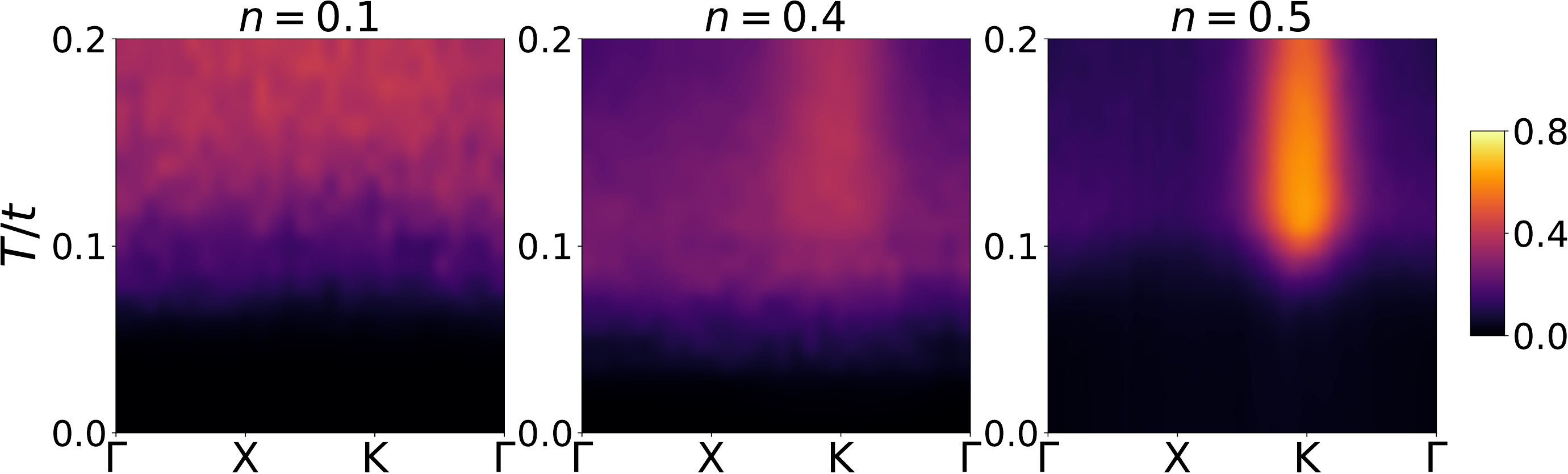}
}
\caption{Maps of low energy weight fraction ($w_{low}$) in
the (${\bf q}, T$) plane for three typical densities-
$n=0.1$ (dilute) , $n=0.4$ (correlated) and $n=0.5$ (ordered).
The x-axis denotes the typical
${\bf q}$ trajectory ($\Gamma-X-K-\Gamma$) and the y-axis
is temperature $T/t$. The parameter points chosen are-
$\lambda/\lambda_{c}=1.2$ for $n=0.1$, $\lambda/\lambda_{c}=2.0$ 
for $n=0.4$, and
$\lambda=0.5$ for $n=0.5$. The crucial feature is the growing
momentum selectivity on moving to higher densities. One
also observes an early onset of increase in $w_{low}$
with respect to $T$ in the correlated case compared
to the dilute and ordered ones.  }
\end{figure}

We comment that similar effects are crucial in other electron
correlation models, like Hubbard, in the strong coupling scenario.
In the auxiliary field language, the `moments' at strong coupling
have a reasonably fixed amplitude. But, large angular fluctuations
dictate the dynamics, instead of `small amplitude' oscillations
as captured within RPA. Hence, the latter method again fails to
describe finite temperature collective mode spectrum.

While it's true that over a wide $T$ window, Langevin approach is
better, the present equation is derived using several drastic
approximations, notably `Ohmic dissipation' and `thermal noise'.
As a consequence, we relegate the phonon variables to the classical
level, which freeze at zero temperature. A more complex, self-consistent
noise and damping could alleviate the issue in principle and
make the method accurate even at low $T$.

\subsection{Low temperature dispersion and perturbation theory}

We compared the phonon dispersions obtained at low $T$ using 
the LD method with a standard perturbative calculation done
in $({\bf q},\omega)$ space in the weak coupling regime. The
results, shown in Figs.14(a)-(d), show a quantitative agreement. The
match is better in the dilute ($n=0.1$) case compared to the
correlated ($n=0.4$) one. We comment that the LD answer contains
the effect of static bare polarizability $\Pi_{0}({\bf q})$, and
not the full frequency dependent $\Pi_{0}({\bf q},\omega)$, as
in the RPA. However, at weak coupling, this distinction is
irrelevant.

\subsection{Anharmonic theory and phonon damping}

Beyond the quadratic phonon theory, one may add cubic 
non-linearities to obtain interaction induced phonon damping.
This is already discussed in our previous paper \cite{sauri}, 
where it's shown that the linear behaviour of phonon linewidth
($\Gamma({\bf q})$) can be explained using a perturbative
approach. The argument doesn't depend on the electron density
or nature of the background state and hence goes through also
in the present case.

\subsection{Low-energy weight in the polaronic phase}

The highlight of the Langevin dynamics method in the strong
coupling phase is the capturing of `thermal tunneling' moves.
These events are rare at low $T$, but become progressively
frequent at intermediate to high temperatures and are crucial
in restoring the `translational symmetry breaking' observed
within the adiabatic limit. The spectral signature, as shown
in detail earlier, is the accumulation of low frequency 
($\omega \ll \Omega$) weight. We choose a cutoff of $0.4\Omega$
and integrate the obtained spectral weight upto this, subsequently
normalizing by the total weight.
As shown in Fig.15, this low energy weight fraction ($w_{low}$)
becomes increasingly momentum
selective as one approaches commensurate filling.
Moreover, one observes (i) quantitatively larger values
in the dilute case compared to the correlated one
beyond a threshold ($T\sim\Omega$), 
related to more room for tunneling, and (ii) a comparatively
earlier onset of increase with respect to $T$ in the correlated
case. The half-filling scenario exhibits a pronounced temperature
sensitivity, owing to critical `domain oscillations'.

\section{Conclusion}

We have studied the phonon dynamics  in the 
two dimensional Holstein model across the polaronic crossover
in three different density regimes: dilute, charge correlated,
and charge ordered. 
We explored two different approaches- 
(i) a generalisation of the traditional RPA to finite temperature,
and (ii)~a Langevin dynamics strategy employing a force
derived from the electronic Hamiltonian.
Both methods reproduce the standard perturbative phonons
at weak coupling.  Approaching the polaronic
transition from this end, the RPA strategy becomes
progressively inaccurate and mainly yields a large 
damping due to the background thermal disorder. 
The Langevin approach, however,
gives more coherent phonons as it takes anharmonic
effects into account correctly.
In the polaronic phase the Langevin approach 
captures the slow polaron tunneling events
and the corresponding low energy spectral weight,
an effect completely missed by the RPA due to its
perturbative `low amplitude' character.
We have quantified these
trends for varying coupling and temperature. 
Our method can handle real lattice geometries, 
strong coupling, finite temperature, and collective effects at
finite electron density, with modest computational effort.
This paper sets the framework based on which
we will discuss inelastic neutron results in 
specific materials in the near future.

{\it Acknowledgement:}
We acknowledge the use of HPC clusters at HRI. SB
acknowledges fruitful discussions with Arijit Dutta and Abhishek Joshi. 
The research of SB was partly
supported by an Infosys scholarship for senior students.
The research of SP was partly supported by the  Olle
Engkvist Byggm\"{a}stare Foundation.

\section{Appendix}

In terms of the fermion eigenvalues and eigenvectors on a given $x_{i0}$ 
background, we may write down the full polarization $\Pi_{ij}(\omega)$ 
as-
\begin{equation}
\Pi_{ij}(\omega)=-\sum_{m}\sum_{n}u_{im}u_{jm}u_{in}u_{jn}\frac{n_{F}
(\epsilon_{m})-n_{F}(\epsilon_{n})}{\omega +i\eta - (\epsilon_{m}-
\epsilon_{n})}
\end{equation}
where the $u$'s are site amplitudes of 
eigenfunctions for a given configuration, $\epsilon$'s being the eigenvalues. 
$n_F$'s are Fermi distribution functions.

This reduces (in the static polarization approximation) to
\begin{equation}
\Pi^{0}_{ij}=\sum_{m}\sum_{n}u_{im}u_{jm}u_{in}u_{jn}\frac{n_{F}
(\epsilon_{m})-n_{F}(\epsilon_{n})}{(\epsilon_{m}-
\epsilon_{n})}
\end{equation}
for $m \neq n$ and\vspace{0.40em}
\begin{equation}
\Pi^{0}_{ij}=-\beta \sum_{m}(u_{im}u_{jm})^{2}n_{F}(\epsilon_{m})
(1 -n_{F}(\epsilon_{m}))
\end{equation}for $m = n$. 

To find the phonon propagator from the full $\Pi_{ij}(\omega)$, 
one has to solve the Dyson's equation 
\begin{equation}
D^{-1}_{ij}(\omega)=D^{-1}_{0,ij}(\omega)+g^{2}\Pi_{ij}(\omega)
\end{equation}
In the approximated theory, one may do a real space diagonalization 
of the matrix-
\begin{equation}
M_{ij} = K\delta_{ij} + {g}^2\Pi_{ij}
\end{equation} to get the phonon eigenmodes. 
The eigenvalues give us the renormalized stiffness constants. Finally, 
in terms of these and the new eigenvectors, the phonon propagator reads-
\begin{equation}
D_{ij} = \sum_{m}V_{im}{V_{jm}}\left(\frac{1}{\omega + i\eta - \xi_{m}} - 
\frac{1}{\omega - i\eta + \xi_{m}}\right)
\end{equation}
where $\xi_{m}$'s are phonon frequencies obtained from the stiffnesses 
and $V$'s denote site amplitudes of the bosonic eigenfunctions. $\eta$ 
is a positive infinitesimal, as usual.

\bibliographystyle{unsrt}

\end{document}